\documentclass[12pt,tightenlines,eqsecnum,floats,showpacs,nofootinbib,amsmath,amssymb,aps,prd]{revtex4-2}

\usepackage{tikz}
\usepackage{graphicx,verbatim}
\usepackage{amsmath}
\usepackage{amsfonts}
\usepackage{amssymb}
\usepackage{psfrag}
\usepackage{accents}
\usepackage{mathrsfs}
\usepackage{enumitem}
\usepackage{hyperref}

\usepackage[title]{appendix}%
\usepackage{multirow}

\usepackage{caption}
\usepackage{subcaption}
\usepackage{placeins}

\usepackage[colorinlistoftodos]{todonotes}
\usepackage{epstopdf}
\usepackage[latin1]{inputenc}
\usepackage{soul}
\usepackage{appendix}
\usepackage{color}

\usepackage{caption}
\def\be{\begin{equation}}
\def\ee{\end{equation}}
\def\ba{\begin{eqnarray}}
\def\ea{\end{eqnarray}}

\def\E{\mathcal{E}}
\def\N{\mathcal{N}}
\def\EIH{{\mathcal{E}_{\IH}}}
\def\Ekerr{\mathcal{E}_{{\rm kerr}}}
\def\Lie{\mathcal{L}}
\def\G{\mathcal{G}}
\def\B{\mathcal{B}}
\def\R{\mathcal{R}}
\def\H{\mathcal{H}}
\def\L{\mathcal{L}}

\def\tq{\tilde{q}}

\def\rmd{\mathrm{d}}

\def\uk{\underline{k}}
\def\ko{\mathring{k}}
\def\uko{\mathring{\underline{k}}}
\def\JIH{J_{\IH}}
\def\RIH{R_{\IH}}

\def\MIH{M_{\IH}}
\def\JDH{J_{{}_{{\mathfrak{D\!h}}}}} %This definition was changed in the revised version
\def\RDH{R_{{}_{{\mathfrak{D\!h}}}}} %This definition was changed in the revised version
\def\MDH{M_{{}_{{\mathfrak{D\!h}}}}} %This definition was changed in the revised version
\def\JNEH{J_{\NEH}}

\def\ub{\underbar}
\def\t{\tilde}
\def\h{\hat}

\def\b{\bar}
\def\ul{\underline}
\def\={\,\hat{=}\,}
\def\f{\frac}

\def\R{\mathcal{R}}

\newcommand{\pullback}[1]{\hbox{\lower0.5ex\hbox{${}_{\leftarrow}$}}\kern-1.9ex{#1}}
\newcommand{\pullbacklong}[1]{\hbox{\lower0.85ex\hbox{${}_{\longleftarrow}$}}\kern-3.0ex{#1}}
\newcommand{\pullbackllong}[1]{\hbox{\lower0.85ex\hbox{${}_{\longleftarrow\!\!-\!\!-\!\!-\!\!-}$}}\kern-6.4ex{#1}}

\newcommand*{\scrip}{\ensuremath{\mathscr{I}^{+}}} 
\newcommand*{\scrim}{\ensuremath{\mathscr{I}^{-}}} 

\def\omegao{\mathring\omega}
\def\qo{\mathring{q}}

\def\vo{\mathring{v}}

\def\ello{\mathring{\ell}}
\def\no{\mathring{n}}
\def\epsilono{\mathring{\epsilon}}
\def\alphao{\mathring{\alpha}}
\def\Omegao{\mathring\Omega}
\def\ko{\mathring{k}}

\def\Do{\mathring{D}}
\def\kappao{\mathring\kappa}
\def\phio{\mathring{\varphi}}
\def\Phio{\mathring\Phi}

\def\to{\mathring{t}}
\def\varphio{\mathring{\varphi}}
\def\Pio{\mathring{\Pi}}
\def\ro{\mathring{r}}
\def\psio{\mathring{\psi}}
\def\epsilono{\mathring{\epsilon}}
\def\omegao{\mathring\omega}
\def\dilaton{\mathring{d}}
\def\etao{\mathring{\eta}}

\def\QLH{\mathfrak{h}}
\def\EH{\mathfrak{eh}}
\def\DH{\mathfrak{Dh}}
\def\KH{\mathfrak{Kh}}
\def\IH{\mathfrak{ih}}
\def\NEH{\mathfrak{nh}}
\def\e{\mathfrak{e}}
\def\n{\mathfrak{n}}

\def\F{\mathfrak{F}}
\def\Fximatt{\mathfrak{F}_{\rm matt}^{{}_{(\xi)}}} %This definition was changed in the revised version
\def\Fxigws{\mathfrak{F}_{\rm gws}^{{}_{(\xi)}}} %This definition was changed in the revised version
\def\Ftgws{\mathfrak{F}_{\rm gws}^{{}_{(t)}}} %This definition was changed in the revised version
\def\Ftmatt{\mathfrak{F}_{\rm matt}^{{}_{(t)}}} %This definition was changed in the revised version
\def\Ftprimegws{\mathfrak{F}_{\rm gws}^{{}_{(t^\prime)}}} %This definition was changed in the revised version
\def\Ftprimematt{\mathfrak{F}_{\rm matt}^{{}_{(t^\prime)}}} %This definition was changed in the revised version

\def\ulxi{\underline{\xi}}
\def\Fulximatt{\mathfrak{F}_{\rm matt}^{{}_{(\ulxi)}}} %This definition was changed in the revised version
\def\Fulxigws{\mathfrak{F}_{\rm gws}^{{}_(\ulxi)}} %This definition was changed in the revised version

\def\q{\mathfrak{q}} %charge

\def\Do{\mathring{D}}

\def\Im{{\rm Im}}

\newcommand\extrafootertext[1]{%
    \bgroup
    \renewcommand\thefootnote{\fnsymbol{footnote}}%
    \renewcommand\thempfootnote{\fnsymbol{mpfootnote}}%
    \footnotetext[0]{#1}%
    \egroup
}%% hack for title page note

\begin{document}

\title[]{Thermodynamics of dynamical black holes\\ beyond perturbation theory}
\author{Abhay Ashtekar}
\affiliation {
Physics Department, Penn State University, University Park, PA 16801}
\author{Daniel E. Paraizo}
\affiliation {
Physics Department, Penn State University, University Park, PA 16801}
\author{Jonathan Shu}
\affiliation{
Physics Department, Penn State University, University Park, PA 16801}

\begin{abstract}
    
The close similarities of the three laws of black hole mechanics, discovered by Bardeen, Carter and Hawking, with the laws of thermodynamics led to the identification of a multiple of the area of the event horizon with entropy. However, developments over the past two decades have shown that this paradigm has some important limitations, especially because of the teleological nature of event horizons. After a brief review of these limitations, we will show that they can be overcome using quasi-local horizons. Specifically, the new first law applies to black holes in general relativity that can be \emph{arbitrarily far from equilibrium} and refers to \emph{finite} changes that occur due to \emph{physical processes} at the horizon. The second law is now a \emph{quantitative} statement that relates the change in the area of a dynamical horizon segment due to fluxes of energy falling into the black hole. Together, they lead one to identify black hole entropy with the area of marginally trapped surfaces in quasi-local horizons, generalizing recent {perturbative} findings that it should be identified not with the area of the event horizon but with the area of a marginally trapped surface inside it.%
\footnote{Dedicated to the memory of Jurek Lewandowski.}
\end{abstract}
%\pacs{04.25.dg, 04.70.Bw, 04.70Dy, 04.25.dg, 04.30.-w, 04.60.-m, 04.60.Pp}

\maketitle
\tableofcontents

\section{Introduction} 
\label{s1}

Over fifty years ago, Bardeen, Carter and Hawking (BCH) showed that black holes in general relativity obey relations that have striking similarities with the laws of thermodynamics \cite{Bardeen:1973gs,LesHouches:1973}. The conceptual economy underlying these laws is remarkable. For the zeroth and the first law, one only uses the Einstein equations satisfied by space-time geometries of stationary, axisymmetric black holes in asymptotically flat space-times. The second law, which had been discovered by Hawking two years earlier, only needed the null energy condition and  asymptotic predictability in addition to the Einstein equations \cite{Hawking:1971vc}. These findings, together with Hawking's subsequent startling discovery that black holes emit quantum radiation \cite{Hawking:1974sw}, have provided an anchor for investigations in quantum gravity in the subsequent decades.

Recall that the zeroth law says that surface gravity $\kappao$ is constant on the horizon of a stationary black hole (BH) just as the temperature $T$ of a body is constant in thermodynamical equilibrium. The first law of thermodynamics $\delta E = T\delta S + p \delta V + \mu \delta N$ constrains the permissible changes in the extensive observables $(E, S, V, N)$ of a system in the passage from one equilibrium state to a nearby equilibrium state. Similarly, the BCH first law,
\be \label{BCH} \delta M = \f{\kappao}{8\pi G}\, \delta A + \Omegao\, \delta J + \Phio\, \delta \q\ee
governs the relation between $(M, A, J, \q)$ of two nearby stationary, axisymmetric BHs in general relativity (GR). Here $M, J$ are the total --or, the  Arnowitt-Deser-Misner (ADM)-- mass and angular momentum of the system, evaluated at spatial infinity. On the other hand, the area $A$, the angular velocity  $\Omegao$ and the electrostatic potential $\Phio$ refer to the horizon. (The electric charge $\q$ can be evaluated either at the horizon or at spatial infinity since it is conserved). Since the zeroth and the BCH first law require stationarity, the horizon in their statements can be taken to be the Killing horizon $\KH$ defined by the time translation isometry. On the other hand, Hawking's second law $\Delta A \ge 0$ applies to BHs that can be far from stationarity, just as the second law $\Delta S \ge 0$ of thermodynamics applies to systems that can be far from equilibrium. Space-times encompassing such dynamical BHs do not admit KHs. The\, $\Delta A$\, that features in the statement of the second law is the change in the area of cross-sections of the \emph{event horizon} (EHs) $\EH$. Since $\EH$ is the future boundary of the causal past of $\scrip$, in contrast to KHs, it is a global notion.
% \cite{Hawking:1971vc}. 

Together, these three laws suggest that a multiple of $\kappao$ could be identified with the temperature of a stationary BH and a corresponding multiple of the horizon area $A$ would then be identified with the BH entropy. It is clear already on dimensional grounds that this identification cannot be made in classical GR. Hawking's detailed calculation of quantum radiation showed how $\hbar$ enters in these relations: $(\hbar/2\pi)\, \kappao$ corresponds to the temperature $T$ and $A/4G\hbar$ to the entropy $S$. Nonetheless, in much of the mathematical physics literature, both in the GR and string theory communities, it has become customary to refer to $A/4G$ as entropy. An example of the former can be found in the very title of some early foundational papers (e.g. \cite{Wald:1993nt,Hawking_1996}) and the usage has continued (e.g.  \cite{Jacobson_1995,Chakraborty_2015,PhysRevD.108.044069,hollands2024entropydynamicalblackholes,visser2025dynamicalentropychargedblack}). Similarly, in the string theory literature  $\kappao/2\pi$  is often referred as the temperature and the $\kappao\rightarrow 0$ limit as  the ``zero temperature limit" (e.g. \cite{Emperan-SGP}). Recently, the gravitational wave community has also used this terminology, and its thermodynamics connotation, to investigate the final stages of BH mergers (e.g., \cite{rinconramirez2026maximumentropyconjectureblack}). To facilitate the exchange of ideas with these diverse communities we will use thermodynamical terminology, in contrast to the ``horizon mechanics" nomenclature used in the literature on quasi-local horizons. 

The diverse developments relating black hole mechanics to thermodynamics are striking. However, as we discuss below, a closer examination reveals some conceptual tensions. They suggest that this traditional formulation of BH thermodynamics can be improved upon, especially in dynamical situations. The purpose of this paper is to present a candidate that is free of the deficiencies of the standard treatment. In this improved paradigm, the three laws of BH thermodynamics are obtained using quasi-local horizons (for a recent review, see \cite{Ashtekar:2025LivRev}). More precisely, KHs $\KH$ are replaced by Isolated Horizon Segments (IHSs)\, $\IH$ \cite{Ashtekar:2000sz} in the zeroth law, as well as in the passive form of the first law. More importantly, there is a  new and active form of the first law involving physical processes across finite segments of Dynamical Horizon Segments (DHSs) $\DH$, and  EHs\, $\EH$ in Hawking's second law are replaced again by DHSs\, $\DH$. Main results were reported in \cite{aps-letter}. In this paper we provide the detailed arguments.  
\vskip-0.7cm

\subsection{Limitations of BCH-type frameworks}
\label{s1.1}
 
Consider first the equilibrium situations that the first law (\ref{BCH}) refers to. As we already noted, the ADM quantities, $M$ and $J$, that appear in the first law are evaluated as 2-sphere integrals at infinity, far from the BH. They represent the \emph{total} mass and angular momentum of the given  space-time, including contributions from what is outside the BH. If one were to consider a space-time in which there is a \emph{physical process} that sends a BH from one equilibrium state to another, the ADM quantities would be absolutely conserved, but the mass and angular momentum of the BH itself would change. Therefore, we need to have expressions of mass and angular momentum that can be calculated directly using just the fields on the horizon, without references to their values at infinity. Indeed, in thermodynamics extensive observables that appear in the first law refer \emph{only} to the system and not what is outside. 

A second drawback of the BCH analysis is that it assumes that the entire space-time is stationary, not just the BH. By contrast, in thermodynamics one only assumes that the system under consideration is in equilibrium; not the whole universe. This suggests that a better correspondence with the zeroth and the first laws of thermodynamics would result if one assumes that only the horizon is isolated --i.e., ensuring that the BH itself is a closed system-- while allowing for radiative processes outside \cite{Ashtekar:2000hw,Ashtekar:2001is,Ashtekar:2004cn}. These situations cannot be incorporated using KHs\, $\KH$. By contrast, IHSs $\IH$ are tailored to incorporate them. Hence they provide a more appropriate setting for the equilibrium forms of the zeroth and the first law.  

Furthermore, it should be possible to establish an active form of the first law, governing the changes in the mass and angular momentum of the black hole due to fluxes across finite portions of the horizon. Using KHs,\, $\KH$, one cannot incorporate such physical processes beyond perturbation theory around stationary space-times . One has to use instead DHSs\, $\DH$,\, across which there can be arbitrarily large fluxes of energy and angular momentum.  
\begin{figure}[]
  \begin{center}
    \includegraphics[width=0.3\columnwidth,angle=0]{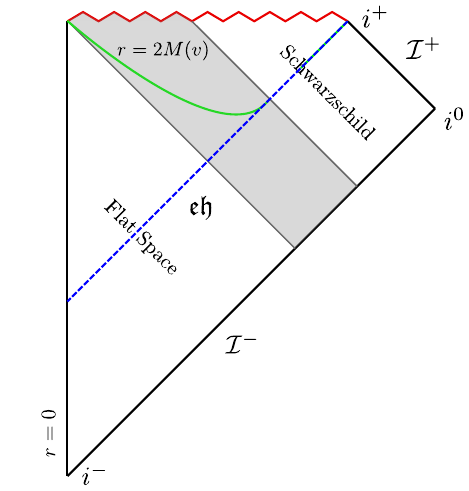} 
  \caption{\footnotesize{\emph{Vaidya Space-time:} Collapse of a null fluid, incident from $\scrim$. To the past of the null fluid, the space-time metric is Minkowskian. The EH $\EH$ forms and grows already in this flat region although there is nothing happening there. To the future of the null fluid, the space-time metric is isometric to the Schwarzschild metric. The $r=2M(v)$ surface is the spherical dynamical horizon segment that forms in the fluid region, grows in area, and joins the EH once all of the null fluid has fallen in. None of it lies in the region where the metric is flat.}}
\end{center}
\label{fig1}
\end{figure}

The third limitation concerns the identification of the entropy of dynamical BHs with the area of their EHs, suggested by the second law, and is far more serious. First, as noted above, the notion of an EH requires the space-time to admit future infinity, $\scrip$. A spatially closed universe, for instance, does not admit EHs but physically there is no reason to preclude dynamical BHs in them. There are also spatially non-compact space-times with positive cosmological constant admitting trapped regions that can be physically thought of as representing BHs. But they do \emph{not admit EHs} because all of $\scrip$ is replaced by a future singularity (although they do admit a complete $\scrim$) \cite{Senovilla:2022bsn}. More generally, to determine if a given space-time contains an EH, one needs to know the metric all the way to the \emph{infinite future}, construct $\scrip$, check that it is complete \cite{Geroch:1978ub}, and examine its past. Therefore, the use of EHs in \emph{dynamical situations} leads to serious difficulties, both at practical and conceptual levels. An illustration of practical difficulties is provided by numerical simulations of binary BH mergers: one cannot use EHs in the course of a numerical evolution to locate the progenitor BHs, nor the remnant; EHs can only be identified at the end of the simulation, as an `afterthought'. In the investigation of conceptual issues  --such as cosmic censorship in classical GR \cite{Andersson:2007fh}, or the endpoint of the evaporation process of BHs in quantum gravity \cite{Ashtekar_2020}-- one cannot make \emph{a priori} assumptions about global properties of space-time that are required to determine if it admits an EH. Indeed, Haji\v{c}ek pointed out already in 1987 that the EH may be completely removed by changes in the space-time geometry in a Planck scale neighborhood of the singularity that could be induced by quantum gravity effects \cite{PhysRevD.36.1065}. Hence, one does not know whether EHs even exist in physically interesting situations \cite{afshordi2024blackholesinside2024}. 

Even when they do exist, EHs have a severe limitation: \emph{they are teleological.} In particular, they can form and \emph{grow} in flat regions of space-time in which nothing at all is happening (see, e.g.,
\cite{Ashtekar:2004cn,kehle2024extremalblackholeformation,afshordi2024blackholesinside2024}). 
This teleology is illustrated in Fig.~1, representing a Vaidya space-time depicting the spherical collapse of a null fluid incident from $\scrim$. In this case, the space-time metric in the triangular region to the past of the infalling null fluid (depicted by the unshaded region) is flat, it becomes dynamical in the shaded region, and finally it is a Schwarzschild metric to the future of that region. Note that the EH $\EH$ \emph{forms and grows} in the flat region --where nothing  is happening-- \emph{in anticipation} of a collapse that is to occur occur sometime in the distant future! 
\footnote{Recent results of Kehle and Unger show that the same phenomenon occurs if one uses Vlasov fluids (that emerge from $i^-$ rather than $\scrim$, see Fig. 1 of \cite{kehle2024extremalblackholeformation}); null fluids incident from $\scrim$ are not essential.} 

Such teleological behavior is normally regarded as `spooky' and is not attributed physical significance. Therefore, from a physical perspective, the growth in area of the EH --and hence the EH itself-- cannot have any entropic connotation in \emph{fully} dynamical situations, such as the one depicted in Fig. 1. What then, should replace EHs in these entropic considerations? We address this question in the next subsection. \vskip0.4cm

\subsection{Overcoming the limitations: Quasi-local horizons} 
\label{s1.2}

Let us begin by noting that a host of difficulties immediately arises when considering fully dynamical situations in place of stationary axisymmetric ones. First, we no longer have Killing fields. What would then be the analogs of the extensive observables $M$ and $J$ that feature in the first law (\ref{BCH})? Since they have to refer only to the individual BHs in any given space-time, they cannot be the ADM quantities. A lack of obvious candidates is widely regarded as a key difficulty in incorporating dynamical situations beyond perturbation theory around stationary BHs (see, e.g., \cite{hollands2024entropydynamicalblackholes}). The absence of direct analogs of intensive parameters $\kappao$ and $\Omegao$ seems even more serious. Indeed, already for familiar thermodynamical systems, while extensive observables such as energy and volume are generally well-defined away for equilibrium, we do not have globally defined intensive parameters such as $(T,\,p,\,\mu)$. Severe as they seem, we will be able to surmount these obstacles using the known results on quasi-local horizons and black hole uniqueness theorems. A key element is the introduction of a new conceptual arena that intertwines dynamical black holes with those in equilibrium, enabling one to assign intensive parameters to BHs that are far from equilibrium.

The underlying ideas can be summarized as follows. Quasi-local horizons can be located without reference to $\scrip$, knowing space-time geometry only in their neighborhoods. In this sense, they are free of teleology. In the quasi-local paradigm, BHs in equilibrium are represented by IHSs\, $\IH$. While KHs\,$\KH$ constitute an important special case, isolated horizon segments can exist in space-times without any Killing vector. As mentioned above, they encapsulate the idea that the BH itself is isolated and in equilibrium, allowing for radiation and dynamical processes elsewhere. Nonetheless, one can use results on IHSs \cite{Ashtekar:2000sz}  to show that the zeroth and the first laws of BH mechanics admit a natural generalization to IHSs. In this extension, the ADM observables $(M, J)$ in (\ref{BCH}) that refer to the global Killing fields $\to^a,\, \varphio^a$ are replaced by the $\IH$-observables $(\MIH, \JIH)$, that refer to the $\IH$-symmetry group $\G$ --the analog of the BMS group at $\scrip$-- and are defined using only local fields on $\IH$. More importantly, one can also extend  black hole mechanics to fully dynamical situations in GR in which BHs are represented by {DHSs}\, $\DH$ \cite{Ashtekar:2025LivRev}. Their geometry is time-dependent because there are fluxes of energy and angular momentum across them. Our first law for DHSs will refer to changes that extensive observables associated with $\DH$ undergo in response to these physical fluxes. Key differences from the BCH first law are the following:\vskip0.05cm 
\noindent(i) Extensive observables are defined \emph{at} $\DH$, without reference to infinity. Indeed, the ADM mass and angular momentum cannot feature in the generalization of the first law to dynamical situations, first because they are absolutely conserved, and second because changes should refer to values of these observables associated with individual BHs without reference to other far away BHs, stars, and radiation in the space-time.\\  
(ii) The changes in physical observables are \emph{finite}, rather than infinitesimal; and, \\
(iii) These changes are caused by \emph{physical processes in space-time}, not results of infinitesimal displacements from one stationary BH to another in the space of solutions.\vskip0.05cm
\noindent Nonetheless, if one restricts oneself to infinitesimal processes on a dynamical horizon segment, one recovers the first law in its familiar BCH form (but changes in values of extensive observables are again due to physical processes at the horizon). Similarly, the second law of DHSs is a quantitative relation between the change in the area of the DHS and the \emph{local} energy flux across it \cite{Ashtekar:2003hk}. It holds both for classical processes by which BHs form and grow and for the quantum process by which they evaporate. It is not just a qualitative statement $\Delta A \ge 0$,\, nor a teleological effect, as with EHs. The two laws imply that in dynamical situations \emph{BH entropy should be associated with the area of cross-sections of DHSs rather than cross-sections of EHs.} 
 
\vskip0.4cm

\subsection{Organization}
\label{s1.3}

We have attempted to make this paper reasonably self-contained. It is organized as follows. Since the literature on quasi-local horizons is copious, for convenience of the reader, we recall the basic underlying notions in section \ref{s2}. In section \ref{s3} we first summarize the relevant features of IHSs, and then use them to generalize the zeroth and the first law from KHs $\KH$ to general IHSs $\IH$. In section \ref{s4} we recall salient properties of DHSs that will be directly useful to our subsequent analysis. Next we turn to thermodynamical considerations for systems that may be far from being in equilibrium. Even for simple systems, one encounters certain difficulties in passing from the well understood equilibrium situations to non-equilibrium ones. In particular it is not obvious how to assign intensive thermodynamic parameters $(T,\, p,\,\mu)$ to non-equilibrium states. In section \ref{s5.1} we examine these difficulties in the context of BHs in GR and provide a strategy to overcome them by introducing a canonical projection map from the space $\N$ of non-equilibrium states to the space $\E_{\rm kerr}$ of equilibrium states, made possible by the black hole uniqueness theorems. This map enables one to assign intensive thermodynamical parameters $(\kappao,\,\Omegao)$ to non-equilibrium states, which are now \emph{time dependent} just as one would expect. However, for standard thermodynamical systems one has a natural notion of energy even in fully dynamical situations, because the underlying space-time has a time translation isometry. We no longer have this isometry in dynamical situations in GR. Therefore, to extend the first law to dynamical BHs, extra input is needed to speak of an `energy-charge' and an `energy-flux' at the horizon. We complete this task in section \ref{s5.2} using the machinery set up in sections \ref{s2} and \ref{s4}.  

In section \ref{s6.1} we use these charges and fluxes to extend the standard first law to BHs that can be \emph{far from equilibrium}. For completeness, in section \ref{s6.2}, we recall the quantitative form of the second law from \cite{Ashtekar:2002ag,Ashtekar:2003hk}. It directly relates the monotonic change in the area of DHSs to the local flux of energy influx into them.% 
\footnote{The BCH paper was entitled  \emph{``The four laws of BH mechanics",} while in this paper we discuss only the first three. It was recently shown that there is a precise sense in which the fourth of the BCH laws --the analog of the third law of thermodynamics-- does not hold in GR \cite{kehle2024extremalblackholeformation}. However, this law has not played a notable role in the extensive literature at the interface of BHs and quantum physics. Furthermore, it has been argued \cite{Emperan-SGP} that: (i) extremal BHs are quantum mechanically unstable; (ii) the limits $\hbar \rightarrow 0$ and ${\rm Temp} \rightarrow 0$ do not commute; and, (iii) since Nature is inherently quantum mechanical, the extremal BHs are not physically significant.}
In section \ref{s6.3}, we discuss the recent perturbative results from the QLH perspective. Up to this point we will have focused on space-like dynamical horizon segments for simplicity of presentation. We discuss the time-like case (that is directly relevant to the quantum evaporation process) in  section \ref{s6.4}. Section \ref{s7} summarizes the main results emphasizing assumptions, compares them to those in the extensive literature on quasi-local horizons, and suggests some directions for future work. 

For completeness, we have added two appendices. In the first we summarize the procedure, based on results of \cite{Korzynski:2007hu,akkl1,Ashtekar:2021kqj}, that enables one to define angular momentum on quasi-local horizons even in the absence of a rotational symmetry. Appendix \ref{a2} addresses a different issue. In the main body of the paper we use phase space methods largely to make contact with the recent literature on BH thermodynamics. On the other hand, the BCH analysis was based on symmetries and the associated (Komar) charges that are conserved due to field equations, without any reference to Hamiltonian frameworks. In Appendix \ref{a2} we present an alternate derivation of the first law using a similar approach based on field equations. This formulation will be useful for mathematical relativists with expertise on horizons and field equations, who may not be as familiar with the intricacies associated with phase spaces of GR.

Our conventions are the following. We work in 4 space-time dimensions with metrics $g_{ab}$ of signature -,+,+,+. For simplicity all fields will be assumed to be smooth (although our results will go through if $g_{ab}$ is only $C^3$).
The torsion-free derivative operator compatible with the metric is denoted by $\nabla$ and curvature tensors are defined by $R_{abc}{}^d\, V_d := 2 \nabla_{[a} \nabla_{b]}\, V_c; \, R_{ac} :=  R_{abc}{}^{b}$ and $R= g^{ab}\, R_{ab}$.  The symbol $\=$ is used to denote equality that holds only at points of the given quasi-local horizon.  In this paper we wish to focus on new conceptual ingredients. Therefore inclusion of charge and cosmological constant will be discussed elsewhere. 
\vskip0.4cm

\section{Quasi-local horizons: A brief overview}
\label{s2}

In contrast to EHs, Quasi-local horizons (QLHs) \cite{Hayward:1993wb,Ashtekar:2000sz,Ashtekar:2000hw,Ashtekar:2001is,Ashtekar:2002ag,Ashtekar:2003hk,Ashtekar:2004cn,Booth:2005qc,Gourgoulhon:2005ng,Hayward:2009ji,Jaramillo:2011zw,Ashtekar:2025LivRev}) do not refer to $\scrip$. Hence one does not need to know the space-time geometry all the way to the infinite future to locate them. Since they refer only to space-time geometry in their immediate vicinity, they are free of teleology. Therefore QLHs are routinely used in numerical simulations of BH mergers to locate BHs, to study  their evolution, and to investigate correlations between happenings at horizons and fluxes at $\scrip$, especially in the postmerger phase \cite{Prasad:2020xgr,Prasad:2021dfr,PhysRevD.100.084044,Chen_2022,Khera:2023oyf,RibesMetidieri:2024tpk,aa-nk,Ashtekar:2025LivRev}. Their properties have been investigated using geometric analysis (see, e.g.,\cite{Ashtekar:2005ez,Bartnik:2005qj,Andersson:2005gq,Andersson:2007fh,Booth_2007,Senovilla_2011,Bengtsson_2011,senovilla2012stabilityoperatormotscore}) 
% Here Booth_2007 is new; added because of referee's comment.
as well as numerical methods (see, e.g., \cite{Pook-Kolb:2018igu,PhysRevLett.123.171102,pookkolb2020horizonsbinaryblackhole,Mourier:2020mwa,PhysRevD.100.084044,Booth:2021sow,Pook-Kolb:2021gsh,Pook-Kolb:2021jpd}). They have also been used to analyze the BH evaporation process (see, e.g.,
\cite{Ashtekar:2005cj,Sawayama:2005mw,Ashtekar_2011a,Ashtekar_2011b,ashtekar2023regularblackholesloop,Agullo_2024,Varadarajan:2024clw}), where one finds that what forms in a gravitational collapse and evaporates due to quantum radiation are QLHs.
For convenience of the reader, in this section we recall the basic definitions that are used in the rest of the paper. Table I summarizes various notions and symbols used throughout the paper.\vskip0.1cm

Let us fix a 4-dimensional globally hyperbolic space-time $(M, g_{ab})$. Every 2-dimensional compact sub-manifold $S$ of $M$ admits precisely two null normal directions, whence we can choose \emph{future pointing} null normals $k^a$ and $\uk^a$ to $S$ such that $k^a\uk_a =-1$.  This normalization still allows rescalings $k^a\, \rightarrow\, k^{\prime\,a} = e^\alpha k^a,\,\, \uk^a \rightarrow\, \uk^{\prime\,a} = e^{-\alpha}\,\uk^a$, where $\alpha$ is a real function. $S$ is said to be a \emph{marginally trapped surface} (MTS) if the expansion of one of the null normals, say $k^a$, vanishes, i.e. if $\theta_{(k)} := \tq^{ab} \nabla_a k_b =0$, where $\tq^{ab}$ is the projection operator into $S$ and $\nabla_a$ is the derivative operator defined by $g_{ab}$. (Note that if $\theta_{(k)}$ vanishes on $S$, so does $\theta_{(k^\prime)}$.) An MTS $S$ is said to be \emph{strictly stable} \cite{Ashtekar:2025LivRev} (or, \emph{strictly stably outermost} \cite{Andersson:2007fh}) if there exists a space-like vector field $s^a$ defined on $S$ with $s_ak^a >0$ such that a  displacement along $s^a$ makes $S$ untrapped. In the simplest example --the Schwarzschild-Kruskal space-time--  the two future horizons that enclose the black hole singularity are strictly stable, while the two past horizons that enclose the white hole singularity are not. Since our focus is on black hole thermodynamics, unless otherwise stated, \emph{we will assume that the MTSs $S$ under consideration are strictly stable. 
\footnote{Whether a given MTS $S$ is strictly stable is determined just by space-time geometry in an infinitesimal neighborhood of $S$. It is a geometric notion, not directly related to the notion of stability under perturbations, that involves field equations. Nonetheless, in Reissner-Nordstr\"om and Kerr space-times, MTSs on inner horizons fail to be strictly stable, and are therefore excluded from our considerations.} 
}

With this concept at hand, we can introduce our basic notions:
\vskip0.1cm
\emph{Definition 1:} A QLH $\QLH$ is a 3-dimensional sub-manifold of $(M, g_{ab})$ that admits a foliation by a 1-parameter family of MTSs. 
\vskip0.05cm
These horizons are `quasi-local' rather than local because the definition requires that $\theta_{(k)}$ should vanish on \emph{entire} compact surfaces $S$; the knowledge of the metric $g_{ab}$ in an open neighborhood of a point does not suffice to ascertain that there is a quasi local horizon passing through it. Since they are world-tubes of MTSs, QLHs have also been called \emph{marginally trapped tubes} (MTTs) in the literature \cite{Ashtekar:2005ez,Booth:2005qc,Andersson:2007fh,Bengtsson_2011,Senovilla_2011,senovilla2012stabilityoperatormotscore}. 
\begin{figure}[] 
\vskip-0.5cm
 \includegraphics[width=2.0in,height=2.2in]{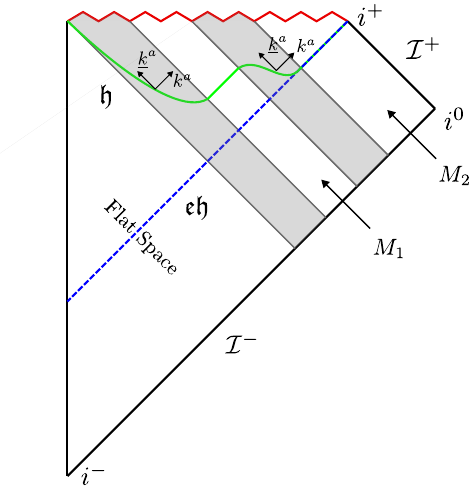}
 \hskip2.5cm 
 \raisebox{0.25\height}
  {\includegraphics[width=2.7in,height=1.5in,angle=0]{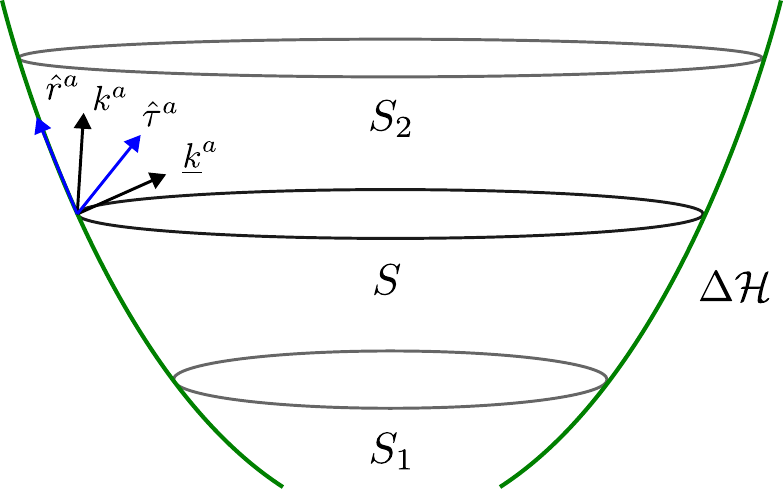}} 
  \caption{\footnotesize{\emph{Left Panel: A double Vaidya null fluid collapse.}\, Again, the event horizon $\EH$ forms and grows in the flat region of space-time. The QLH (in green) lies entirely in the curved region. It starts out as a (space-like) DHS that grows in area in response to infalling matter, then settles down to a (null) IHS in the space-time region $M_1$, only to become a DHS because of the second infall and finally settles to an IHS in $M_2$, the only segment that coincides with the EH. Our thermodynamic considerations apply to all segments of the QLH.\,\,\, {\emph{Right Panel: A DHS.}}\, The portion $\Delta \mathcal{H}$ of a space-like DHS is bounded by two MTSs $S_1$ and $S_2$. $k^a,\,\uk^a$ are the two future directed null normals to MTSs $S$, with $\theta_{(k)}=0$. $\h\tau^a$ is the unit normal to the DHS and $\h{r}^a$ is the unit normal to MTSs within the DHS.}} 
\vskip-0.5cm
 \label{fig:2} 
\end{figure}

As the left panel of Fig. 2 illustrates, a BH can have several evolutionary phases, in some of which it is in equilibrium, and in others, dynamical (because of influx of matter or gravitational waves). DHSs constitute the segments of the QLH across which there is an energy flux.
\vskip0.05cm
\emph{Definition 2:} An open connected component $\DH$ of a QLH $\QLH$ is said to be a \emph{dynamical horizon segment} (DHS) if\\
(i) $\DH$ is nowhere null. Being connected, $\DH$ is either a space-like or a time-like component of the QLH $\QLH$;\\
(ii) The expansion $\theta_{(\uk)}$ of the other null normal is nowhere zero. Again, since $\DH$ is  connected, $\theta_{(\uk)}$ is either positive or negative on each DHS; and\\
(iii) $\mathcal{E} := \t{q}^{ac}\,\t{q}^{bd}\,\sigma_{ab} \sigma_{cd} + R_{ab} k^a k^b$ does not vanish identically on any MTS $S$ in $\DH$. Here $\sigma_{ab}= {\rm TF}\,\t{q}_a{}^c \t{q}_b{}^d\, \nabla_c k_d$ is the shear of the null normal $k^a$ evaluated on $S$, and ${\rm TF}$ stands for `trace-free part'.\\ 
For motivation behind these conditions, see \cite{Ashtekar:2025LivRev}. \vskip0.05cm
\noindent The right panel of Fig. 2 depicts a space-like DHS, foliated by MTSs $S$. In our discussion of the generalization of the first law to dynamical situations, we will consider portions $\Delta \H$ of DHSs, bounded by two marginally trapped surfaces $S_1, S_2$. Changes in values of energy and angular momentum charges associated with $S_1$ and $S_2$ will be related to fluxes of these quantities across $\Delta \H$, carried by matter fields and gravitational waves in physical processes.
\vskip0.1cm

\emph{Remark:}\, The quantity $\mathcal{E}$ can be thought of as a `null energy flux' crossing the QLH, defined by the vector field $k^a$ (that includes both matter and gravitational wave contributions). Conditions (i) - (iii) have been chosen to incorporate the most common situations one encounters in practice both in classical GR and quantum investigations (although  in BH mergers one does encounter very short periods near the merger during which the QLH violates these conditions \cite{Ashtekar:2025LivRev,pookkolb2020horizonsbinaryblackhole,Mourier:2020mwa,PhysRevD.100.084044,Booth:2021sow,Pook-Kolb:2021gsh,Pook-Kolb:2021jpd}). The third condition eliminates \cite{Ashtekar:2005ez} some exceptional situations with high degree of symmetries where the first two conditions are satisfied but space-time does not admit any trapped region \cite{Senovilla_2003} (and hence there is no BH in a physically reasonable sense). Finally, since EHs are null, a DHS cannot be a portion of an EH. Since a DHS is foliated by MTSs, in classical GR it lies inside the EH (if there is one), and can approach it, e.g., when a dynamical BH reaches its final equilibrium state. 
\vskip0.1cm
 
Intuitively, the third condition implies that there is a non-vanishing flux of energy across every DHS. When this flux vanishes, the BH enters an equilibrium phase (which can be temporary; see the left panel of Fig. 2). To distinguish between equilibrium and dynamical settings, it will be convenient to use a\, $\mathring{}$\, over equilibrium quantities. In particular, we will denote the null normals to MTSs during equilibrium phases by $\ko^a$ and $\uko^a$ (and continue to use $k^a$ and $\uk^a$ for null normals during dynamical phases).

\setlength{\tabcolsep}{5pt}
\renewcommand{\arraystretch}{1.3}
\begin{table}
    \centering
    \begin{tabular}{|c|c|c|c|c|}
         \hline \textit{Acronym}& \textit{Notation} & \textit{Causal Nature} & \textit{Primary Fields} & \textit{Defining Properties} 
         \\\hline\hline
         QLH& $\mathfrak{h}$ & General & $\ko^a, \uko^a, \qo_{ab}, \omegao_a$ & Foliated by MTSs ($\theta_k=0$)  \\\hline
         DHS& $\mathfrak{Dh}$ & Nowhere Null & $\h\tau^a,\,\h r^a,\,q_{ab},\,\t\omega_{a},\,\t\sigma_{ab},\,\t\zeta^a$ & $\theta_{(\uk)}\neq0$, $\mathcal{E}\not\equiv0$  \\\hline \hline
         NEHS& $\mathfrak{nh}$ & Null &$\{\ko^a,\, \uko^a\},\,\,\qo_{ab},\,\{\omegao_a\}$ & $\ko^a$ normal, $R_a{}{^b}\ko^a\propto \ko^b$  \\\hline
         IHS& $\mathfrak{ih}$ & Null & $\ko^a,\,\uko^a,\,\qo_{ab}$,\,$\omegao_a$ & $[\Lie_{\ko},\Do_a]v^b \,\hat{=}\,0,\,\,\, \forall\,v^b$ tangent to $\IH$
         \\\hline
         KH& $\mathfrak{Kh}$ &Null  & $\ko^a=\to^a+\Omegao\varphio^a$ & $\nabla_{(a}\ko_{b)}=0$  \\\hline
    \end{tabular}   
    \caption{\footnotesize{\emph{Various Horizons:} A QLH incorporates both equilibrium and non-equilibrium phases of BHs, while a DHS represents a non-equilibrium phase. In both cases, during its dynamical phase, the BH can be arbitrarily far from being in equilibrium. The NEHSs, IHSs and KHs represent equilibrium phases in an increasingly stronger sense. With an NEHS or an IHS, there can be dynamical processes and radiation in space-time but the BH itself is isolated and in equilibrium. A KH requires a stationary space-time and represents a BH in \textit{global} equilibrium.  As for defining properties, the one for QLHs is shared by all subsequent cases; they are all QLHs. Similarly, every  IHS is a NEHS and therefore only the additional properties are given explicitly. While in the first four cases, all conditions are  required to hold only to the respective horizon, for KHs, $\ko^a$ is a global Killing field in space-time.}}
\label{table}
\end{table}

\emph{Definition 3:} A connected component $\NEH$ of a QLH $\QLH$ is said to constitute a \emph{non-expanding horizon Segment} (NEHS) of $\QLH$ if:\\ (i) it is null, with $\ko^a$ as a normal; and,\\ 
(ii) the Ricci tensor of the 4-metric $g_{ab}$ satisfies $R_a{}^b \ko^a\, \propto\, \ko^b$ on the QLH. \\
(Given (i), condition (ii) is implied by the Dominant Energy Condition.) \vskip0.05cm

Via the Raychaudhuri equation, the conditions in \emph{Definition 3} imply that $\mathcal{E} \=0$ on every $\NEH$, distinguishing NEHSs from DHSs. Since $\NEH$ is null, the `metric' $\qo_{ab}$ induced on it by $g_{ab}$ is degenerate and satisfies $\Lie_{\ko} \qo_{ab} \= 0$. Therefore the area of any cross-section of an $\NEH$ is the same.  This property is shared by \emph{all} null normals $\ko^a$ to $\NEH$. Even though $\qo_{ab}$ is degenerate, every $\NEH$ inherits from the space-time derivative operator $\nabla$ a \emph{canonical} intrinsic derivative operator $\Do$ \cite{Ashtekar:2000hw}. The triplet $(\ko^a,\qo_{ab}, \Do)$ is said to constitute the NEH geometry. Although $\Lie_{\ko} \qo_{ab} \=0$, on a general $\NEH$, the derivative operator $\Do$ is not Lie-dragged by $\ko^a$ (because $\Do$ is not uniquely picked out by the condition $\Do_a \qo_{ab} \=0$ since $\qo_{ab}$ is degenerate). Thus, generically, $\ko^a$ is not a symmetry of the NEH geometry. \vskip0.1cm

\emph{Definition 4:} An NEHS is said to be an \emph{isolated horizon segment} (IHS) if it admits a null normal $\ko^a$ that is a symmetry of the NEH geometry, i.e., also satisfies $[\Lie_{\ko},\, \Do_a]v^b \=0$ for all vector fields $v^b$ tangential to the NEHS.
\vskip0.05cm

As noted in section \ref{s1}, IHSs will be denoted by $\IH$. They can be open or closed sub-manifolds of $M$. Every KH\, $\KH$\, is, in particular, an\, $\IH$, but an\, $\IH$ need not be a\, $\KH$. For example, the Robinson-Trautman solution admits an $\IH$ \cite{pc,Podolsky:2009an}; \emph{although there is gravitational radiation in every neighborhood of this $\IH$, it grazes along and none of it traverses $\IH$}. An analysis using the initial value problem based on two null surfaces shows that there is an infinite-dimensional class of other examples \cite{Lewandowski:1999zs}. (This generality is illustrated by the fact that while the Newman-Penrose component $\Psi_4$ of the Weyl tensor is necessarily `time-independent' on any $\KH$, it can have an arbitrary `time-dependence' on a generic $\IH$.)

In the evolutionary history of the BH, IHSs represent phases in which the BH itself is isolated and in equilibrium, allowing for  fully dynamical processes and radiation in the rest of the space-time. By contrast, such processes are \emph{not} allowed in paradigms that use KHs of stationary space-times. For example, the BCH framework assumes that the entire space-time under consideration is stationary. Other discussions of the first law are anchored in Hamiltonian frameworks. Typically, they allow for radiation on stationary backgrounds but only perturbatively (see e.g. \cite{wald2001}). Moreover, they assume that the background space-time admits a bifurcate horizon. As the simple example depicted in the left panel of Fig 2 (and its more sophisticated variants) show, although fully dynamical BH space-times can admit IHSs, they are unlikely to admit bifurcate horizons. In spite of this generality of IHSs, as we discuss in sections \ref{s3.1} and \ref{s3.3}, the zeroth law and the standard infinitesimal version of the first law extend to IHSs. Thus, for these laws one does not need KHs; \emph{it suffices that the BH itself is isolated,} just as in ordinary thermodynamics it suffices that only the system under consideration is in equilibrium --and not the whole universe. 
\vskip0.1cm

\emph{Remarks:} \vskip0.05cm

1. Note that QLHS $\QLH$ are 3-dimensional sub-manifolds of space-time in their own right, satisfying certain conditions on fields defined on them intrinsically. In numerical relativity, in practice they are located using a space-time foliation and finding MTSs on each partial Cauchy surface of this foliation. However, although important, this is only a tool to find them. Neither the notion of a QLH nor its properties refer to these space-time foliations, or indeed any structure away from the 3-manifold $\QLH$.\vskip0.05cm

2. Recall that the null normals to our MTSs are \emph{future} directed. On KHs of the Schwarzschild-Kruskal space-time, the static Killing field $\to^a$ is generally taken to be future directed on KHs of the right region. It is then past-directed on left KHs. Because of the `strict stability' requirement on MTSs we imposed in this paper, only the future KHs in both regions qualify as IHSs, and the vector field $\ko^a$ is future directed and $\kappao$ is positive on both of them.

3. Vaidya space-time is known to admit MTSs that partly lie in the flat regions of space-time (see, e.g., \cite{Schnetter:2005ea}). This feature is considered as surprising (or even mildly disturbing) in some of the literature.  But note that even Minkowski space-time admits 2-spheres on parts of which $\theta_{(k)}$ vanishes,  and one is not surprised by this. As remarked immediately after \emph{Definition 1}, the notion of a MTS is only `\emph{quasi}-local' in that $\theta_{(k)}$ has to vanish on an \emph{entire} 2-sphere. The key point is that Vaidya space-time does not admit any (marginally) trapped surface $S$ that lie entirely in the flat region of space-time. By contrast, as we saw, there are many 2-sphere cross-sections of the EH $\EH$ that do.

\section{Mechanics of Isolated Horizon Segments.}
\label{s3}

As we just discussed, IHSs represent BHs that are themselves in equilibrium, while allowing for dynamical processes arbitrarily close to them. In the first part of this section we provide a brief summary of key properties of IHSs that will be used in our thermodynamic considerations, in particular recalling from \cite{Ashtekar:2000hw} that the zeroth law holds on IHSs even when they are not KHs. 

The first law requires a notion of the horizon angular momentum.  As we discuss in Appendix \ref{a1}, this notion is well-defined on all IHSs (indeed, on all NEHSs) even when the intrinsic metric is not axisymmetric. However, that construction involves a long detour. Therefore, for simplicity of presentation, in the second part of this section we restrict ourselves to those IHSs $\IH$ that admit a rotational vector field $\varphio^a$ as a symmetry of the IHS geometry $(\ko^a, \qo_{ab}, \Do)$ and discuss angular momentum. (Note that the space-time metric $g_{ab}$ need not admit a rotational Killing vector even in a neighborhood of $\IH$.)

In the third part we extend the first law to IHSs. Using the notion of angular momentum from Appendix \ref{a1}, mentioned above, our discussion of the first law extends to general IHSs,  modulo the caveat discussed in Remark 2 of section \ref{s3.1}. It paves the way to generalize the first law to fully dynamical situations, discussed in section \ref{s6}.

\subsection{Salient Properties of IHSs, Including the Zeroth Law}
\label{s3.1}

We will begin with NEHSs $\NEH$ that are more general than IHSs, and then specialize to IHSs. (For details, see \cite{Ashtekar:2000hw}.) Every $\NEH$ has the following properties: \vskip0.1cm
\noindent(i) The pullback to any cross-section of the NEHS of $\qo_{ab}$ provides the intrinsic positive definite 2-metric $\t{q}_{ab}$ thereon.  On every NEHS there is a well-defined 2-form $\epsilono_{ab}$ satisfying $\epsilono_{ab} \ko^b \=0$, whose pull-back to any cross-section provides the area 2-form $\t\epsilon_{ab}$ compatible with $\t{q}_{ab}$. \\
(ii) Because an NEHS is a null surface, any null normal to it satisfies $\ko^a \nabla_a \ko^b \= \kappao\, \ko^b$. The acceleration $\kappao$ is referred to as \emph{surface gravity}.\\  
(iii) While the definition only asks that a QLH should admit a foliation by MTSs, \emph{every} cross-section of an NEHS is both expansion-free and shear-free. \\ 
(iv) The space-time Ricci tensor satisfies $R_{ab}\ko^a v^b \= 0$, for all vector fields $v^a$ tangential to the NEHS. Therefore by Einstein's equations $T_{ab} \ko^a v^b \=0$: there are no matter fluxes across NEHSs. Since $\sigma_{ab}$ also vanishes on each MTS, it follows that there is no flux of gravitational energy either.\\
(v) The Weyl tensor satisfies $C_{abcd}\, \ko^a v_1^b \ko^c v_2^d \=0$ for all vector fields $v_1^a, v_2^b$ tangential to the NEHS. Hence $\ko^a$ is a repeated principal null direction of the Weyl tensor and, in the Newman Penrose notation, $\Psi_0 \= \Psi_1 \=0$ and $\Psi_2\, \=\, \f{1}{2}\, \big(C_{abcd} + i {}^\star C_{abcd}\big)\, \ko^a \uko^b \ko^c \uko^d$ is gauge invariant, i.e. insensitive to the the choice of the cross-section used to define $\uko^b$, as well as the rescaling freedom in $(\ko^a, \uko^a)$. \\ 
(vi) The canonical derivative operator on the NEHS satisfies $\Do_a \ko^b\, \=\, \omegao_a \ko^b$. The surface gravity can be expressed in terms of this 1-form as $\kappao \,\=\, \omegao_a\,\ko^a$. Under rescalings $\ko^a \mapsto e^{\alpha}\,\ko^a$ we have $\omegao_a \mapsto \omegao_a + \Do_a \alpha$. Thus $\omega_a$ is a connection 1-form (on a vector bundle over $\IH$ whose fibers are spanned by $(\ko^a,\uko^a)$.) Its curvature is given by $\Im\,\Psi_2$: $\Do_{[a} \omegao_{b]} = \Im\, \Psi_2\, \epsilono_{ab}$. \\
\vskip-0.3cm

Let us now turn to IHSs $\IH$. The first major difference from NEHSs arises directly from the fact that $\ko^a$ is a symmetry of the full $\IH$-geometry including the derivative operator $\Do$. As we noted in section \ref{s2}, while every null normal $\ko^a$ satisfies $\Lie_{\ko} \qo_{ab} \=0$, a generic NEHS does not admit any null normal $\ko^a$ that Lie-drags $\Do$, i.e.,  satisfies $[\Lie_{\ko}, \Do_a] v^b \= 0$. Hence the passage from an NEHS to an IHS is a genuine restriction that has an important consequence: the available rescaling freedom in the choice of the null normal is drastically reduced. If an NEHS is an IHS with respect to two null normals $\ko^{\prime\, a}$ and $\ko^a$, then $\ko^{\prime\, a} \= e^{\alphao}\, \ko^a$ for some \emph{constant} $\alphao$.  (For details, see \cite{Ashtekar:2001jb}.) Therefore one introduces an equivalence class $[\ko^a]$ of null normals, where any two are equivalent if and only if one is obtained from the other by a rescaling by a positive constant, and denotes IHS geometries by the triplets $([\ko^a],\,\qo_{ab},\, \Do)$. Since the permissible rescaling factor is a constant on $\IH$, all null normals $\ko^a\, \in\, [\ko^a]$ define the same the rotation 1-form $\omegao_a$. \vskip0.1cm

The fact that $\ko^a$ is a symmetry of the IHS geometry leads to two interesting results. First following Ref. \cite{Ashtekar:2000hw}, one can show that the zeroth law holds on all IHSs . One starts with the observation that, an immediate consequence of the property  $[\Lie_{\ko},\, \Do_a]\, \ko^b \=0$  (of $\ko^a$) is  $\Lie_{\ko} \omegao_a \=0$. By the Cartan identity, we also have 
\ba \label{0law}\Lie_{\ko}\, \omegao_b \,\=\,& 2&\, \ko^a\Do_{[a} \omegao_{b]} + \Do_b (\ko^a \omegao_a) \nonumber\\
         \= \, &2& \,\Im\,\Psi_2\,\, \ko^a \epsilono_{ab} + \Do_b \kappao \, \=\, \Do_b \kappao\, , \ea
since $\ko^a \epsilono_{ab} =0$. Hence, $\Do_b \kappao \, \=\, 0$: surface gravity $\kappao$ is constant on $\IH$. 
This constancy holds in spite of the fact that $\IH$ can be highly distorted due to an infinite dimensional freedom in the higher shape moments of IHSs \cite{Ashtekar:2004gp}. Thus, \emph{the zeroth law of black hole mechanics is generalized from KHs to IHSs.} 
\vskip0.1cm

The second result concerns charges associated with the $\IH$-symmetry $\ko^a$. Let us begin by recalling a fact about asymptotically flat space-times. If such a space-time admits a Killing field, it automatically preserves the universal structure at $\scrip$ and is therefore also a BMS symmetry. Therefore one can associate with the Killing field the charge associated with the BMS symmetry. Similarly, the exact symmetry $\ko^a$ of an $\IH$ preserves the universal structure shared by all NEHSs and defines an infinitesimal element of the $\NEH$-symmetry group $\G$ \cite{akkl1} (see Appendix \ref{a1}). Associated with each generator of $\G$ there is a conserved charge on the covariant phase space of all solutions of Einstein's equations that admit an NEHS as its inner boundary \cite{Ashtekar:2021kqj}. For the symmetry field $\ko^a$ this phase space charge is given by :
\be \label{ko-charge1} Q_{{}_{\rm PS}}^{{}^{(\ko)}}\,[S]\, \=\, \f{1}{8\pi G}\,{\oint_S} \omegao_a\,\ko^a\, \rmd^2 V\, 
\=\, \f{1}{8\pi G}\, \kappao\,A[S]\,, \ee
which is manifestly independent of the 2-sphere cross-section $S$ of $\IH$ used in the evaluation. This calculation requires the first derivative of $\ko^a$ away from the IHS. However, in this phase space framework, the first derivative of  any symmetry vector field is completely determined by (the universal structure and) the restriction of the vector field to $\IH$ itself. 

Eq. (\ref{ko-charge1}) holds for all $\ko^a\, \in [\ko^a]$ because if $\ko^a \mapsto e^{\alphao}\,\ko^a$, then $\kappao \mapsto e^{\alphao}\,\kappao$\, so that both sides scale in the same way. But one can single out a \emph{canonical} charge associated with the $\IH$ by fixing this freedom of rescaling  $[\ko^a]$. A natural strategy is to use some properties of Kerr space-times.  Recall that, in the standard convention, the null normal $\ko^a$ to the Kerr KH $\KH$  is taken to be the restriction to $\KH$ of the linear combination 
\be \label{kerr-ko} \ko^a\, \=\,  \to^a + \Omegao_{{}_{\rm kerr}}\, \varphio^a\, ,\ee
of the time translation Killing field $\to^a$ that is normalized to be unit at spatial infinity, and the rotational Killing field $\varphio^a$. The coefficient $\Omegao_{{}_{\rm kerr}}$ is the Kerr angular velocity (which is generally expressed as a function of the ADM mass $M$ and angular momentum $J$). This procedure makes a strong reference to spatial infinity. However, one can pick out the \emph{same} $\ko^a$ by working just at the Kerr $\KH$. Note first that since $\kappao$ scales with $\ko^a$, one can pick out a unique $\ko^a$ in $[\ko^a]$ by fixing the value of its surface gravity. Therefore, the $\ko^a$ of (\ref{kerr-ko}) can be singled out by first expressing  $\kappao_{{}_{\rm kerr}}$ in terms of $J$ and the areal radius $R$, which are intrinsically defined on the horizon,
\be \label{kerr-kappa} \kappao_{{}_{\rm kerr}}\, \=\, \big(2 R^3\,(R^4 + 4 G^2 J^2)^{\f{1}{2}}\,\big)^{-1}\,(R^4 - 4 G^2 J^2)\,,  \ee
and then requiring $\kappao\, \=\, \kappao_{{}_{\rm kerr}}$. ($R$ is given by $A = 4\pi\, R^2$.)  Since $J$ and $R$ are determined by geometrical fields on the  Kerr $\KH$, this procedure enables one to pick out the `correctly scaled' null normal $\ko^a$ on the Kerr $\KH$ without having to refer to properties of fields at infinity.

This strategy immediately generalizes to all IHSs by replacing $R$ by $\RIH$ and $J$ by $\JIH$, the angular momentum that can be defined intrinsically on $\IH$ (see Eq. (\ref{J-IH1}) below). From now on, given any IHS \emph{we will restrict ourselves to the null normal $\ko^a \in [\ko^a]$ for which the surface gravity $\kappao$ is given by (\ref{kerr-kappa})}. Then $\IH$ will have a specific surface gravity that has the same functional form as in the Kerr family \emph{in terms of} $\RIH$ and $\JIH$. Clearly, this can be done without loss of generality on \emph{any} IHS. For this null normal $\ko^a$, on the Schwarzschild horizon we obtain and $Q_{{}_{\rm PS}}^{{}^{(\ko)}} = M/2$.  Therefore, we will define the isolated horizon charge associated with $\ko^a$ to be 
\be \label{ko-charge4}  Q_{\IH}^{{}^{(\ko)}} :=   2 Q_{{}_{\rm PS}}^{{}^{(\ko)}}\,.\ee 
To summarize, on each $\IH$ one can single out a preferred symmetry vector field $\ko^a$ of its geometry by replacing $(R, J)$ in (\ref{kerr-kappa}) by $(\RIH,\, \JIH)$. The associated phase space charge is given by (\ref{ko-charge1}). We emphasize that the expression of this $\ko^a$-charge is arrived at using only the intrinsic structure on $\IH$.  The distinguished horizon charge  $Q_{\IH}^{{}^{(\ko)}}$ is obtained by a simple rescaling to match the standard conventions. Because there can be gravitational radiation and matter fields outside $\IH$, its value is  different from the corresponding ADM charge at spatial infinity. It is this charge, defined entirely at $\IH$, that will feature in our generalization of the first law.
\vskip0.1cm

\emph{Remarks:} \vskip0.1cm 

1. Our derivation of the zeroth law make a crucial use of  $\Lie_{\ko} \omegao_a\, \=\,0$. This condition does not hold on a general NEHs, whence the zeroth law \emph{does not hold on general NEHSs either}. But the necessary condition,\, $\Lie_{\ko} \omegao_a\, \=\,0$,\, is equivalent to\, $[\Lie_{\ko},\, \Do_a] \ko^b\, \=\,0$,\, which is weaker than the IHS requirement: $[\Lie_{\ko},\, \Do_a] v^b\, \=\,0$ for \emph{all} vector fields $v^a$ on $\IH$. NEHSs satisfying the weaker  condition are called \emph{weakly isolated horizons} and, in the literature on quasi-local horizons, the zeroth law was derived for WIHs, without reference to the stronger notion of IHSs (for details, see \cite{Ashtekar:2000hw}). 
However, our discussion of the first law in section \ref{s3.3} uses IHSs.

2. The prescription (\ref{kerr-kappa}) to select a preferred $\ko^a$ from the equivalence class $[\ko^a]$ of null normals to an IHS would provide a non-negative $\kappao$ only if $R^2 \ge 2JG$. Interestingly, this condition is always met by IHSs considered in this section, thanks to the available area-angular momentum inequalities on axisymmetric MTSs \cite{Jaramillo:2011pg,Dain:2011pi,Dain:2010qr,Clement_2012,Clement_2013}. 
%Here, Dain:2011kb. Was removed because it referred only to charge. Clement_2012,Clement_2013 were added as suggested by the referee.
%
Now, as discussed in Appendix \ref{a1}, thanks to results in Refs. \cite{akkl1,Ashtekar:2021kqj,Korzynski:2007hu}, angular momentum is well-defined also on non-axisymmetric QLHs. However, in this case we do not yet know the status of the area-angular momentum inequality. Of course, this cannot occur for stationary BHs in vacuum which, by uniqueness theorems, would be in the Kerr-Newman family. But, as we have emphasized, IHSs can exist in wider contexts in which there are dynamical fields and gravitational waves. The discussion of the initial value problem based on two null surfaces \cite{Lewandowski:1999zs} shows that there is a very large class of IHSs on which the inequality $\RIH^2 \ge 2G\JIH$ will continue to hold. The simplest strategy would be to restrict ourselves to this class. For further discussion, see Remark 1 in section \ref{s5.1}.

\subsection{Angular momentum}
\label{s3.2}

Let us now consider axisymmetric IHSs with a rotational symmetry $\varphio^a$ of its intrinsic geometry. One can again use the  covariant phase space framework \cite{Ashtekar:2021kqj} to compute the angular momentum charge associated with $\varphio^a$. It has the same form as the $\ko^a$-charge of Eq. (\ref{ko-charge1}), but with $\ko^a$ replaced by $\varphio^a$: 
\be Q_{{}_{\rm PS}}^{{}^{(\varphio)}}\,[S]\, \=\, \f{1}{8\pi G}\,{\oint_S} \omegao_a\,\varphio^a\, \rmd^2 V\, . \ee
It is just the negative of the $\IH$-angular momentum charge $\JIH$ that was introduced using geometrical considerations sometime ago \cite{Ashtekar:2000sz,Ashtekar:2001is,Gourgoulhon_2005}: 
\be \label{J-IH1} \JIH\,  \=\, -\f{1}{8\pi G} {\oint_S} \omegao_a \,\varphio^a \, \rmd^2 V\, . \ee
Since any divergence-free vector field $\varphio^a$ on $S$ has the form $\varphio^a \= \epsilono^{ab}\, D_b \mathring{f}$ and $\epsilono^{ab} \Do_a \omegao_b \,\=\, 2\, \Im [\Psi_2]$, $\JIH$ can also be expressed as 
\be \JIH\, \,\=\, -\f{1}{4\pi G} \oint_S \mathring{f}\, \Im [\Psi_2]\, \rmd^2V\, ,\ee
Note that his form makes it manifest that angular momentum remains unchanged if one adds a gradient to $\omegao_a$, i.e. even if one were to rescale $\ko^a \rightarrow e^f\, \ko^a$ by a function. (Thus the notion of angular momentum is well-defined on any axisymmetric NEH $\NEH$; and indeed, any $\NEH$ using the construction summarized in  Appendix \ref{a1.2}.)
%
%For later considerations it is useful to relate $\JIH$ to other angular momentum charges used in the literature, using Hamiltonian frameworks. Let us begin with the covariant phase space framework constructed from space-times that admit such an NEHS as an inner boundary \cite{Ashtekar:2021kqj}, used in section \ref{s3.1} to obtain $Q_{{}_{\rm PS}}^{{}^{(\ko)}}$. This Hamiltonian analysis requires the knowledge of the first order derivative of $\varphio^a$ away from $\IH$ to first order but, as noted in section \ref{s3.1}, it is completely determined by the restriction of the symmetry vector field to $\IH$. With this extension, one obtains: 
%%
%\be  Q_{{}_{\rm PS}}^{{}^{(\varphio)}}[S]\, =\, - \f{1}{16\pi G}\, {\oint_S}  \epsilon_{ab}{}^{cd}\,\, \nabla_c \varphio_d \,\,\rmd S^{ab}\, , \ee
%%
%which is again insensitive to the choice of the cross-section $S$ of $\IH$. Furthermore, using the extension one can show that $ Q_{{}_{\rm PS}}^{{}^{(\varphio)}}[S] = - \JIH$.

For later considerations it is useful to relate $\JIH$ to the angular momentum charge obtained using the canonical phase space (that will be used for DHSs). Let us consider a space-like 3-surface $\Sigma$ extending to spatial infinity with an interior boundary, S at which it intersects $\IH$ and to which $\varphio^a$ is tangential. This framework also provides an angular momentum charge $Q_{{}_{\rm Can}}^{(\varphio)}\,[S]$ at $S$ as the generator of the canonical transformations induced by $\varphio^a$, now expressed in terms of the metric $\ub{q}_{ab}$ and extrinsic curvature $\ub{k}_{ab}$ of $\Sigma$:
\be  \label{J-IH2} Q_{{}_{\rm Can}}^{(\varphio)}\,[S]  = \f{1}{8\pi G}\,{\oint_S} \ub{k}_{ab}\,\varphio^a {\h{\ub{r}}}^b\, \rmd^2 V\, , \ee
where ${\h{\ub{r}}}^b$ is the unit outgoing normal to $S$ within $\Sigma$. In this calculation, one can extend $\varphio^a$ from $S$ to \emph{any} smooth vector field on $\Sigma$; the result depends only on the restriction of $\varphio^a$ to $S$. By using the definition of the extrinsic curvature $\ub{k}_{ab}$ one can express the integrand as $\varphio^a\,{\h{\ub{r}}}^b\, \nabla_a \h{\ul{\tau}}^a$, where  $\h{\ul{\tau}}^a$ is the unit normal to $\Sigma$. Then writing ${\h{\ub{r}}}^b$\, and $\h{\ul{\tau}}^a$ as linear combinations of the two null normals and using the fact that $\varphio^a$ is divergence-free, one obtains $Q_{{}_{\rm Can}}^{(\varphio)}\,[S] = Q_{{}_{\rm PS}}^{{}^{(\varphio)}}[S]$, the charge on the covariant phase space, just as one would expect. Thus, we have the following relation between $\JIH$ and the (covariant or canonical) phase space angular momentum charges 
\be \label{J-IH4} \JIH\, \=\,- Q_{{}_{\rm PS}}^{{}^{(\varphio)}} .\ee
which will be useful in sections \ref{s3.3} and \ref{s6.1} to generalize the BCH first law to IHSs and DHSs respectively.

\subsection{Extension of the BCH First Law to IHSs}
\label{s3.3}

Following BCH \cite{Bardeen:1973gs,LesHouches:1973}, standard treatments of the first law start by considering  asymptotically flat space-times $(M, g_{ab})$ that admit a time translation Killing field $\to^a$  and a rotational Killing field $\varphio^a$. Next, one assumes that the space-time admits a Killing horizon $\KH$ whose null normal is given by a (constant) linear combination $\ko^a := \to^a +\Omegao\, \varphio^a$ of the two Killing fields. The surface gravity $\kappao$ is given by the acceleration of $\ko^a$ evaluated at $\KH$: \, $\ko^a\nabla_a \ko^b \= \kappao\,\ko^b$. Note that $\KH$ is also a KH with respect to $\ko^\prime = e^{\alphao}\, \ko^a$ where $\alphao$ is a constant, and the two surface gravities are related by $\kappao^\prime = e^{\alphao}\, \kappao$. On the other hand, in the BCH form of the first law, $\delta M = (\kappao/8\pi G) \delta A + \Omegao\, \delta J$, we have a specific $\kappao$. As discussed earlier, in the standard treatment a preferred $\ko^a$ is selected by demanding that \emph{$\to^a$ be unit at infinity}.% 
\footnote{The rescaling freedom in $\varphi^a$ is removed, as always, by asking that the affine parameter of $\varphio^a$ range from $0$ to $2\pi$. Note that this condition does not make reference to infinity since it can be imposed on any one orbit of $\varphio^a$ which can, in particular be chosen to lie on $\KH$.} 
This normalization of $\to^a$ also fixes the ADM energy associated with $\to^a$\, (so that $E^{(\to)} = M$),\, and also the horizon angular velocity $\Omegao$. 

This procedure seems natural in the stationary, axisymmetric sector. However, as discussed in section \ref{s1}, in  generalizations to non-equilibrium situations --e.g. involving multiple black holes-- ADM quantities (and the normalization condition on $\to^a$) that refer to infinity cannot feature in the laws of horizon mechanics. And indeed, even in the discussion of equilibrium, one should only assume that the BH is isolated, allowing for non stationary processes outside. Can one even formulate the BCH first law using quantities that are intrinsically defined at the horizon, without any reference to spatial infinity? Our discussion in section \ref{s3.1} suggests that this should be possible; first by replacing KHs $\KH$ by  IHSs $\IH$, and then working with fields that live on $\IH$ itself, thereby avoiding any reference to their behavior far away. We will now show that this expectation is correct. The resulting reformulation will then serve as a point of departure for the desired generalization to non-equilibrium situations.  

The areal radius $\RIH$ is intrinsically defined on any $\IH$, and as we noted in section \ref{s3.2}, so is the horizon angular momentum $\JIH$ on any axisymmetric $\IH$. For Kerr BHs, these extensive quantities can be used to parametrize the Kerr IHSs uniquely. (Indeed, Kerr solutions themselves can be unambiguously labeled by the pair $(\RIH, \JIH)$ in place of the ADM pair $(M, J)$.) Next, as we noted at the end of section \ref{s3.1}, we can single out the `correct' null normal $\ko^a$ to the Kerr IHS by demanding that its surface gravity --evaluated using only the intrinsically defined  derivative operator $\Do$ on the IHS-- be given as the function of $(\RIH,\, \JIH)$ specified in Eq.(\ref{kerr-kappa}). On any $\IH$, this procedure provides us with a canonical null normal $\ko^a \in [\ko^a]$.

The mass $M$ that features in the BCH first law is the Komar charge associated with the stationary Killing field $\to^a$, normalized to be unit at infinity. Therefore to generalize the notion of mass to IHSs $\IH$, we need to find the symmetry vector field analogous to the Kerr $\to^a$ on the generic axisymmetric $\IH$ now under consideration. To carry out this task, we first note that the angular velocity $\Omegao$ of the Kerr IHS can also be expressed as a function of $(\RIH, \JIH)$ in place of the commonly used ADM pair $(M, J)$,
\be \label{kerr-Omega} \Omegao{{}_{\rm kerr}}\, \=\, 2\, \big(\RIH\,\, (\RIH^4 + 4 G^2 \JIH^2)^{\f{1}{2}}\,\big)^{-1}\, \JIH\,,\ee
and then {\it define} $\to^a$ on \emph{any} axisymmetric IHS via $\to^a \,\=\, \ko^a - \Omegao\, \varphio^a$ with $\Omegao$ given by (\ref{kerr-Omega}). One can check that in the Kerr family, this $\to^a$ is the restriction to the $\KH$ of the correctly normalized stationary Killing field. However, in the above construction one never refers to infinity; every step is carried out intrinsically at the IHS. 

This $\to^a$ is also a symmetry of the intrinsic geometry $(\ko^a, \qo_{ab}, \Do)$ on \emph{any} axisymmetric IHS  because both $\varphio^a$ and $\ko^a$ are. Since the phase space charges are linear in their dependence on space-time vector fields, and since  $\to^a\,\=\, \ko^a - \Omegao\,\varphio^a$, we have:
\be Q_{{}_{\rm{ PS}}}^{{}^{(\to)}}\, \=\, Q_{{}_{\rm{PS}}}^{{}^{(\ko)}} - \Omegao\, Q_{{}_{\rm{PS}}}^{{}^{(\varphio)}}\ee
Therefore, relations (\ref{ko-charge4}) and (\ref{J-IH4}),
\be \MIH \,\=\,  2 Q_{{}_{\rm{ PS}}}^{{}^{(\to)}}\, \qquad{\rm and} \qquad \JIH\, =\, - Q_{{}_{\rm{ PS}}}^{{}^{(\varphio)}}\ee
between the phase space and horizon charges immediately imply 
\be \label{M-IH1} \MIH \,\=\, \f{\kappao\, A[S]}{4\pi G}\, +\, 2\Omegao\, \JIH\,. \ee
on every $\IH$.  On Kerr IHSs, we automatically have $\MIH = M$, the ADM mass parameter. Thus, just as (\ref{J-IH1}) provides an intrinsic definition of the $\IH$-angular momentum that yields the correct $J$ on Kerr IHSs, (\ref{M-IH1}) provides a definition of the $\IH$-mass, again in terms fields defined intrinsically on $\IH$ which nonetheless agrees with the standard ADM $M$ in the Kerr family.

With this explicit expression at hand, one can check that the extensive quantities $\MIH,\, A[S]$ and $\JIH$ associated with two nearby solutions admitting IHSs satisfy the first law 
\be \label{1law1} \delta \MIH\,  \= \,\f{\kappao}{8\pi G} \delta A + \Omegao\, \delta \JIH \, , \ee
because, thanks to their functional dependence on $\RIH, \JIH$, the intensive parameters  $\kappao$ and $\Omegao$ satisfy the identity
\be \label{identity1} \f{A}{4\pi G}\, \delta\kappao\, +\, 2\JIH \, \delta\Omegao \,\=\, -\Big(\f{\kappao}{8\pi G}\, \delta A\, +\, \Omegao \,\delta \JIH \Big).\ee 
For KHs in space-times satisfying source-free Einstein's equations, $\MIH$ and $\JIH$ equal the ADM mass and angular momentum $M,J$ and Eqs. (\ref{M-IH1}) and (\ref{1law1}) reduce to the Smarr relations and the BCH first law respectively. However, \emph{for general IHSs $\IH$, the equality with ADM quantities does not hold} because gravitational waves and matter fields outside $\IH$ also contribute to the ADM quantities. Thus (\ref{M-IH1}) and (\ref{1law1}) are \emph{genuine generalizations} of the familiar expressions from the 1970s, in that they refer only to quantities defined intrinsically \emph{at} the IHS $\IH$. Finally, although to define $\JIH$ we restricted ourselves to axisymmetric IHSs, as explained in the beginning of this section, this is not necessary. As summarized in Appendix \ref{a1}, one can define angular momentum on a generic $\IH$ in an invariant fashion, and if the $\IH$ happens to be axisymmetric, the general definition reduces to $\JIH$  {\smash{(of Eq. (\ref{J-IH1})).}} Consequently, the first law we discussed extends to \emph{all} IHSs, including those that are not axisymmetric. 

Although the form of Eq. (\ref{1law1}) is the same as that of the BCH first law (\ref{BCH}), there are important conceptual differences. As pointed out earlier, in the BCH-type discussions one starts with a stationary Killing field $\to^a$ that is normalized at infinity, and a rotational Killing field $\varphio^a$ on the entire space-time. The vector field $\ko^a$, with respect to which $\KH$ is a Killing horizon, is \emph{secondary}, defined in terms of $\to^a$ and $\varphio^a$ via $\ko^a\, = \to^a + \Omegao \varphio^a$. Secondly, the ADM mass $M$ (or energy $E^{(\to)}$ associated with $\to^a$), and the ADM angular momentum $J$ are the primary observables. Instead, our emphasis is on \emph{general} axisymmetric IHs $\IH$, and primary objects are the ones defined using just the $\IH$-geometry, without reference to what is outside the BH. Thus, to begin with, the symmetry vector fields are the ones that preserve the IHS geometry --the rotation $\varphio^a$ and the null normals to the IHS $[\ko^a]$-- and the primary observables are the area radius $\RIH$ and the horizon angular momentum $\JIH$. We use them first to pick out a unique $\ko^a \in [\ko^a]$ by asking that its surface gravity be given by (\ref{kerr-kappa}) as a function of $(\RIH, \JIH)$. This condition can be imposed on any IHS and, in the case when $\IH$ is the Kerr IHS, we recover the standard $\ko^a$. Finally, now $\to^a$ is a \emph{secondary} vector field, obtained from the symmetries $\varphio^a$ and $\ko^a$ of $\IH$ via $\to^a \,=\, \ko^a -\Omegao \varphio^a$, where the constant $\Omegao$ is a function of $\RIH, \JIH$ given by (\ref{kerr-Omega}). Again this procedure selects a unique symmetry vector field $\to^a$ on any axisymmetric IHS, with a guarantee that this $\to^a$ agrees with the standard time translation on any Kerr IHS. With these choices, we have the extension (\ref{1law1}) of the first law to any IHS. The geometry of these general IHSs can be quite different from that of any of the Kerr IHs. Nonetheless (\ref{1law1}) holds on them.
\vskip0.1cm

\emph{Remarks:} 
\vskip0.05cm

1. Angular momentum is well-defined on general NEHSs as well. But since the acceleration of the null normal $\ko^a$ is not a constant on a general NEH, neither the zeroth nor the first law holds on them. \vskip0.05cm

2. The older literature on quasi-local horizons contains versions of the zeroth and the first laws that were established for weakly isolated horizons \cite{Ashtekar:2000hw,Ashtekar:2001is}. Our discussion of the zeroth law is taken directly from \cite{Ashtekar:2000hw}. However, there are some key differences from the previous discussions of the first law. These are discussed in Remark 2 of section \ref{s7.2}.

\section{Salient properties of Dynamical Horizon Segments}
\label{s4}

In the rest of this paper, we will extend the BCH first law \cite{Bardeen:1973gs} and  Hawking's second law \cite{Hawking:1971vc} to black holes that may be very far from being in equilibrium. As we saw in section \ref{s2}, such BHs can be represented by  DHSs. Therefore, we will begin in this section by summarizing their properties that will be important for our extension. 
\vskip0.1cm

1. First, one can show that on space-like DHSs, the topology of MTSs is highly restricted if the cosmological constant $\Lambda$ is non negative \cite{Ashtekar:2003hk}. If $\Lambda >0$, then the topology has to be $S^2$, and if $\Lambda =0$ then the topology is either $S^2$ or, in a very degenerate case, $T^2$. The degenerate case is physically uninteresting. These topological restrictions are similar to those on EHs. Since our focus is on extending BH thermodynamics to non-equilibrium situations represented by DHSs, \emph{in the rest of the paper we will assume that all MTSs are topologically $S^2$}. Thus, henceforth all QLHs will be assumed to have topology $S^2\times R$. \vskip0.1cm

2. Second, in striking contrast to the IHSs, \emph{the foliation of DHSs by MTSs is unique}. The fact that $\DH$ is not null plays an important role in this analysis. (Indeed, as we noted in section \ref{s3.1}, this uniqueness does not hold for IHSs.) For space-like $\DH$, using an argument closely related to the  geometric maximum principle for minimal surface theory in Riemannian geometry \cite{andersson1997}, it was shown in \cite{Ashtekar:2005ez} that if a $\DH$ admits a sub-manifold $\t{S}$ that is an MTS, then $\t{S}$ must necessarily coincide with one of the MTSs $S$ in the foliation that $\DH$ comes equipped with. For time-like DHSs,  the uniqueness result continues to hold \cite{Galloway:2025} using the time-like version of the maximum principle proved in \cite{Galloway:1999ny}.\vskip0.1cm

3. Third, a number of fields that vanish on IHSs are non-zero now, and furthermore all fields of physical interest \emph{vary from one MTS to another} on any given DHS $\DH$. Since the foliation by MTSs is unique, one can think of each MTS as representing an `instant of time' and write down `evolution equations' for fields on $\DH$.
\footnote{In the case when $\DH$ is space-like, this `time dependence' can be understood as follows. Consider a foliation of space-time by partial Cauchy surfaces, each of which intersects $\DH$ in a MTS. Then, each MTS carries the time-coordinate $t$ that labels the leaf of the foliation and thus fields on $\DH$ acquire a $t$-dependence.}
We will denote the intrinsic (non-degenerate) metric on $\DH$ by $q_{ab}$, its volume form by $\epsilon_{abc}$, its derivative operator by $D$. 

The detailed properties of DHSs differ depending on whether they are space-like or time-like and whether the expansion $\theta_{(\uk)}$ is negative or positive. (For a discussion of various possibilities, see section 3 of \cite{Ashtekar:2025LivRev}.) For definiteness we will now consider the case that is of main interest for physical applications in classical GR: A space-like $\DH$ with $\theta_{\uk} <0$ so that $\DH$ separates an untrapped region from the trapped region, i.e. we have a $\DH$ adapted to BHs rather than white holes. (Time-like DHSs are discussed in section \ref{s6.3}.) When $\DH$ is space-like, the unit normal $\h\tau^a$ to it is time-like and the extrinsic curvature $k_{ab}$ of $\DH$ is given by $k_{ab} = q_a{}^c q_b{}^d\, \nabla_c \h\tau_d$. Fields intrinsic to MTSs $S$ will generally carry a `tilde'. Thus, the intrinsic 2-metric on $S$ will be denoted by $\t{q}_{ab}$, its area 2-form by $\t\epsilon_{ab}$, and its derivative operator by $\t{D}$. The unit space-like normal to $S$ within $\DH$ will be denoted by $\h{r}^a$. With these conventions,
\be k^a\, \=\, \f{1}{\sqrt{2}}(\h\tau^a + \h{r}^a)\quad {\rm and}\quad  \uk^a\, \=\, \f{1}{\sqrt{2}}(\h\tau^a - \h{r}^a)\ee
can be taken as the two null normals to MTSs. We will do so, keeping in mind that one has the freedom to perform a rescaling by a function, $(k^a,\, \uk^a) \mapsto (e^\alpha k^a,\, e^{-\alpha} \uk^a)$. While expansions and shears of $\ko^a$ vanish on IHSs, now only $\theta_{(k)}$ vanishes. Of particular interest is the shear of $k^a$:
\be \t\sigma_{ab}\, :\!\hat{=} \,\,\t{q}_{a}{}^{c} \t{q}_{b}{}^{d}\, \nabla_c k_d\, , \ee
which is trace-free since $\theta_{(k)} =0$. This field is associated with gravitational waves falling across $\DH$. In  perturbation theory off stationary space-times, $\t\sigma_{ab}$ captures the 2-radiative degrees of freedom of the gravitational field on the KH. However, while the gravitational field in GR carries only two true phase space degrees of freedom on null surfaces, on space-like surfaces there are four. On a DHS, two of these are encoded in $\t\sigma_{ab}$, while the other two in a 1-form \cite{Ashtekar:2025LivRev}
\be \t{\zeta}_a\, \=\, \t{q}_{ac}\, \h{r}^d \,\nabla_d k^c\,.\ee 
Both $\t\sigma_{ab}$ and $\t{\zeta}_a$ are `transverse' fields on $\DH$ in the sense that they are orthogonal to $\h{r}^a$. Finally, DHSs also admit a rotational 1-form $\t\omega_a$:
\be \label{omegaDH} \t\omega_a \,\=\, - \t{q}_a{}^{c}\uk_b \, \nabla_c k^b\,\, \=\,  \, \t{q}_{a}{}^{c} \h{r}^b \, k_{bc}\, ,\ee
that is tangential to each MTS $S$ in $\DH$.
\vskip0.05cm

\emph{Fields on a given MTS $S$ of any DHS $\DH$ determine the instantaneous non-equilibrium state of a BH.} It evolves as one passes from one MTS to another on $\DH$. In sections \ref{s6} we will show that this evolution leads us to a generalization of the BCH first and second laws to non-equilibrium situations
\vskip0.1cm

4. The fourth feature of DHSs concerns angular momentum. Since $\DH$ is a space-like manifold, it is simplest to use the canonical phase space.  As usual we can use the momentum constraint, 
\be D_a(k^{ab} - k q^{ab}) - 8\pi G\, T_{ac}\, q^{ab}\h\tau^c\,\=\,0\, , \ee
to associate charges, fluxes and balance laws with vector fields on $\DH$. Let  $v^a$ denote a vector field on $\DH$ that is tangential to each MTS. Then, by contracting the momentum constraint with $v^a$ and integrating over a portion $\Delta {\H}$ bounded by two MTSs $S_1$ and $S_2$ we have a balance law
\be \label{Jbalance1} {\Big(} \oint_{S_2} - \oint_{S_1}{\Big)}\, k_{ab}\, v^a \h{r}^b\,\rmd^2 V\, \=\, \f{1}{2}\,\int_{\Delta\H} (k^{ab}- k q^{ab})\,\Lie_{v} q_{ab}\,\rmd^3 V\, +\, 8\pi G\,\int_{\Delta\H} T_{ab}\, v^a \h\tau^b \rmd^3 V\, . \ee
relating the differences between charges associated with $S_2$ and $S_1$ and a flux across $\Delta\H$. The symplectic structure on the canonical phase space provides the overall normalization factor in the definition of charges and fluxes (see, e.g., \cite{Ashtekar3_2024}). Using (\ref{omegaDH}) one can write the phase space charge associated with $v^a$ as
\be Q^{{}^{(v)}}_{{}_{\rm PS}}\, [S]\, \=\, \f{1}{8\pi G}\, \oint_{S} \, k_{ab}\, v^a \h{r}^b\,\rmd^2 V\, \=\, \, \f{1}{8\pi G}\, \, \oint_{S} \, \t\omega_a v^a\, \rmd^2 V\,. \ee 
$Q{{}^{(v)}} [S]$ would correspond to an angular momentum charge if the vector field $v^a$ represents a rotation. Now,
if the metric $q_{ab}$ on $\DH$ admits a rotational \emph{Killing} field $\varphi^a$, then by the uniqueness result on the foliation of $\DH$ by MTSs, it follows that $\varphi^a$ has to be tangential to each MTS. Hence the angular momentum evaluated at the MTS $S$ is given by substituting $v^a$ with $\varphi^a$. Since $\varphi^a$ is a Killing field, the gravitational wave contribution to the angular momentum flux would vanish in this case. \vskip0.05cm

{However, one can also consider generic DHSs that are not axisymmetric. Korzynski \cite{Korzynski:2007hu} has provided an invariant construction to assign angular momentum $\JDH[S]$ to each MTS. (For a brief summary, see Appendix \ref{a1}.)  The procedure uses only the geometry on the MTS $S$ and exploits the fact that the group of conformal isometries on $S$ is (isomorphic to) the Lorentz group. It selects from the conformal Killing fields on $S$ a preferred one, $\varphi^a$.% 
\footnote{The construction assumes a genericity condition that excludes spherically symmetric MTSs. But in that case $J$ vanishes identically and one does not need to single out a $\varphi^a$ to perform the integral.}
\,\,It is a rotation in the sense that: (i) it has closed orbits; (ii) exactly two zeros; and, (iii) the range of its affine parameter is $[0, \, 2\pi)$. The $\varphi^a$ picked out by this procedure depends on $\qo_{ab}$ and $\omega_a$ --i.e. on the specifics of the geometry of the MTS $S$ under consideration--  just as one would physically expect. The angular momentum of the MTS is given by 
\be \label{J-NEH2} \JDH [S]  \=\, -\f{1}{8\pi G} {\oint_S} \t\omega_a \,\varphi^a \, \rmd^2 V\,. \ee
This $\JDH [S]$ is time-dependent, i.e. varies as we change the MTS S, satisfying the balance law (\ref{Jbalance1}). 

To summarize, we have: }%
\be \label{JDH1} \JDH^{{}^{(\varphi)}} [S] \,\equiv \, -\,Q^{{}^{(\varphi)}}_{{}_{\rm PS}}\, [S]\, \=\,
 -\f{1}{8\pi G}\, \, \oint_{S} \, \t\omega_a \varphi^a\, \rmd^2 V\, , \ee
satisfying the balance law
\be \label{JDH2} \JDH^{{}^{(\varphi)}} [S_2]\, -\, \JDH^{{}^{(\varphi)}}[S_1] \, \,\=\, -\int_{\Delta\H} \, \Big[T_{ab}\varphi^a \hat{\tau}^b\,+\, \f{1}{16\pi G} \big(k^{ab} - kq^{ab}\big)\, \Lie_{\varphi} q_{ab} \Big]\, \rmd^3 V\,\, \equiv\,\, \mathcal{F}^{{}^{(\varphi)}} [\Delta\H]\, .  
\ee
Because $\Lie_{\varphi}  q_{ab}\, \not= 0$ on a general DHS, there is a non-trivial flux $\mathcal{F}^{{}^{(\varphi)}}$ of angular momentum carried by gravitational waves across $\Delta\H$, in addition to the matter angular momentum flux.  Finally, one can also define the Komar integral for $\varphi^a$ by extending it infinitesimally away from $\DH$ in a natural manner%
\footnote{\label{fn4} If the DHS $\DH$ is a leaf of a foliation given by $\tau = const$ in its neighborhood, one only needs to assume that the extension of $\varphi^a$ to the neighborhood is tangential to the foliation.}
and one finds: $\JDH^{{}^{(\varphi)}} [S] \,\equiv \, -\f{1}{2}\,Q^{{}^{(\varphi)}}_{{}_{\rm K}}\, [S]\,$. 
Thus, various angular momentum charges on DHSs are related by
\be \label{JDH3} \JDH\, \=\, - Q_{{}_{\rm PS}}^{{}^{(\varphi)}}[S]\, \=\, - \f{1}{2}\, Q_{{}_{\rm K}}^{{}^{(\varphi)}}\,[S]  .\ee
These relations parallel those for IHSs (see Eq. (\ref{J-IH4})). In both cases, we have spelled out the details because the relative minus signs and factors of 2 that arise from different frameworks can be confusing. The angular momentum $\JDH^{{}^{(\varphi)}}$ and its flux $\mathcal{F}^{{}^{(\varphi)}} [\Delta\H]$ will play an important role in sections \ref{s5.2} and \ref{s6.1}.  
 
\vskip0.15cm 
\goodbreak

\noindent\emph{Remarks:} \vskip0.05cm 

1. \emph{Approach to equilibrium:} Of particular interest to thermodynamic considerations is the process by which a dynamical BH approaches equilibrium. In our framework, the non-equilibrium phase is described by a DHS, and the equilibrium state by an IHS. Both are parts of a QLH. As mentioned above, the approach can be asymptotic --as in a binary black hole merger-- or may occur at a finite instant of time --as in the left panel of Fig. 2-- when an open DHS approaches a MTS which is a boundary of an IHS. In these situations, the metric $\t{q}_{ab}$ on the MTS and the curl of the rotational 1-form $\t\omega_a$ on the DHS side tend to those on the IHS side (and $\varphi^a$ does so by its construction). Therefore, the area and the angular momentum on the DHS side approach those on the IHS. This matching holds for other observables as well --for example, once $(k^a, \uk^a)$ are rescaled appropriately so that they are smooth on the entire QLH, the shear $\t\sigma_{ab}$ and the vector field field $\t\zeta_a$ that capture the radiative degrees of freedom on the DHS side tend to zero in the limit. (For details, see \cite{Ashtekar:2013qta}.) The fact that fields on $\DH$ smoothly extend to those on $\IH$ as equilibrium is reached will provide motivation for a key construction in section \ref{s5.2}. \vskip0.1cm 

2. \emph{Existence and uniqueness:} While QLHs are 3-manifolds defined in their own right, to explore the issues of existence and uniqueness one has to use the Cauchy problem. If a Cauchy slice admits a MTS $S$, would $S$ yield a unique QLH under evolution? Local existence has been shown if the initial MTS $S$ is strictly stable (or, strictly-stably-outermost) \cite{Andersson:2007fh}. Furthermore, in this case the QLH is either space-like (if $\mathcal{E}$ is non-zero somewhere on $S$) or null (if $\mathcal{E}$ vanishes identically on $S$). Let us consider the case when the QLHs so obtained are DHSs. Then, if space-time satisfies the NEC, one also has the following uniqueness theorem: If at any time $t_\circ \in (t_1, t_2)$, the MTSs of  $\DH$ and $\widetilde{\DH}$ determined by a space-time foliation agree, then $\DH = \widetilde{\DH}$ \cite{Ashtekar:2005ez}. Thus, DHSs obtained via Cauchy evolution cannot bifurcate.  They can of course merge as in binary black hole coalescences. As with restrictions on the topology of MTSs discussed above, this behavior mimics that of EHs, but now the analysis is quasi-local. Finally, as discussed above, on any given DHS $\DH$, the foliation by MTSs is unique. It provides a canonical notion of `time evolution' for the non-equilibrium state of the BH defined by fields on the MTSS of $\DH$.

There is also a broader uniqueness issue: As discussed in \cite{Ashtekar:2005ez}, associated with any given dynamical phase of a BH, there can be many DHSs that ``interweave". The generalized first and second laws discussed in section \ref{s6} hold on each of them. If the BH reaches equilibrium, as is the case, e.g., for the remnant after a binary merger, all these DHSs approach the same IHSs and, as we just discussed, the extensive and intensive observables defined on any one of these DHSs approach those on the IHS. Thus, there is overall coherence. Nonetheless, at least at first, the fact that there can be several DHSs associated with the same dynamical phase of a BH seems surprising and somewhat unsettling. We will discuss this issue in section \ref{s7.2}.

\section{Thermodynamics of BHs far from Equilibrium: Strategy} 
\label{s5}

Already for ordinary thermodynamical systems, one faces a key conceptual obstacle in extending the first law to non-equilibrium situations due to lack of intensive parameters. In section \ref{s5.1} we first state these difficulties and then present a strategy, motivated by constructions introduced in section \ref{s3.3}, to overcome them. In this respect, BHs are simpler than ordinary thermodynamic systems. However, they are also more complicated in that, unlike in systems normally considered in thermodynamics, for dynamical BHs, we do not have a natural notion of instantaneous energy. In section \ref{s5.2} we show that this difficulty can be overcome using a natural additional input.

\subsection{Overcoming a Key Conceptual Obstacle: Intensive Parameters}
\label{s5.1}

For ordinary thermodynamical systems, the extensive quantities such as the total energy $E$, volume $V$ and particle number $N$ are well-defined also when the system is far from equilibrium. However, one cannot assign intensive quantities such as the temperature $T$, pressure $p$  or the chemical potential $\mu$ to the entire system in general. 
This is a key obstacle in extending the first law to systems that are far from equilibrium since $T, p, \mu$ enter in its statement. In this subsection we will show that BHs are quite special in that this obstacle can be removed.

As we saw in section \ref{s3.1}, on a general axisymmetric IHS, the rescaling freedom in $\ko^a \in [\ko^a]$ can be naturally eliminated by requiring that the surface gravity $\kappao$ on $\IH$ be given by the function (\ref{kerr-kappa}) of $(\RIH, J_\IH)$. Similarly while any (constant) linear combination of $\ko^a$ and $\varphio^a$ is an $\IH$-symmetry, one can naturally pick out one, $\to^a = \ko^a - \Omegao \varphio^a$, where $\Omegao$ is given by (\ref{kerr-Omega}). Properties of Kerr solutions were brought-in \emph{only} to ensure that the procedure yields the standard $\ko^a$ and $\to^a$ for globally isolated BHs represented by Kerr space-times. But this procedure has an interesting consequence: Now every $\IH$ acquires a surface gravity $\kappao$ and a rotational velocity $\Omegao$ --the `intensive parameters' that enter (\ref{1law1})--  even if the space-time metric is not stationary or axisymmetric. The functional dependence (\ref{kerr-kappa}) and (\ref{kerr-Omega}) of the intensive parameters $(\kappao,\, \Omegao)$ on $(\RIH, J_\IH)$ is the same as in Kerr space-times. But in Kerr space-times, $(\kappao,\, \Omegao)$ are generally expressed as functions of the ADM quantities $(M, J)$ and this relation will \emph{not} hold on a general axisymmetric $\IH$ since matter and radiation outside the $\IH$ also contribute to $(M,J)$. 

\begin{figure}[]
  \begin{center}
    \includegraphics[width=0.8\columnwidth,angle=0]{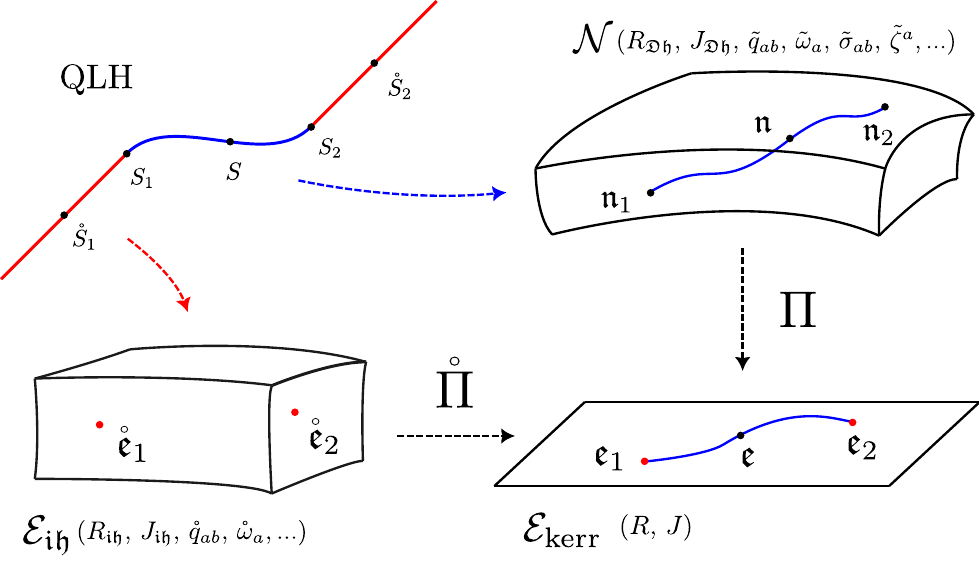}   
  \caption{\footnotesize{\emph{Projection maps:}} The upper-left part of the figure depicts a QLH which has a DHS (shown in blue), bounded by MTSs $S_1$ and $S_2$, and sandwiched between an IHS to the past of $S_1$ and another to the future of $S_2$ (shown in red). An equilibrium state $\mathring{\e}$ corresponds to \emph{time-independent} fields $(\qo_{ab}, \omegao_{a})$ on an IHS $\IH$, constructed from its geometry. The space $\EIH$ of these $\IH$-states --depicted in the lower-left part of the figure-- is infinite dimensional.  A non-equilibrium state $\n$ is specified by the values of an infinite number of fields on a MTS $S$ of a DHS $\DH$ constructed from the initial data $(q_{ab}, k_{ab})$ on $\DH$ and the null normals $(k^a,\, \uk^a)$ to its MTSs. The space $\N$ of non-equilibrium states $\n$ much larger than $\EIH$ because more fields are needed to characterize $\n$. From these fields, one can construct $(\RIH,\, \JIH)$ on $\IH$ and $\RDH [S],\,\JDH [S]$ on $\DH$. Kerr Killing horizons $\KH$ constitute a 2-dimensional space $\Ekerr$ of states $\e$, each representing a BH in global equilibrium. Each $\e$ is characterized by a pair of numbers $(R,\, J)$. The projection maps $\Pio$ and $\Pi$ (respectively) send any given $\mathring{\e}$ and $\n$ to a global equilibrium state $\e$ such that the 2-parameters $(R,\, J)$ labeling $\e$ equal $(\RIH, \JIH)$ of $\mathring{\e}$ and $(\RDH [S],\, \JDH [S])$ of $\n$. Thus, $\Pio$ and $\Pi$ ignore all the rich information contained in $\mathring{\e}$ and $\n$ except for these two numbers. Via $\Pi$, the 1-parameter family of non-equilibrium states defined by $\DH$ project down to a trajectory in $\Ekerr$ along which $R$ changes monotonically. }
\end{center}
\label{fig3}
\end{figure} 

This structure can be re-expressed more formally as follows. Let $\EIH$ denote the space of all axisymmetric IHS geometries. A point $\mathring{\e}$ in $\EIH$ represents a BH that is itself in equilibrium, but allowing for possible dynamical processes outside. $\EIH$ is an infinite dimensional space since $\mathring{\e}$ is labelled by $(\qo_{ab}\, \omegao_a)$ which, although time-independent, are \emph{fields} on $\IH$. (Alternatively, $\mathring{\e}$ can be labelled by $\RIH$, and an infinite tower of shape and spin multipoles, $\JIH$ being the spin dipole \cite{Ashtekar:2004gp, Ashtekar:2025LivRev}). Let $\Ekerr$ denote the 2-parameter family of Kerr solutions, labelled by $(R,\, J)$. A point $\e$ in $\Ekerr$ represents a BH that is in \emph{global} equilibrium. Then, there is a natural projection map $\Pio$ from $\EIH$ to $\Ekerr$:\, $\Pio\,(\EIH) = \e$, where $\e$ is the Kerr solution labelled by $(R,\, J) = (\RIH, \JIH)$, evaluated at the state $\mathring{\e}$. (Thus, $\Pio$ ignores all the information contained in higher multipoles of $\mathring{\e}$, being sensitive only to its pair $(\RIH, J_\IH)$. The map is surjective but not injective.) Then the intensive parameters\, $(\kappao, \Omegao)$\, on $\EIH$ are the pull-backs under $\Pio$ of the intensive parameters of the Kerr family: $(\kappao,\,\Omegao)|_{\EIH}\, =\, \pullback{\Pio}\, [(\kappao,\,\Omegao)|_{\E}]$. This association of intensive parameters to general equilibrium states $\mathring{\e}$ was a key step in our generalization of the first law to all IHSs $\IH$, allowing for dynamical processes in the exterior region.
\vskip0.1cm 

We will now show that, using our discussion in section \ref{s4}, one can extend this strategy to assign intensive parameters also to fully dynamical BHs represented by DHSs $\DH$. Recall that a non-equilibrium state $\n$ of a BH is represented by fields on a MTS $S$ of any DHS $\DH$, constructed from the Cauchy data $(q_{ab},\, k_{ab})$ on $\DH$, and the unit normals $\h\tau^a$ to $\DH$, and $\h{r}^a$ to $S$ within $\DH$. These fields include, in particular, the metric $\t{q}_{ab}$ on $S$, the 1-form $\t\omega_a$ that determines the angular momentum on $S$, as well as the pair $(\t\sigma_{ab},\, \t\zeta_a)$ that encode the true (i.e., radiative) degrees of freedom in GR, associated with gravitational waves traversing $S$. Denote by $\N$ the infinite dimensional space of non-equilibrium states of dynamical BHs in GR. This space is much larger than the space $\EIH$ associated with IHSs discussed above, first because fields such as $(\t\sigma_{ab},\, \t\zeta^a)$ that vanish on IHSs are now non-zero, and second because fields such as $(\t{q}_{ab},\, \t\omega_a)$ that were time independent on IHSs are now time dependent. Again, from the metric $\t{q}_{ab}$ on an MTS $S$, one can construct the area radius $\RDH [S]$, and from $\t\omega_a$, the angular momentum $\JDH [S]$. Therefore, as in the discussion of IHSs, there is a natural projection $\Pi$ from $\N$ to the 2-dimensional space $\Ekerr$ of Kerr BHs: $\Pi (\n) = \e$, where $\e$ is the Kerr solution labeled by  $(R, J) = (\RDH [S], \JDH [S])$, the pair associated with $\n$. This map ignores all the rich information contained in the (time-dependent) fields encapsulated in $\n$, except for two numbers. Again, we can pull-back $(\kappao,\,\Omegao)$ from $\Ekerr$ to $\N$, thereby assigning an effective `surface gravity' and `angular velocity' to each $\n$ defined by the MTS $S$ in $\DH$:\, $(\kappao,\,\Omegao)|_{\n}\, =\, \pullback{\Pio}\, [(\kappao,\,\Omegao)|_{\e_{\rm kerr}}]$.   

Recall that under the projection map $\Pio$, each IHS is mapped to a single point of the space $\Ekerr$ of states in global equilibrium. But now, under the projection $\Pi$, each DHS defines a \emph{trajectory} in the space $\Ekerr$ of equilibrium states, as shown in Fig. 3. Thus, upon a coarse graining that ignores observables on $\DH$ other than $(\RDH [S],\,\JDH[S])$, each dynamical BH can be envisaged as passing through a continuous family of equilibrium states $\e(t)$, where $t$ labels the MTSs $S$ on $\DH$.  Along this trajectory\, $\e(t)$\, in\, $\Ekerr$,\, the extensive observables $(R\, , J)$ labeling the Kerr solutions change, capturing the `time dependence' of $(\RDH [S],\,\JDH [S])$ due to physical fluxes traversing $\DH$. Similarly, the intensive parameters $(\kappao,\, \Omegao)$ also change, just as one would expect for a system that is not in equilibrium. In section \ref{s6.1} we will see that this time dependence leads to a further generalization of the first law, now encompassing dynamical BHs in which the extensive observables on $\DH$ change in response to active, physical processes in space-time.
\vskip0.1cm 
\hfill\break

\emph{Remarks:} 
\vskip0.05cm
1. The projection map $\Pi$ would be well-defined only if each MTS on the DHS satisfies the inequality $\RDH^2[S] \ge 2G\JDH [S]$. This is the case if the MTS is axisymmetric \cite{Jaramillo:2011pg}. Now, as discussed in Appendix \ref{a1}, angular momentum is well-defined also on non-axisymmetric QLHs. In numerical simulations, the DHSs approach a Kerr IHS, whence there is always a regime in which we have $\RDH^2[S] \ge 2G\JDH$. But at this stage, we do not know if in the fully dynamical regime some exotic situations can arise in which the inequality is violated; this issue is being investigated in collaboration with J. L. Jaramillo. If it is violated, the simplest strategy would be to restrict our thermodynamical considerations to DHSs on which it does hold. 

But it is worth keeping a second possibility in mind. One may consider extending $\Ekerr$ to the full half plane $R >0$. Then the map $\Pi$ would be well-defined but, on the portion $\RDH^2[S] < 2G\JDH$, one would have $\kappao<0$. At first this seems totally inadmissible because from the BCH 1st law (\ref{BCH}) we have\, $ \partial M/ \partial (A/4G) \,\=\,(\kappao/2\pi)$\,, and in the familiar thermodynamical systems, $\partial E/ \partial S$ is always positive. But, from a statistical mechanical perspective,  $\partial E/ \partial S$ can also be negative. Indeed, systems with ``negative temperature" --or, more precisely, in which the entropy decreases as energy increases-- are \emph{not forbidden} theoretically and, furthermore, they \emph{exist in Nature}. Therefore, it may be worth exploring consequences of allowing $\kappao$ to become negative, especially while investigating the statistical mechanical origin of entropy of dynamical BHs in quantum gravity. There is a precedent for the necessity of such unforeseen extensions. In familiar thermodynamical systems specific heat, $\partial E/\partial T$, is positive --one normally needs to add energy to raise the temperature. But the statistical mechanical micro-canonical ensemble allows for both signs and negative values do occur in Nature in the presence of long range interactions. Indeed, for Kerr black holes, specific heat can have either sign.
\vskip0.05cm

2. The trajectory defined by a DHS could join on to two IHSs (as in the left panel of Fig. 2), one to the past of an MTS $S_1$ and one to the future of an MTS $S_2$. Or, the  BH could reach equilibrium only in the asymptotic future, in which case the DHS would be non-compact in the future, approaching an IHS only asymptotically. Since the IHSs and the DHS both belong to a smooth QLH, one can show that the pair $(\RDH [S], \JDH[S])$ associated with the DHS tend to the labels $(\RIH, J_\IH)$ as the BH reaches equilibrium in a smooth manner \cite{Ashtekar:2013qta}, whence the intensive parameters on the DHS match with those on the IHSs.

\subsection{Overcoming a Key Conceptual Obstacle: Notion of Energy}
\label{s5.2}

The assignment of intensive parameters $\kappao$ and $\Omegao$ to non-equilibrium states $\n$ defined by any $\DH$ removes a key  obstacle faced in extending equilibrium thermodynamics to non-equilibrium situations encountered in standard thermodynamical systems. However, as we indicated in section \ref{s1}, for BHs one encounters a new obstacle that is absent in familiar thermodynamical systems. The Hamiltonian or energy of these systems is well-defined even when they are far from equilibrium because one works in flat space-time and has access to time-translation Killing fields. On the other hand, space-times representing dynamical BHs  do not admit such Killing fields. Recall, however, that our generalization of the first law to IHSs in section \ref{s3.3} required, to begin with, only the vector field $\ko^a$ that is defined just \emph{at} the IHS. Furthermore, the canonical phase space framework of full GR enables us to associate energy $E^{(\xi)}$ with any causal vector field $\xi^a$. Therefore, to extend the first law to dynamical BHs it is natural to seek a causal vector field $\xi^a$, defined just on $\DH$, that is the analog of the $\ko^a$ on $\IH$. In this sub-section we show that the desired analog exists and is unique on any $\DH$, and obtain the balance law between the charges and fluxes associated with this $\xi^a$. In section \ref{s6.1} we will use this balance law to arrive at the first law for DHSs. \vskip0.1cm

As we saw in section \ref{s4}, for space-like DHSs under consideration one can use the canonical phase space framework based on the $\DH$ itself; we don't need an auxiliary space-like surface $\Sigma$ used in the discussion of IHSs. Let us begin by recalling the relevant results from this framework \cite{Ashtekar:2003hk}. The constraint equations on $\DH$ are:
\ba \label{constraints1} H &:=& \mathcal{R} - k^2 +k^{ab}k_{ab} - 16\pi G\, T_{ab}\, \h\tau^a \h\tau^b\, \=\,0, \quad {\rm and}\nonumber\\ 
H^a &:=& D_b(k^{ab} - k q^{ab}) -8\pi G\, T_{bc}\, \h\tau^b q^{ac}\,\=\,0\, , \ea
where, as before, $D$ and $\mathcal{R}$ are the intrinsic derivative operator and scalar curvature of $\DH$ defined by its metric $q_{ab}$, and $k_{ab}$ its extrinsic curvature, and indices are raised and lowered using $q_{ab}$.
Given a suitable space-time vector field $\xi^a$ on $\DH$, one can smear these constraints with the lapse and shift fields provided by $\xi^a$ to obtain a 2-sphere charge, a 3-dimensional flux integral and a balance law (such that the charges and fluxes are linear in $\xi^a$). We saw an explicit example of this structure in section \ref{s4}: When smeared with the vector field $\varphi^a$, the  diffeomorphism constraint $H^a \=0$ leads us to the angular momentum charge $\JDH^{(\varphi)}[S]$ on each MTS $S$ and a balance law relating the difference between charges associated with MTSs $S_1$ and $S_2$ and the flux of angular momentum carried by matter and gravitational waves across the DHS $\Delta \H$ bounded by $S_1$ and $S_2$ (see Eqs. (\ref{JDH1}) and (\ref{JDH2})). In this section we will extend these considerations to energy. 

In our extension of the first law to IHSs, we began by associating energy with the null vector field $\ko^a$ which is a symmetry of the IHS-geometry. We do not have such a symmetry vector field on a DHS precisely because fields on $\DH$ are time-dependent. But in the Hamiltonian framework, one can smear constraints with lapse-shift pairs constructed from any regular vector fields; they don't have to be symmetries. Now, DHSs do admit a preferred causal \emph{direction} field: the one provided by null vector fields  $k^{\prime\,a} = e^\alpha k^a$ for which $\theta_{(k^\prime)} \=0$. Furthermore if a dynamical BH reaches equilibrium, then this direction field on $\DH$ becomes parallel to the symmetry vector field $\ko^a$ on $\IH$ representing the equilibrium state. Therefore, it is natural to replace $\ko^a$ on IHSs by a vector field $\xi^a = F\, k^a$ for a suitable function $F$ on $\DH$. 

Now, in this section, and most the next, we work exclusively with DHSs. For notational simplicity, we will drop the suffix $\DH$ on the areal radius $\RDH$ when the context makes it clear that we are dealing only with DHSs. Then, the volume element on $\DH$ is given by $\rmd^3 V = |DR|^{-1}\, \rmd{R}\,\rmd^2V$ (where $\rmd^2V$ is the area element on MTSs $S$). We will set $F = \sqrt{2} |DR| f$ for some smooth function $f$, where the factor of $\sqrt{2}$ has been inserted just to simplify subsequent equations. It turns out  that requiring $f$ to be a function only of $R$ leads to balance laws involving charges $Q_{\DH}^{(\xi)}$ and fluxes $\F^{(\xi)}$ constructed from local fields \cite{Ashtekar:2003hk}.%
Therefore we are led to smear the constraints by causal vector fields %
\be \label{xi} \xi^a = \sqrt{2}\, |DR|\, f(R)\, k^a = |DR|\, f(R)\,  (\h\tau^a\, +\, \h{r}^a)\, , \ee
so that the combination $(G_{ab} - 8\pi G\, T_{ab})\xi^a \h\tau^b\, \=\, 0$ of constraints becomes 
\be \label{constraints2} |DR| f(R)\, \big(H + 2\h{r}^a H_a\big)\, +\,16\pi G\, T_{ab}\, \xi^a\, \h\tau^b \, \=\, 0\,. \ee
As we now show, this combination leads to a balance law for any $f(R)$.

The first observation is that, using the fact that $\theta_{(\xi)}\, \=\,0$, the linear combination (\ref{constraints2}) of constraints yields \cite{Ashtekar:2003hk}:
\be \label{keyconstraint} |DR|\, f(R)\,\Big[ \t{\mathcal{R}} -\t{\sigma}_{ab}\, \t\sigma^{ab}\, -\, 2 \t\zeta_a \t\zeta^a \, + \, 2\,\t{D}_a\t\zeta^a\Big] = 16\pi G\, T_{ab}\, \h\tau^a \xi^b\,. \ee
Here $\t{\mathcal{R}}$ is the scalar curvature of the intrinsic 2-metric $\t{q}_{ab}$ on MTSs and, as in section \ref{s4}, indices of fields with a tilde are all tangential to the MTSs $S$. Let us integrate this equation over a portion $\Delta\H$ of $\DH$ bounded by two cross-sections $S_1$ and $S_2$. Using $\rmd^3 V = |DR|^{-1}\, \rmd{R}\,\rmd^2V$ and the fact that the integral of $\t{\mathcal{R}}$ over any $S$ is the Gauss invariant, $\oint_S \t{\R}\, \rmd^2V = 8\pi$,  and then writing $f(R)$ as $f(R) = 2G\, \big({\rmd Q^{(\xi)}_{{}_{\rm PS}} (R)}/{\rmd R}\big)$  one obtains a balance law:\vskip0.05cm
\ba \label{balance1} Q^{(\xi)}_{{}_{\rm PS}} [S_1] - Q^{(\xi)}_{{}_{\rm PS}}  [S_2] &=& \, \f{1}{8\pi G}\,\int_{\Delta\H}\! |DR| f(R) \,\,\big(|\t{\sigma}|^2 + 2|\t\zeta|^2\big) \rmd^3 V \, +\, \int_{\Delta\H} T_{ab}\,\xi^a \h\tau^b\, \rmd^3 V\, \nonumber\\
&\hat{\equiv}& \,\,\,\Fxigws [\Delta\H]\,+\, \Fximatt [\Delta\H] \, \equiv\, \mathfrak{F}^{{}^{(\xi)}} [\Delta\H]
\ea
\vskip0.05cm
\noindent involving charges $Q^{(\xi)}_{{}_{{}_{\rm PS}}} [S_1]$ and $Q^{(\xi)}_{{}_{{}_{\rm PS}}} [S_2]$ defined on any two MTSs, and fluxes $\Fxigws [\Delta\H]$ and $\Fximatt [\Delta\H]$ of $\xi$-energy carried by gravitational waves and matter fields across the portion $\Delta\H$ of $\DH$ bounded by those MTSs.%
\footnote{From its form, it is manifest that $\Fximatt$ represents the flux of matter $\xi^a$-energy. The detailed justification for interpreting $\Fxigws$ as the $\xi^a$-energy carried by gravitational waves is given in section III.B of \cite{Ashtekar:2003hk}.}
This law holds for all vector fields $\xi^a = |DR|\, f(R)\, k^a$, i.e. for all functions $f(R)$ on $\DH$. Each choice of $f(R)$ leads to a charges and a flux. 

To obtain a canonical energy we need to find a \emph{preferred} vector field $\xi^a$ --the $\DH$-analog of our $\ko^a$ on $\IH$. That is, we need to constrain the function $f(R)$ using physical considerations. Let us suppose that the fluxes across a $\DH$ terminate smoothly at a cross-section $S_2$ so that the QLH has an IHS $\IH$ to the future of $S_2$ (see the top left part of Fig. 3). In this case, as remarked in section \ref{s4}, we know that the limit of the angular momentum charges (\ref{JDH3}) on the DHS agree with the corresponding angular momentum charges (\ref{J-IH4}) on the IHS. We would like this agreement to hold also between the energy charge $Q^{{}^{(\xi)}}_{{}_{\rm PS}} [S_2]$ on $\DH$ and $Q^{{}^{(\ko)}}_{{}_{\rm PS}}$ on $\IH$, so that we have a notion of energy that seamlessly continues from the DHS of the QLH to the IHS. The key question then is: Is there is a vector field $\xi^a$ --i.e., a function $f(R)$-- on $\DH$ that would satisfy this consistency condition? A priori it is not clear that such a $\xi^a$ would exist.

To analyze possibilities, let us first recall from section \ref{s5.1} that each MTS $S$ on $\DH$ is assigned a surface gravity $\kappao [S]$ via the projection map $\Pi$:
\be \label{dhkappa} \kappao [S]\, \=\, \big(2R^3[S]\,(R^4[S] + 4 G^2 \JDH^2[S])^{\f{1}{2}}\,\big)^{-1}\,(R^4[S] - 4 G^2 \JDH^2 [S])\,.  \ee
(See Eq.(\ref{kerr-kappa}).) Now, on any given $\DH$, the areal radius $R$ of MTSs is a monotonic function, whence each MTS $S$ is unambiguously labeled by the value of its $R[S]$. Hence for any given $\DH$, the surface gravity $\kappao$ assigned to its MTSs is a function of $R$ alone. Therefore, if  we choose
\be \label{f(R)} f(R)\, \=\,\, { \f{1}{4\pi}}\,\f{\rmd {(\kappao A)}}{\rmd R}
 \equiv\,  2\,G\,\f{\rmd Q^{{}^{(\ko)}}_{\rm PS}}{\rmd R}\, ,\ee
we are, in particular, guaranteed that the desired agreement between $Q^{(\xi)}_{\rm PS}[S_2]$ and $Q^{(\ko)}_{\rm PS}$ would hold anytime when the infalling flux ends at a MTS $S_2$ and the DHS joins on to an IHS there, as in the left panel of Fig. 3. \emph{From now on, in this sub-section (as well as in section \ref{s6.1}) we will  restrict ourselves to vector fields $\xi^a = |DR| f(R) k^a$ where $f(R)$ is given by (\ref{f(R)}).} With this choice, we have the charge
\be \label{Qxi} \quad Q^{{}^{(\xi)}}_{\rm PS}\, [S]\, \=\, \f{(\kappao A)}{8\pi G}\,[S] \ee
associated with any MTS $S$, and the balance law
\be \label{balance2}  \Delta\,\Big[\f{\kappao A}{8\pi G}\Big]\,\,\hat{\equiv}\,\, \Big[\f{\kappao A}{8\pi G}\Big]_{S_1}^{S_2} \, \=\,\,\, \Fxigws [\Delta\H]\,+\, \Fximatt [\Delta\H]
\ee
associated with any portion $\Delta\H$ of $\DH$ bounded by the MTSs $S_1$ and $S_2$. Thus, the requirement that there be a seamless matching between the phase space charges associated with $\xi^a$ on $\DH$ and $\ko^a$ on $\IH$ as equilibrium is reached uniquely fixes $f(R)$.  The resulting $\xi^a$ is the DHS analog of the vector field $\ko^a$ on IHSs.
\vskip0.05cm

Next, recall that on IHSs, motivated by the relation between the canonical charge $Q^{{}^{(\ko)}}_{{}_{\rm PS}}$ and mass in the Schwarzschild space-time, we were led to introduce $Q^{{}^{(\ko)}}_{{}_{\IH}} = 2 Q^{{}^{(\ko)}}_{{}_{\rm PS}}$. To make the comparison transparent, let us introduce the analogous $\xi$-charge 
\be \label{relation} Q^{{}^{(\xi)}}_{{}_{\mathfrak{D\!h}}} = 2 Q^{{}^{(\xi)}}_{{}_{\rm PS}} \ee  
on DHSs, so that if the DHS segment joins on to the Schwarzschild IHS at a cross-section $S$, then  $Q^{{}^{(\xi)}}_{{}_{{\mathfrak{D\!h}}}}$ equals $M$ on $S$. Then if a non-equilibrium state $\n$ projects down to a Kerr equilibrium state $\e$ under the map $\Pi$ of section \ref{s5.1}, we find that the values of the charge  $Q^{{}^{(\xi)}}_{{}_{{\mathfrak{D\!h}}}} \big{|}_\n$ equals the value of $Q^{{}^{(\ko)}}_{\IH}$ at ${\e}$. (Similarly $J_{\DH}\big{|}_\n$ equals the value of $\JIH$ at ${\e}$, by the very definition of the map $\Pi$). These equalities hold along the entire trajectory $\e(R)$ defined by any given $\DH$. The non-triviality lies in the fact that the difference between charges $Q^{{}^{(\ko)}}_{\IH}\big{|}_{\e}$ (and $\JIH\big{|}_{\e}$) evaluated at any two equilibrium states on the trajectory in $\Ekerr$ is given by a flux integral on the corresponding $\DH$.  Furthermore, the integrand consists of \emph{local fields} $|\sigma|^2,\, |\zeta|^2, q_{ab}, k_{ab}$ and $T_{ab}$ on $\DH$, that know nothing  about the projection map $\Pi$, or the space of Kerr equilibrium states $\Ekerr$. This structure brings out an unforeseen synergy between non-equilibrium and equilibrium states of BHs. 

In section \ref{s6.1} we will show that this choice of $\xi^a$ also leads to the physical process version of the first law on any $\DH$ by itself.
\vskip0.1cm

\section{Thermodynamics of BHs far from Equilibrium}
\label{s6}

We will now use results from sections \ref{s4} and \ref{s5} to arrive at the first and second laws of BH thermodynamics for black holes that may be arbitrarily far from equilibrium.  As mentioned in section \ref{s1}, the first law will be in an active form, adapted to physical processes at $\DH$: It will relate changes in the key extensive observables on any two cross-sections $S_1$ and $S_2$ of a $\DH$ to fluxes of energy and angular momentum passing through the dynamical horizon segment $\Delta\H$ joining them.  The second law will provide a quantitative relation between the increase in the area of the DHS and the flux of energy falling into it.

\subsection{The Physical Process Version of the First Law for DHSs}
\label{s6.1}

Recall that, since the BCH first law (\ref{BCH}) refers to the energy $Q^{(\to)}$ defined by the stationary Killing field on the Killing horizon $\KH$, to obtain its generalization (\ref{1law1}) to IHSs we introduced the analogous symmetry vector field $\to^a \= \ko^a - \Omegao\,\varphio^a$ on $\IH$ and the corresponding charge $Q^{(\to)}_\IH$. In this construction, the required $\kappao$ and $\Omegao$ were provided by the projection map $\Pio$. On DHSs, we can follow the same procedure, now using the projection map $\Pi$ that associates, via pull-back, surface gravity $\kappao [S]$ and angular velocity $\Omegao [S]$ to each MTS $S$ on $\DH$. Thus, we are led to set $t^a = \xi^a - \Omegao \varphi^a$.%
\footnote{This vector field $t^a$ is space-like if $\JDH[S]$ is non-zero and null if it vanishes, mimicking the behavior of  $\to^a$ on IHSs, and in particular, on the Kerr $\KH$.} 
Then the charge and flux associated with $t^a$ can be expressed as linear combinations of charges and fluxes associated with $\xi^a$ and $\Omegao\varphi^a$. 

Let us begin with angular momentum.  Since $\Omegao$ is a constant on each MTS $S$,\, $\Omegao\varphi^a$ is also (tangential to and) divergence-free on each $S$. Therefore it can be used in place of $\varphi^a$ in the charge-flux balance law (see Eq. (\ref{Jbalance1})). It follows from Eq.(\ref{JDH1}) that the angular momentum associated with $\Omegao\varphi^a$  is given by:
\be \label{JDH4} \JDH^{{}^{(\Omegao\varphi)}} [S] \,\= - \, Q_{{}_{\rm PS}}^{{}^{(\Omegao\varphi)}}\,[S]\, \=\
-\f{1}{8\pi G}\, \, \oint_{S} \, \t\omega_a\, (\Omegao\varphi^a)\, \rmd^2 V\, \, \= \,\, \Omegao [S]\,  \JDH^{{}^{(\varphi)}} [S]\,  \ee 
since $\Omegao$ is constant on $S$. Next, let us use the definition $t^a \,\=\, \xi^a - \Omegao \varphi^a$. The phase space charges of the three vector fields have the same linear relation. Following the definition $M_{\IH} = 2 Q_{{}_{\rm PS}}^{{}^{(\to)}}$ on IHSs, let us set $\MDH [S] = 2 Q_{{}_{\rm PS}}^{{}^{(t)}} [S]$. Then, from (\ref{Qxi}) and (\ref{JDH4}) we have
\be \label{t-charge}  \MDH [S] \,\=\,  \big(\f{\kappao A}{4\pi G}\big)\,[S] \,+\, 2\big(\Omegao \JDH^{{}^{(\varphi)}}\big) [S]\, .\ee
Since the right side of (\ref{t-charge}) only involves fields on the MTS $S$, the charge $\MDH$ is defined intrinsically on each MTS of $\DH$. If the DHS reaches equilibrium --i.e., if we have a QLH on which a DHS joins on smoothly to an IHS at a MTS $S$--  then $M_{\DH}[S]$ on the DHS joins on smoothly to $M_{\IH}$ at $S$ because every quantity on the right side of (\ref{t-charge}) does.  Just as $\JDH [S]$ represents the `instantaneous angular momentum' of the DHS, $\MDH$ can be thought of as representing the `instantaneous mass'.
\vskip0.1cm

Thus, given any portion $\Delta H$ of $\DH$ bounded by two MTSs $S_1$ and $S_2$, we have:
\be \label{dh1law1} \Delta \big[M_{{}_{\DH}}\big]\, \,\=\, \Delta\, \big[\f{\kappao A}{4\pi G}\big]\, +\, 2 \Delta\, \big[\Omegao \JDH \big]\,, \ee
where, as before we have used the notation $\Delta \big[ M_{{}_{\DH}} \big]\, :=\, \big[M_{{}_{\DH}}\big]_{S_1}^{S_{2}}$. This is our \emph{dynamical} generalization of the first law (\ref{1law1}) for IHSs, which itself was a generalization of the BCH first law (\ref{BCH}) for KHs. \vskip0.05cm

While the form of (\ref{dh1law1}) is similar to the BCH 1st law (\ref{BCH}), or its generalization (\ref{1law1}) to IHSs, there are important conceptual differences between them. First, $\Delta$ in (\ref{dh1law1}) refers to \emph{finite} changes while $\delta$ in (\ref{1law1}) refers to infinitesimal ones. Second, $\delta M_{{}_{\IH}},\, \delta A$ and $\delta\JIH$ in (\ref{1law1}) represent infinitesimal changes in mass, area and and angular momentum of the IHS as one moves from a given space-time with an IHS to a nearby one. This is a passive version of the first law in that there is no physical process that is responsible for this change. By contrast, the finite change $\Delta M_{{}_{\DH}}$ in (\ref{dh1law1}) is caused by physical fluxes of energy and angular momentum across the segment $\Delta \H$ of a given DHS  :
\ba \label{balance3}\Delta {\MDH} &\,\=\,&  2 \,\int_{\Delta\H}\f{1}{8\pi G}\,\big[|DR|\, f(R)\,\,(|\t\sigma|^2+2|\t\zeta|^2)\,+\, (k^{ab}- k q^{ab})\, \Lie_{\Omegao\varphi}\,q_{ab}\,\,\big] +\,  T_{ab}\, t^a\h\tau^b \rmd^3V \nonumber\\
&\=&\,\,\, \Ftgws [\Delta\H]\, +\, \Ftmatt [\Delta\H] \,\,\,\hat\equiv\,\,\, \mathfrak{F}^{{}^{(t)}} [\Delta \H]
\ea
so that the first law can be rewritten as:
\be \label{dh1law2}\,\, \mathfrak{F}^{{}^{(t)}} [\Delta \H]\,\=\, \Delta\,\big[\f{\kappao A}{4\pi G}\big]\, +\, 2 \Delta\, \big[\Omegao \JDH \big]\,.\ee
(The factor of 2 in front of the integral on the right side of (\ref{balance3}) descends from the factor of 2 relating $\MDH$ and phase space charges.)\vskip0.1cm

The third difference between (\ref{1law1}) and (\ref{dh1law1}) is that $\kappao$ and $\Omegao$ in (\ref{dh1law1}) are `time-dependent' --they change from one MTS to another-- and therefore do not come out of the flux integrals. The fourth difference is that the right hand sides of (\ref{1law1}) and (\ref{dh1law1}) differ by a factor of 2. Finally, the fifth difference is that while in (\ref{dh1law1}) we have $\Delta(\kappao\, A)$ and $\Delta(\Omegao \JDH)$, in (\ref{1law1}) we have $\kappao\delta A$ and $\Omegao\delta \JIH$ (i.e., $\kappao,\, \Omegao$  are inside $\Delta$ on DHSs, while they are outside $\delta$ on IHSs).

It turns out that the last three  differences are intertwined. To make this transparent, let us consider the infinitesimal form of the integral version, obtained by letting $S_2$ approach $S_1$ so that they are infinitesimally separated. Then the DHS first law (\ref{dh1law1}) reduces to 
\be \label{dh1law3} \delta M_{{}_{\DH}} \, \=\, \delta [\f{\kappao A}{4\pi G}]\, +\,  2 \delta [\Omegao\, \JDH]\, ,\ee
where $\delta$ denotes infinitesimal changes at $S_1$, along $\DH$. Next, recall that every DHS defines a trajectory\, $\e(R)$\, on the space $\Ekerr$ of equilibrium states and $S_1$ corresponds to a point $\e_1$ in $\Ekerr$. Thanks to the properties of  the projection $\Pi$, the infinitesimal changes $\delta \MDH,\, \delta [\f{\kappao A}{4\pi G}]$ and $\delta [\Omegao \JDH]$ on $\DH$ equal infinitesimal changes $\delta M_{{}_{\IH}},\, \delta [\f{\kappao A}{4\pi G}]$ and $\delta [\Omegao \JIH]$  at $\e_1$, along $\e(R)$. Recall from section \ref{s3.3} that the Smarr relations imply an identity on $\EIH$:
\be \label{identity2} \f{A\, \delta \kappao}{4\pi G} + 2\JIH \delta\Omegao \= -\big(\,\f{\kappao\, \delta A}{8\pi G} + \Omegao \delta \JIH\,\big)\, .\ee
By pulling it back to $\N$ via $\Pi$, the infinitesimal version of (\ref{dh1law3}) can be rewritten as
\vskip0.2cm
\be \label{dh1law4}\delta M_{{}_{\DH}} \= \f{\kappao}{8\pi G}\,\delta A\,+\, \Omegao\, \delta \JDH\, ,\ee 
which has exactly the same form as the first law (\ref{1law1}) on IHSs. But, as discussed above, on DHSs the infinitesimal changes are caused by physical fluxes into $\DH$ (near $S_1$) in the given physical space-time; it is not a passive change from one equilibrium state to a nearby one in the space of solutions. Finally, it is interesting and important to note that the finite version (\ref{dh1law1}) of the first law cannot be rewritten in this form:  since $(\kappao, \Omegao)$ are dynamical on $\DH$, they vary along the trajectory\, $\e(R)$\, in $\Ekerr$, whence the identity (\ref{identity2}) does not hold beyond infinitesimal displacements along $\DH$.  Put differently, in the flux expression on the right side of (\ref{dh1law2}) $(\kappao, \Omegao)$ cannot be pulled outside the integral. ($\kappao$ enters through $f(R)$ in the flux integrand.) 

This concludes our discussion of the generalization of the first law to dynamical BHs in general relativity. 
\vskip0.1cm

\emph{Remarks:} 
\vskip0.1cm

1. The rich literature on DHSs contains a number of relations that carry the flavor of the first law (see, in particular, \cite{Ashtekar:2003hk}). They all follow from Einstein's equations. However, in all but one of these relations, individual terms do not have physical dimensions of energy as in the BCH first law (and the first law of thermodynamics.) The one with the same physical dimensions as the BCH first law featured the Hawking mass $M_{\rm Haw}$ in place $\MDH[S]$. While that relation is also an exact consequence of Einstein's equations, it does not reduce to the first law (\ref{1law1}) when a BH reaches equilibrium unless the equilibrium state is a Schwarzschild BH. Hence, it is not a satisfactory generalization of the BCH first law to dynamical situations. In those investigations, two key ingredients introduced in this paper were missing: (i) The protection map $\Pi$ of section \ref{s5.1}; and, (ii) the preferred vector field $\xi^a$ introduced in section \ref{s5.2} ensuring the seamless matching between the notions of energy on DHSs with those on IHSs when equilibrium is reached. This is why the previously known relations did not contain the first law (\ref{dh1law2}). 
\vskip0.1cm

2. To gain intuition for the flux integrals, it is instructive to consider the case when the angular momentum $\JDH[S]$ vanishes identically on a DHS. Of course this can happen in a spherical collapse where the area increase is entirely due to collapsing matter. But it can also occur in much more interesting dynamical situations in which there is gravitational radiation. Consider, in particular, a head-on collision of two BHs. In this case, the angular momentum vanishes both on the DHSs of the progenitors and the DHS of the remnant. But there is energy flux into the DHSs. In fact in this case, $\Fximatt$ vanishes and the growth in the horizon area is \emph{entirely} due to $\Fxigws$! 

More generally, when $\JDH=0$, we have $\xi^a = t^a$ and the `time-dependent' surface gravity $\kappao [S]$ on $\DH$, provided by the projection map $\Pi$  is given simply by $\kappao[S] = 1/2R$ (see Eq. (\ref{kerr-kappa})). Hence our $f(R)$ of Eq. (\ref{f(R)}) simplifies: $f(R) =\f{1}{2}$ and the first law (\ref{dh1law2}) reduces to:
\be \label{dh1law5}  \frac{1}{8\pi G}\,\int_{\Delta\H} \, \big[ |DR|(|\t\sigma|^2 \,+\, 2 |\t\zeta|^2)\big]\, \rmd^3V \, + 2 \int_{\Delta H} T_{ab}\, t^a \hat{\tau}^b \rmd^3 V\, \=\, \f{\kappao A}{4\pi G} [S_2]\, - \,\f{\kappao A}{4\pi G} [S_1]\,.\ee
The time dependence of $\kappao$ manifests in the fact that the change in $\MDH$ caused by the infalling flux of energy is \emph{not} just proportional to the change in the area $\Delta A$ (as it is in the infinitesimal version) because $\kappao [S_2] \not= \kappao[S_1]$. 

Finally note that, because $\JDH \=0$,\, $Q^{{}^{(\xi)}}_{\DH} = Q^{{}^{(t)}}_{\DH}$, and this horizon charge reduces to the Hawking mass: $M_{\rm Haw} = R/2G$. Hence the integrand of the left hand side represents infalling the `Hawking flux density' on $\Delta\H$. We will use this fact at the end of section \ref{s6.2}.

\subsection{A Sharper Version of the Second Law}
\label{s6.2}

Recall from section \ref{s5} that associated with every vector field $\xi^a = F\, k^a\, \equiv \sqrt{2} f(R) |DR|\, k^a$ on $\DH$ there is a charge $Q^{{}^{(\xi)}}[S]$, and  fluxes $\Fxigws [\Delta\H]$ and $\Fximatt [\Delta\H]$ on the canonical phase space, and they satisfy a balance law. For the choice (\ref{f(R)}) of $f(R)$ this general structure leads one to the integral form  (\ref{dh1law2}) of the first law. For completeness, we will now recall a result from \cite{Ashtekar:2003hk} showing that a different choice of $f(R)$ leads one to a sharper, quantitative version of Hawking's second law. 

To avoid confusion with symbols used in section \ref{s6.1}, we will use an underbar to refer to these choices. Let us set  $\ub{f}(R) := 4\pi R$\, and\, $\ulxi^a = \ub{F}\, k^a \,\equiv \,\sqrt{2}\, \ub{f}(R) |DR| k^a$. Then considerations that led to the balance law (\ref{balance1}) imply that the charge associated with $\ulxi^a$  is given by 
\be Q^{{}^{(\ulxi)}}_{{}_{\DH}} [S] = \f{A[S]}{4G}\,, \ee
and the corresponding balance law is:
\ba \label{dh2law1} \Delta \Big[ \f{A}{4G}\Big] \,\,\,\hat{\equiv}\,\,\, \f{A[S_2]}{4G} - \f{A[S_1]}{4G}\, \,&\=&\,\, \f{1}{8\pi G}\,\int_{\Delta\H}\! |DR|\,\ub{f}\,\,\big(|\t{\sigma}|^2 + |\t\zeta|^2\big) \rmd^3 V \, \, + \int_{\Delta\H} T_{ab}\,\ulxi^a \h\tau^b\, \rmd^3 V\, \nonumber\\
&\hat{\equiv}& \,\,\,\Fulxigws [\Delta\H]\,+\, \Fulximatt [\Delta\H]\, . \ea
Note that the contribution due to gravitational waves is always positive on a DHS and the matter contribution is non-negative if the dominant energy condition is satisfied. Hence $\Delta A$ is positive in classical GR. This is the second law for DHSs. 

As we noted in section \ref{s1}, Hawking's result \cite{Hawking:1971vc} on the growth of the area of EHs is a qualitative statement. Furthermore its area can grow along a segment that lies in a flat region of space-time because the growth is teleological. By contrast, on DHSs, (\ref{dh2law1}) provides an explicit, quantitative relation between the growth in the area of any portion of $\DH$ and the energy carried by gravitational waves and matter across that portion. There is no teleology and in particular there are no DHSs in flat regions of space-time. 

Historically, the BCH first law \cite{Bardeen:1973gs}, Hawking's second law \cite{Hawking:1971vc} and Bekenstein's thought experiments \cite{Bekenstein:1973ur,Bekenstein:1974ax} led to the identification of the area of cross-sections of EHs with a multiple of the BH entropy. But as we discussed in section \ref{s1}, this interpretation has severe limitations for dynamical BHs. In light of our generalizations (\ref{dh1law2}) and (\ref{dh2law1}) of the first and the second law to DHSs, the intuitive arguments that led to the identification of entropy with the area of EHs now imply that the left side of Eq. (\ref{dh2law1}) -- $A [S]/4G$-- should be interpreted as the entropy of a dynamical BH at the time instant represented by the MTS $S$ on it.  The integrand in the right side of (\ref{dh2law1}) can be interpreted as the entropy current.

In classical GR, for physical matter fields normally considered, the area always increases. As we discuss in section \ref{s6.4}, in the analysis of BH evaporation, what forms due to gravitational collapse is a DHS and what evaporates due to quantum processes is also a DHS. During evaporation the dominant energy condition is violated and the area of the evaporating DHS decreases in time.
\vskip0.1cm 

\emph{Remarks:} \vskip0.1cm

1. The first and the second laws resulted from the same linear combination (\ref{keyconstraint}) of constraint equations, but with different choices for the smearing function $f(R)$ and $\ub{f}(R)$. For the first law, our choice of $f(R)$ endows the vector field $\xi^a$ with the same dimensions as a time-translation in Minkowski space (i.e., $L^0 M^0$, in the c=1 units; it is dimensionless). Therefore the charge $Q^{{}^{(\xi)}}_{\rm PS}$ and fluxes $\Fximatt$ and $\Fxigws$ have dimensions of energy. On the other hand, to arrive at the second law we chose of $\ub{f}(R)$ such that the vector field {$\ulxi^a$} has dimensions of length, so that its charge $Q^{{}_{(\ulxi)}}_{{}_{\DH}}$, and fluxes $\Fulximatt,\, \Fulxigws$  have physical dimensions $ML$, i.e., dimensions of $A/4G$, rather than of energy. It is quite striking that replacing $\xi^a$ by $\ulxi^a$ to smear the same linear combination of constraints leads to quite different physical consequences. In the first case, we obtain the relation (\ref{dh1law1}) between the intensive parameters $\kappao, \Omegao$ and extensive observables $\MDH,\, \JDH,\, A$ that generalizes the first law, while in the second case, we obtain the relation (\ref{dh2law1}) that spells out how the area changes in response to the flux $\Fulximatt$ and $\Fulxigws$, providing us with a quantitative version of the second law. This is possible because, on DHSs, the first as well as the second law are ``balance laws" for appropriate charges . By contrast, in the first treatments of BH thermodynamics, the two laws had a different character: the BCH first law is an equality, while Hawking's second law is a qualitative inequality.
\vskip0.05cm

2. As in section \ref{s6.1}, it is instructive to consider the case when the angular momentum $\JDH[S]$ vanishes identically on a DHS. As noted there, this can happen in a spherically symmetric collapse in which case the change in area is caused only by the matter flux across $\Delta\H$, or in a head-on collision of BHs, in which case the growth is only due to the flux carried by gravitational waves. Since $\kappao[S] = 1/2R$ when $\JDH =0$,\,\, $\ub{f}(R)$ used in (\ref{dh2law1}) can be written as $2\pi/\kappao(R)$, and the second law takes the form
\ba \label{dh2law2} \Delta \Big[ \f{A}{4G}\Big] \,\,&\=&\,\, \f{1}{8\pi G}\,\int_{\Delta\H}\! \Big(\f{2\pi}{\kappao (R)}\Big)\, \,|DR|\,\big(|\t{\sigma}|^2 + 2 |\t\zeta|^2\big) \rmd^3 V \, +\, 2 \int_{\Delta\H} \Big(\f{2\pi}{\kappao(R)}\Big)\, T_{ab}\, t^a \h\tau^b\, \rmd^3 V \,   \nonumber\\
&\hat{\equiv}& \,\,\,\Fulxigws [\Delta\H]\,+\, \Fulximatt [\Delta\H]\, . \ea
where $t^a$ is the vector field we used in Eq. (\ref{dh1law5}) (i.e for $f(R)=\f{1}{2}$). Thus, we see that the integrand on the right side is $\big(\f{2\pi}{\kappao(R)}\big)$ times the flux-density of Hawking energy considered there. Note, however, that $\big(\f{2\pi}{\kappao(R)}\big)$ cannot be taken outside the integral, again because $\kappao$ is time-dependent on $\DH$. 

However, a simplification occurs if we consider `slowly evolving' DHSs \cite{Booth:2003ji} on which $T_{ab}t^a \hat\tau^b,$ \,$|\t\sigma_{ab}|,\, |\t\zeta_a|$ are small (compared to the scale set by the horizon radius) so that the horizon geometry changes adiabatically. To make this explicit, we let us introduce a smallness parameter $\delta$ to track the orders in smallness at which various terms contribute. Then the matter term on the right hand side contributes to order $\delta^{(1)}$ and the gravitational waves, to order $\delta^{(2)}$. Now, because $R$ is slowly varying on $\Delta \H$, the surface gravity $\kappao$ is constant to leading order. Hence, keeping terms only to first order, we have
\be \label{dh2law3}  \f{\kappao}{8\pi G}\,\, \Delta [\delta^{(1)} A]\, \=\, \int_{\Delta\H} \, (\delta^{(1)} T_{ab})\, t^a \h\tau^b\, \rmd^3 V\,\, \equiv \,\,\delta^{(1)} \Ftmatt\, , \ee
If there is no matter flux, as, for example in the head-on collision of black holes, to first order the area does not change  while to second order one has
\be \label{dh2law4}  \f{\kappao}{8\pi G}\,\, \Delta [\delta^{(2)} A]\, \=\, \delta^{(2)}\,\Ftgws \ee
However, on generic DHSs that are not slowly evolving the factor $(2\pi/\kappao)$ cannot come out of the integral in Eq. (\ref{dh2law2}). \vskip0.05cm

\subsection{Relation to Perturbative Approaches}
\label{s6.3}

Recently, generalizations of the first law have appeared in the perturbative context where one perturbs around stationary, non-extremal BHs \cite{PhysRevD.108.044069}, \cite{hollands2024entropydynamicalblackholes,visser2025dynamicalentropychargedblack}. Since BHs are non-extremal, the Killing horizon $\KH$ can be extended to the past until one arrives at a bifurcate 2-sphere $S_\circ$ on which horizon Killing vector $\ko^a$ (with surface gravity $\kappao$) vanishes. One considers affinely parametrized null geodesic vector fields $\b{k}^a$ on $\KH$ with affine parameter $\b{v}$ with $\b{v} =0$ on $S_\circ$ so that $\ko^a = \kappao\,\b{v}\,\b{k}^a$. We will denote the $\b{v} = {\rm const}$ cross-sections of $\KH$ by $S$. 

With this structure at hand, one considers perturbations on the background stationary solution. Let us denote the EH of the \emph{perturbed BH} by $\EH$. It is a teleologically defined null surface, that lies close to the KH $\KH$ of the background. By performing a diffeomorphism, one makes $\EH$  coincide with $\KH$. Then $\KH$ inherits a new metric from that on the perturbed horizon, and each $S$ acquires a new, slightly larger area. Let us denote this change by $\delta A_{\EH} [S]$. Using the Hamiltonian framework one can find the dynamical change  $\delta Q^{(\ko)} [S]$ in the value of the charge associated with the horizon Killing field $\ko^a$, and the corresponding flux $\delta \mathcal{F}^{(\ko)} (\Delta \KH)$ across any portion of $\KH$ bounded by two cross-sections $S_1$ and $S_2$. This balance law relates changes in areas to the flux, along the same lines as section \ref{s6.2} on DHSs. Keeping terms to \emph{first order} in perturbations, one obtains \cite{PhysRevD.108.044069,hollands2024entropydynamicalblackholes,visser2025dynamicalentropychargedblack}:
\be \label{p1law1} \f{\kappao}{8\pi G}\, \Delta [\delta^{(1)} \mathcal{A}]\, \=\, \delta^{(1)}\,\mathfrak{F}^{(\ko)}_{\rm matt}\, [\Delta \KH]\, , \ee
with 
\be \label{p1law2} \Delta [\delta^{(1)} \mathcal{A}]\, \=\, \Big[(1 - {v} \partial_{\b{v}}) \delta^{(1)} {A}_{\EH}\Big]_{S_1}^{S_2} \qquad {\rm and} \qquad 
\mathfrak{F}^{(\ko)}_{\rm matt} [\Delta \KH]\,\=\, \int_{\Delta \KH}\!\! (\delta^{(1)} T_{ab}) \ko^a \ko^b \rmd^3 V\, .\ee
If the perturbation is purely due to gravitational waves, then the area $\mathcal{A}[S]$ changes only to second order and one obtains:
\be \label{p1law3} \f{\kappao}{8\pi G}\, \Delta [\delta^{(2)} \mathcal{A}]\, \=\, \delta^{(2)}\,\mathfrak{F}^{(\ko)}_{\rm gws}\, [\Delta \KH], \, \ee
where 
\be \label{p1law4} \Delta [\delta^{(2)} \mathcal{A}]\, \=\, \Big[(1 - \b{v} \partial_{\b{v}}) \delta^{(2)} {A}_{\EH}\Big]_{S_1}^{S_2}\,\,\,, \qquad \qquad 
\mathfrak{F}^{(\ko)}_{\rm gws} [\Delta \KH]\,\=\, \int_{\Delta \KH}\!\! \kappao\, |(\delta^{(1)} \t\sigma|^2 \rmd^3 V\,, \ee
and $\delta^{(1)} \t\sigma_{ab}$ is the shear to first order in perturbations. These are the perturbative generalizations of the BCH first law to dynamical situations. Note that the changes in area $\delta^{(n)} \mathcal{A}$ that feature in these laws are \emph{not} just the changes $\delta^{(n)} A_{\EH}$ in the EH-area. There is an extra term, bringing out the fact that already in the perturbation theory \emph{entropy is not given by the area of the EH!} Indeed, the changes $\Delta [\delta^{(n)} {A}_{\EH}]$ of the event horizon area are teleological; they depend on the energy flux also to the future of $S_2$. A priori one might think that $\Delta [\delta \mathcal{A}]$ is also teleological because it is built from $\delta {A}_{\EH}$. But this is not the case since the flux $\mathfrak{F}^{(\ko)} [\Delta \KH]$ on the right side depends only on what happens between $S_1$ and $S_2$.

Could we have avoided the introduction of the EH altogether? In the QLH framework the answer is in the affirmative. 
As we emphasized earlier, in this framework one does not need a stationary background metric; in place of KHs $\KH$ one uses IHSs $\IH$, and replaces the $\EH$ of the perturbed metric with a perturbed-$\IH$. Thus, one has to extend our discussion of  sections \ref{s2} and \ref{s3} to a perturbed-$\IH$. This procedure was carried out in detail in \cite{Ashtekar:2021kqj}. One can construct the phase spaces of solutions to Einstein's equations that admit an IHS, and a perturbed IHS, as their inner boundary. The null normals $\ko^a$ of IHSs are symmetry vector fields that preserve the universal structure at the boundary, just as Killing symmetries of an asymptotically flat space-time preserve the universal structure at $\scrip$ (see appendix \ref{a1.1}). Therefore there is a 2-sphere charge and 3-surface flux associated with $\ko^a$, subject to a balance law. Since the focus of \cite{Ashtekar:2021kqj} is on vacuum solutions, the fluxes vanish to the first order in perturbation theory and a non-trivial balance law emerges at the second order. It has the same form as  Eq.(\ref{p1law3}), and  $\mathfrak{F}^{(\ko)}_{\rm gws}$ is in fact the same as in (\ref{p1law4}) (with $\Delta \KH$ replaced by $\Delta \IH$). But now the expression of $\Delta [\delta^{(2)} \mathcal{A}]$ is different:
\be  \label{p1law5} \Delta [\delta^{(2)} \mathcal{A}]\, \=\, \Big[ (1 - \kappao^{-1} \ko^a \partial_a)  \delta^{(2)} {A}_{\IH} \Big]_{S_1}^{S_2} \ee
where $\delta^{(2)} A_{\IH}[S]$ is simply the change in the area of the cross-section $S$ of $\IH$ induced by the presence of gravitational waves (see Eqs. (4.8) and (4.10) of \cite{Ashtekar:2021kqj}). On a Killing horizon we trivially have $\kappao^{-1} \ko^a \partial_a\, \=\, \b{v}\partial_{\b{v}}$. Thus (\ref{p1law5}) is a generalization of (\ref{p1law4}) with two simplifications: (i) One does not have to introduce $\EH$ at all; one works directly with the perturbations on the inner boundary; and (ii) one does not need the bifurcate 2-sphere.\vskip0.1cm

Next, let us return Eqs (\ref{p1law2}) and (\ref{p1law4}) and examine the expression  {\smash{$\f{1}{4 G}\,\Delta [(1 - \b{v} \partial_{\b{v}} ) \delta^{(n)} {A}_{\EH}]$}} of the change  in entropy to better understand the underlying structure. This function turns out to have a geometrical interpretation. Each $S$ on $\KH$ has an associated MTS $\bar{S}$, defined by the perturbed metric. $\b{S}$ is close to $S$ and lies inside the perturbed $\EH$. The expression under consideration  represents the difference between the areas of the MTSs $\b{S}_1, \, \b{S}_2$ defined by the perturbed metric \cite{hollands2024entropydynamicalblackholes}. Thus, \emph{the world-tube of these MTSs $\b{S}$ is a QLH, and the entropy refers to the area of cross-sections of this QLH!} Assuming that the null energy condition holds, Eq. (\ref{p1law1} - (\ref{p1law4}) imply that the area of these MTSs increases monotonically. Therefore it is natural to expect that world tube to be a space-like, \emph{slowly evolving} DHSs $\DH$ \cite{Booth:2003ji} and the area $\mathcal{A}$ that directly enters the first law (\ref{p1law1}) to be just the area of the MTSs that foliate this $\DH$. 

This expectation is borne out in detail in a simple example: A Vaidya collapse of null fluid on a pre-existing Schwarzschild space-time of mass $M_1$, ending with a Schwarzschild solution of mass $M_2 >M_1$. Let us use the Eddington-Finkelstein coordinates of the Vaidya metric and restrict ourselves to a mass function $M(v)$ that is changing adiabatically from $M_1$ to $M_2$, so that we have a slowly evolving $\DH$. Then, given any two MTSs $S_1$ and $S_2$ on $\DH$, corresponding to $v=v_1$ and $v=v_2$,  Eq. (\ref{dh2law3}) implies 
\be  \label{p1law6}\Delta [\delta^{(1)} A_{\DH}]\, \=\, \f{8\pi G}{\kappao}\, \mathfrak{F}^{(\ko)}_{\rm matt}\,\=\, \f{8\pi G}{\kappao}\, (M(v_2) - M(v_1))\, ,\ee
where $\kappao = 1/4M_1$, and $\delta^{(1)} A_{\DH}$ is the change in the area of the MTSs on $\DH$ under the adiabatic evolution. On the other hand, one can regard the full space-time geometry as a perturbation on the initial Schwarzschild solution of mass $M_1$, assuming that $\EH$  of the perturbed space-time is close to the $\KH$ of the initial Schwarzschild space-time. Keeping terms that are only first order in the deviation $\delta M(v)/M_1$, one can calculate the change $\delta^{(1)} A_{\EH}$ in the area  of the spherically symmetric cross-sections of $\EH$. One then finds:
\be  \label{p1law7} \Delta [(1 - \kappao^{-1}\, \ko^a \partial_a) \delta^{(1)} A_{\EH} ] = \f{8\pi}{\kappao}\, (M(v_2) - M(v_1))\, . \ee
Thus the left hand sides of (\ref{p1law6}) and (\ref{p1law7}) --the first referring to the area of MTSs of the slowly-evolving-$\DH$, and the second to the area of cross-sections of the perturbed-$\EH$-- are equal. But, as we noted above, the left hand side of (\ref{p1law7}) has the interpretation of the change in the difference in the area in MTSs $\b{S}_2$ and $\b{S}_1$ of the perturbed metric (that lie inside the EH). Therefore, the QLH defined by the surfaces $\b{S}$ is just the slowly evolving  DHS. (In this discussion we did not need the bifurcate horizon because the preferred cross-sections $\bar{v} = {\rm const}$ on the $\KH$ of the background were picked out by spherical symmetry.) For a generic perturbed $\EH$, some technical assumptions may be necessary to show that the QLH foliated by the MTSs $\bar{S}$  constitute a slowly evolving DHS. This is an interesting open issue. 

Thus, the perturbative calculation can be interpreted in two ways from the QLH perspective. The direct way is in terms of perturbed IHSs. It leads to the desired, perturbative extensions of the first laws. But, given the interpretation of the perturbative expression of entropy in terms of areas of the MTSs $\b{S}$, it is more illuminating to regard these $\b{S}$ as MTSs of a slowly evolving $\DH$, since this identification directly connects the perturbative notion of entropy with the non-perturbative one. These two ways of interpreting perturbative results  are complementary. The perturbed-$\IH$ route starts with an equilibrium state and tries to incorporate the leading effects of the non-equilibrium dynamics perturbatively. The $\DH$ route follows the opposite strategy: It starts with fully non-equilibrium situations and examines the approach to equilibrium by considering slowly evolving DHSs. But the two strategies are rather different --not only conceptually, but also technically. In the first, the emphasis is on \emph{null structures} associated with $\IH$ and perturbed-$\IH$, while in the second, it is on fields on \emph{space-like} surfaces associated with $\DH$ and slowly-evolving-$\DH$. Since the transition from structures associated with space-like surfaces to those associated with null surfaces is delicate, subtleties arise. For example on DHSs, the gravitational field has four phase space degrees of freedom and the energy flux carried by gravitational waves has two fields, $\t\sigma_{ab}$ and $\t\zeta_a$, that persist also on slowly evolving DHSs (see Eq. (\ref{dh2law2}). On null surfaces there are only two degrees of freedom and so only $\t\sigma_{ab}$ appears in the energy flux (see Eq. (\ref{p1law4}). A better understanding of such subtleties would be quite rewarding.

To summarize, while the recent perturbative generalization of the first law \cite{PhysRevD.108.044069,hollands2024entropydynamicalblackholes,visser2025dynamicalentropychargedblack} emphasizes EHs and bifurcate 2-spheres, from a QLH perspective, when restricted to GR, these results are consequences of the known balance laws on perturbed IHSs.  The QLH-analysis does not need EHs, nor the bifurcate surface of KHs. If one does start with EHs, one inherits teleological baggage and some work is needed to show that the expression {\smash{$\f{1}{4 G}\,\Delta [(1 - \b{v} \partial_{\b{v}} ) \delta^{(n)} {A}_{\EH}]$}} of entropy-difference in terms of the area $A_{\EH}$ is free of teleology. However, this seemingly mysterious expression  has been shown to have a nice geometrical interpretation, as areas of certain MTSs inside the perturbed EH. It is likely that the world tube of these MTSs constitutes a slowly evolving   DHSs \cite{Booth:2003ji}. A detailed analysis establishing this result would open a useful corridor between perturbative and fully dynamical treatments of BH thermodynamics.\vskip0.1cm

\emph{Remarks:} 
\vskip0.1cm

1. An intermediate step in arriving at the perturbative first law --either using the perturbed-$\IH$ framework, or the more recent discussions-- involves the Noether charge $\f{\kappao}{8\pi G}\, \mathcal{A}$. This is also the case in older derivations of the first law using a covariant phase space and Killing horizons $\KH$. This has led to an oft-repeated phrase: \emph{Entropy is a Noether charge}. This characterization is misleading because the entropy in question is $\f{\mathcal A}{4G}$, which is \emph{not} a horizon Noether charge associated with any symmetry. Indeed the two have different physical dimensions, and are related by a factor of $\f{\kappao}{2\pi}$. Furthermore, far from equilibrium, since there is no stationary background to anchor the discussion, $\kappao$ is not constant, whence the entropy and the Noether charge are distinct. Mixing the two can cause unnecessary confusion especially  among beginning researchers. \vskip0.2cm

2. Our discussion of the first law using IHSs and DHSs is restricted to general relativity. The perturbative analyses \cite{PhysRevD.108.044069,hollands2024entropydynamicalblackholes,visser2025dynamicalentropychargedblack}, on the other hand, have tremendous generality because they extend to any diffeomorphism-covariant metric theory of gravity. Note however, that the notions of QLHs extend to all these theories and so do Hamiltonian frameworks we used to define charges. Therefore, as will be discussed elsewhere, our derivation of the first law for IHSs (in section \ref{s3.3}) goes through for a large class of relativistic theories of gravity, including the $f(R)$ and scalar-tensor theories theories that have received considerable attention. For DHSs, our discussion of the first and the second laws does not use full Einstein's equations. We only needed  the constraint equations on $\DH$. That is, we only used the Gauss-Codazzi equations --which, being differential geometric identities, hold in any metric theory of gravity-- and then replaced $G_{ab} \xi^a \h{\tau}^b$ by $ 8\pi G\,T_{ab}\xi^a \h\tau^b$. 
In $f(R)$ theories of gravity, our analysis of the first and the second law goes through directly in the Einstein frame, and entropy is again given by the area of the MTSs of the DHS. It would be interesting to investigate if the current analysis can be extended to other admissible metric theories of gravity. 

\subsection{Time-like DHSs} 
\label{s6.4}

We will conclude this section with a discussion of time-like DHSs. In classical GR, one does encounter dynamical situations --e.g. in the Oppenheimer-Snyder collapse of a \emph{homogenous} fluid, depicted in the left panel of Fig. 4-- in which the world-tube of MTSs is time-like. %one does encounters them, e.g. in the Oppenheimer-Snyder collapse of a \emph{homogenous} fluid, depicted in the left panel of Fig. 4. 
It is clear from the figure that displacements in any space-like direction $s^a$, with $k^a s_a >0$, map every MTS on the world-tube of MTSs into the trapped (rather than untrapped) region. Thus, every MTS on this world-tube is unstable, whence, as per our discussion in section \ref{s2}, this case is excluded from our thermodynamical considerations. More generally, if a strictly stable MTS $S$ lies in a QLH and the energy condition $|\sigma_{(k)}|^2 + T_{ab} k^a k^b >0$ holds somewhere on $S$, then the QLH is space-like near $S$ \cite{Andersson:2007fh}. Therefore time-like DHSs are not of interest for our considerations in classical GR.

However, interesting time-like DHSs occur \emph{generically} as the BH evaporates due to quantum radiation \cite{Ashtekar:2005cj,Sawayama:2005mw,Ashtekar_2011a,Ashtekar_2011b,ashtekar2023regularblackholesloop,Agullo_2024,Varadarajan:2024clw}. The semiclassical phase of this process is depicted in the right panel of Fig. 4. In this case, the space-like DHS formed during the null fluid collapse, as well as the time-like DHS that develops during the evaporation process, have MTSs that are strictly stable,  but the energy condition is violated on the time-like portion. Therefore our thermodynamical considerations apply to both these DHSs. (They also apply to the anti-trapping DHS, expected to develop in loop quantum gravity beyond the semiclassical approximation; see Fig. 4b in \cite{ashtekar2025blackholeevaporationloop}.)
\begin{figure}[] 
\vskip-0.5cm
 \includegraphics[width=1.7in,height=2in]{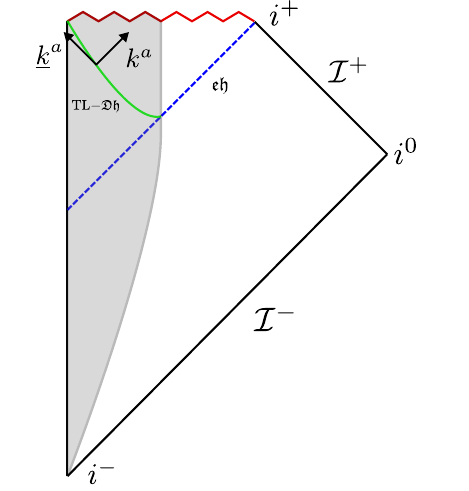}
 \hskip2.5cm 
  \includegraphics[width=1.4in,height=2.1in,angle=0]{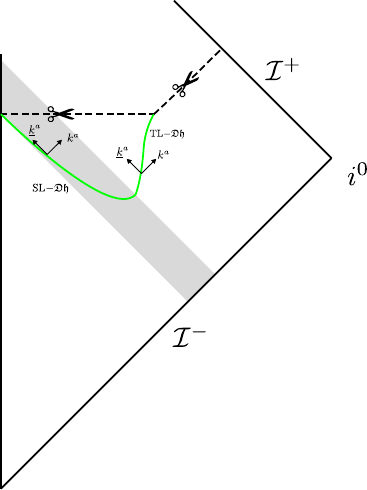} 
  \caption{\footnotesize{\emph{Left Panel: Oppenheimer-Snyder collapse.}\, $\DH$ is time-like if the collapsing star is homogeneous, or nearly homogenous.\,\,\, {\emph{Right Panel: Semi-classical phase of a spherical, evaporating BH.}}\, To investigate the issue of information loss, the incoming state has to include the collapsing matter. Therefore, a stellar collapse is unsuitable but one can consider collapse of a massless scalar field that is in a coherent state on $\scrim$. Then, in the semi-classical approximation, the DHS formed during the collapse is space-like. At the end of the collapse it is instantaneously null and then becomes time-like in response to the infalling negative energy quantum flux. The portion of space-time that lies outside the domain of applicability of the semi-classical approximation has been excised.}} 
\vskip-0.5cm
 \label{fig:4} 
\end{figure}

Let us begin our discussion of these time-like DHSs by introducing some notation. Since the normal to time-like $\DH$ is space-like, let us denote it by $\h{r}^{\prime\,a}$. Similarly, since the normal to the MTSs within $\IH$ is now time-like, let us denote it by $\h{\tau}^{\prime\, a}$. The intrinsic metric $q_{ab}$ on $\IH$ has signature -,+,+ and the extrinsic curvature of $\DH$ is given by $k_{ab}^\prime = q_a{}^c q_b{}^d\, \nabla_c \h{r}^{\prime}_d$. The constraint equations on $\DH$ are now given by (see Appendix B of \cite{Ashtekar:2003hk})
\ba \label{constraints3} H^\prime &:=& -\mathcal{R} - k^{\prime\,2} +k^{\prime\,ab}k^\prime_{ab} - 16\pi G\, T_{ab}\, \h{r}^{\prime\,a} \h{r}^{\prime\, b}\, \=\,0, \quad {\rm and}\nonumber\\ 
H^{\prime\,a} &:=& D_b(k^{\prime\,ab} - k^\prime q^{ab}) -8\pi G\, T_{bc}\, \h{r}^{\prime\,b} q^{ac}\,\=\,0\, . \ea
Thus the main changes are in the sign in front of the scalar curvature term in the Hamiltonian constraint, and the exchange between $\h\tau^a$ and $\h{r}^{\prime\,a}$ in the stress-energy terms. With these changes, our discussion of angular momentum of MTSs discussed in section \ref{s3} goes through in a straightforward manner. However, to our knowledge, investigations of BH evaporation that take into account the back reaction have been carried out only in spherically symmetric cases.  In this case, the gravitational-wave contribution to the flux vanishes and the first law (\ref{dh1law5}) for space-like DHSs is replaced by
\be - 2 \,\int_{\Delta\H}\!\! \big[T_{ab}\, \xi^{a}\h{r}^{\prime\,b}\,\big] \rmd^3V\,\, \=\,\,
 \f{\kappao A}{4\pi G} [S_2]\, - \,\f{\kappao A}{4\pi G} [S_1]\,,\ee
and the infinitesimal form (\ref{dh1law4}) by
\be \delta \MDH \, =\, - \f{\kappao}{8\pi G}\, \delta A[S]\,. \ee
Thus, the form of the first law remains the same but the flux is no longer positive.  The essence of this equality was already known \cite{Ashtekar:2003hk}, but it was phrased in terms of the decrease in the horizon radius. There was no  reference to surface gravity since the strategy of associating intensive thermodynamic parameters to non-equilibrium states discussed in section \ref{s5} was not available at the time.

Finally, let us consider the second law. Eq. (\ref{dh2law2}) for space-like DHSs is replaced by
\be \label{dh2law5} \Delta \Big[ \f{A}{4G}\Big] \,\,\=\,\, \, - \int_{\Delta\H} \Big(\f{4\pi}{\kappao(R)}\Big)\, T_{ab}\, t^a \h{r}^b\, \rmd^3 V \, ,\ee
During the evaporation of a non-rotating BH (as well as in the Oppenheimer-Snyder collapse), the right side is such that the area $A$ decreases to the future.

\section{Discussion}
\label{s7}

This section is divided into two parts. In the first we summarize the main results, emphasizing the new conceptual ingredients. In the second we make a number of remarks, putting our results in a broader context.

\subsection{Summary} 
\label{s7.1}

Einstein's equations imply that there is a multitude of interesting relations between geometrical quantities on the Killing horizons $\KH$ of stationary, axisymmetric BHs. BCH \cite{Bardeen:1973gs} isolated a particularly interesting one, Eq. (\ref{BCH}), that has an astonishing similarity with the first law of thermodynamics. However, from a thermodynamical perspective, the requirement of stationarity and axisymmetry on the entire space-time is over-restrictive; the thermodynamical first law requires only the system to be in equilibrium, imposing no restriction on the rest of the universe. In section \ref{s3.3} we extended the BCH first law to general \emph{isolated horizons} $\IH$ which represent BHs which are themselves in equilibrium as desired, but allowing for dynamical processes and gravitational radiation in the rest of space-time. However, the $\IH$-first law, (\ref{1law1}), is still `passive' in the sense that the changes $\delta$ refer to the properties of nearby IHSs in the \emph{space of solutions} of Einstein's equations, rather than to shifts in the values of BH observables caused by physical processes in a given space-time. Also, these shifts are only infinitesimal. The generalization of the first law of section \ref{s6.1} is more radical. It overcomes both these limitations by replacing IHSs $\IH$ with DHSs $\DH$ representing dynamical BHs that can be very far from equilibrium. The resulting first law (\ref{dh1law2}) refers to \emph{finite} processes in the \emph{physical space-time}, and changes in the values observables are caused by fluxes of energy and angular momentum across $\DH$. 

The BCH analysis makes a heavy use of structures at spatial infinity, far away from the BH, both in its use of ADM charges,  and in its normalization of the stationary Killing field $\to^a$. As discussed in section \ref{s1}, to mimic the familiar laws of thermodynamics, both extensive observables and intensive parameters should refer just to individual BHs and not to what is outside. To achieve this goal we had to introduce of several new ingredients. 

First, already for IHSs we had to shift the starting point from $\to^a$ to the family $[\ko^a]$ of $\IH$ null-normals, that feature in the very definition of an $\IH$ as symmetries of its geometry. Any two null normals in this family are related by a rescaling with a positive constant. Our second observation was that extensive observables $(\RIH, \, \JIH)$ can be introduced using only that structure which is intrinsically available on any $\IH$, and that the rescaling freedom in $[\ko^a]$ can be naturally eliminated by fixing the surface gravity $\kappao$ of $\ko^a$ to be  $\kappao_{\rm kerr} (\RIH, \JIH)$, given by (\ref{kerr-kappa}). This strategy selects a unique null normal $\ko^a$ from $[\ko^a]$ on \emph{any} $\IH$, and has the attractive feature that, on a Kerr $\KH$, it picks out the `correct' horizon null normal using only the fields available at the horizon, without having to refer to spatial infinity to first fix the normalization of $\to^a$. One can calculate the charge $Q_\IH^{(\ko)}$ associated with this symmetry either using Hamiltonian considerations (as in section \ref{s3.1}), or using field equations (as in Appendix \ref{a2}). In both cases, one only uses the structure that is directly available  on $\IH$. The third ingredient was the introduction of the IH-symmetry $\to^a$ as a linear combination (of the primary ones), $\to^a\, \=\, \ko^a - \Omegao\varphio^a$, where $\varphio^a$ is the rotational vector field on $\IH$ (given by the Korzynski construction, discussed in Appendix \ref{a2}), and $\Omegao$ is set equal to $\Omegao_{\rm kerr}$, which is a function of $(\RIH, \JIH)$ specified in (\ref{kerr-Omega}). Thus, in contrast to the BCH analysis, where one begins with $\to^a$, normalized to be a unit time translation at infinity, and then defines $\ko^a$ via $\ko^a := \to^a + \Omegao\,\varphio^a$,\, we adopt a perspective that is intrinsic to the horizon, where $\ko^a$ is primary and $\to^a$, secondary. The charge $\MIH$ associated with $\to^a$ is given simply by the linear combination $\MIH \,\=\, Q^{{}_{(\ko)}}_{\IH} - \Omegao\, Q_{\IH}^{{}_{(\varphio)}}$. With this infrastructure, the $\IH$-first law (\ref{1law1}) followed from the fact that, thanks to the specific functional dependence of $\kappao$ and $\Omegao$ on $(\RIH,\, \JIH)$, we have the identity (\ref{identity1}). Note that these new ingredients were essential to extend (\ref{BCH}) on KHs to (\ref{1law1}) on IHSs.

They also paved the way to the more radical generalization to DHSs $\DH$ in sections \ref{s5} and \ref{s6}. Just as the geometric fields defined on an $\IH$ define the state of a BH in equilibrium, fields defined on any MTS $S$ of a $\DH$ determine the non-equilibrium state $\n$ of the BH at the `instant of time' represented by $S$. In both cases, the space of these fields is infinite dimensional, and furthermore on a $\DH$ these fields are `time-dependent', varying from one MTS to another. Now, these fields determine, in particular, the horizon radius and angular momentum: $(\RIH,\, \JIH)$ on $\IH$,\, and $(\RDH [S],\,\JDH[S])$ on any MTS $S$ of a $\DH$. It is also true that thanks to the BH uniqueness theorem, the $\IH$-geometry of any Kerr horizon is also completely characterized by the pair $(\RIH,\, \JIH)$. Therefore, there are natural projection maps, $\Pio$ from the space $\EIH$ of $\IH$-equilibrium states,\, and $\Pi$ from the space $\N$ of non-equilibrium states,\, to the 2-dimensional space $\Ekerr$ of Kerr equilibrium states. As we summarized above, there is a natural assignment of $(\kappao,\, \Omegao)$ to $\IH$ that uses just the intrinsic structure of $\IH$. It is given by the \emph{pull-back} under $\Pio$ of $(\kappao_{\rm kerr},\, \Omegao_{\rm kerr})$ from $\Ekerr$ to $\EIH$. This pull-back ignores all the rich structure contained in the equilibrium state $\mathring\e$ and focuses only on its two primary observables $(\RIH,\,\JIH)$. This led us to use the same procedure for non-equilibrium states: we used the pull-back under $\Pi$ to associate the intensive parameters $(\kappao,\,\Omegao)$ to each non-equilibrium state $\n$. The two numbers can be thought of as representing an `average temperature' and an `average angular velocity' of the BH, at the `time' represented by $S$. While for a BH in equilibrium, the intensive parameters are associated with its entire $\IH$, for a dynamical BH, they are associated with each MTS $S$ on $\DH$. Thus, they are `time dependent' just as one would physically expect. Still, at first it is surprising that one can assign intensive parameters to non-equilibrium states $\n$. This was made possible by the BH uniqueness theorems. They imply that $(\RIH, \JIH)$ suffice to completely determine the \emph{micro-state} of the Kerr $\IH$-geometry.  Specifically, they determine all the multiple moments that provide a complete, invariant characterization of the micro-state of any given $\IH$. This is in striking contrast to ordinary thermodynamical systems where there are \emph{infinitely many micro-states} compatible with pre-specified values of extensive quantities (such as $(S,\, V,\, N)$) that feature in the first law.

Our first law (\ref{dh1law2}) for DHSs has three notable features. First, the extensive quantities $(\MDH [S],\, \RDH [S],\, \JDH [S])$ that appear in this law are defined intrinsically on each MTS $S$ of any $\DH$, just as the extensive 
quantities $(\MIH,\, \RIH,\, \JIH)$ that appear in the first law (\ref{1law1}) on IHSs are defined intrinsically on any $\IH$, using only the structure that is naturally available there. Yet, each time a BH reaches equilibrium so that its DHS joins on to an IHS at an MTS $S$, the extensive quantities on the DHS agree with those on the IHS. Second, the change $\Delta [\f{\kappao A}{4\pi G}] \,+\, 2\Delta [\Omegao \JDH]$ between two MTSs $S_1$ and $S_2$ is given by the \emph{physical flux} $\mathfrak{F}^{{}^{(t)}} [\Delta \H]$ of $t^a$-energy falling into the segment $\Delta \H$ of $\DH$ bounded by $S_1$ and $S_2$. This flux receives contributions both from matter fields and gravitational waves traversing $\Delta\H$ and its integrand is \emph{locally defined}. Third, the intensive parameters $(\kappao [S],\,\Omegao [S])$ in the flux expression appear inside the integral sign and \emph{can not} be taken out because they are time-dependent. This is precisely the non-triviality one would expect in the passage from equilibrium situations  to non-equilibrium ones. Nonetheless, if $S_2$ is infinitesimally separated from $S_1$ then the integral expression reduces to the familiar form (\ref{dh1law4}) of the first law, with the intensive parameters outside the infinitesimal variations $\delta$. In section \ref{s6.3}, we discussed in detail the relation between these first laws, based on QLHs, and the recent discussions based on perturbations off stationary space-times.

Finally, given \emph{any} $\DH$, since fields on each of its MTSs $S$ define a non-equilibrium state, dynamics on $\DH$ is faithfully represented by a curve in the infinite dimensional space $\N$ of all non-equilibrium states $\n$. We found that, thanks to the projection map $\Pi$, this dynamical trajectory in $\N$ provides an unique shadow trajectory in the 2-dimensional space $\Ekerr$ of Kerr equilibrium states. The projection corresponds to a huge coarse-graining that ignores all the details of the dynamical process (captured, e.g., in the $\DH$-multipole moments) and focuses on the evolution of just two observables --$(\RDH [S],\, \JDH[S])$. Although one loses `most' of the information in the non-equilibrium evolution because of this massive coarse-graining, the projected trajectory is very helpful for thermodynamical considerations since it faithfully captures the evolution of the two extensive observables that feature in the first law: One can represent the non-equilibrium evolution in terms of the shadow trajectory, \emph{but only for thermodynamical considerations}. This interplay is both interesting and subtle. In particular, one can envisage two DHSs $\DH$ and $\DH^\prime$ in two different space-times that happen to have cross-sections $S_1, S_2$ and $S_1^\prime, S_2^\prime$ such that the area and angular momentum on $S_1, S_2$ agree with those on $S_1^\prime, S_2^\prime$, respectively. Then the two DHSs will define distinct trajectories on $\Ekerr$ that intersect at two points (the Kerr states $\e_1,\, \e_2$ corresponding to $\n_1,\,\n_1^\prime$ and  $\n_2,\,\n_2^\prime$, respectively.) The full non-equilibrium dynamics in the two processes can be very different from one another. It is not just that the dynamical fields, such as $\t\sigma_{ab},\, \t\zeta_a, \t\omega_a$, that encode the full dynamical information on each $\DH$ will have no relation to those on $\DH^\prime$, but the evolution of the thermodynamical extensive variables $(\RDH [S],\, \JIH [S])$ on the two projected trajectories in $\Ekerr$ will also be unrelated except at the end points $\e_1,\, \e_2$. Yet, since the first law (\ref{dh1law2}) holds on both DHSs, we will have  $\mathfrak{F}^{{}^{(t)}} [\Delta \H] = \mathfrak{F}^{{}^{(t^\prime)}} [\Delta \H^\prime]$; the \emph{total} fluxes will be equal. Such non-trivial interplays between non-equilibrium evolutions in $\N$ and their shadow trajectories on $\Ekerr$ may well have deeper significance.

For completeness, we also recalled from \cite{Ashtekar:2003hk} the second law on DHSs. Hawking's second law $\Delta A \ge 0$ refers to EHs \cite{Hawking:1971vc}. As we emphasized in section \ref{s1}, since EHs are teleological, the increase in area cannot be attributed to a physical process in the vicinity of the portion of the EH where the area is increasing. Furthermore, in fully dynamical situations, EHs are generically non-smooth because of creases, corners and caustics \cite{gadioux2023creasescornerscausticsproperties,Rojas_2025};
%second reference added as per referee's suggestion.
in fact they can even be nowhere differentiable \cite{Chrusciel:1996tw}! Therefore, one cannot envisage a simple equation governing the dynamics of a generic EH. By contrast, without loss of generality, DHSs can be taken to be smooth. Therefore, it is possible to go beyond the qualitative statement  $\Delta A \ge 0$ and provide an exact equality (\ref{dh2law1}) relating the change $\Delta A$ of the area of MTSs on a DHS to the influx of energy into it. 

Perhaps the central lesson from our generalizations of the first and the second law is that the thermodynamic entropy of dynamical BHs in GR is given, not by the area of cross sections of the EH $\EH$, but rather, by the area of MTSs that foliate DHSs $\DH$.

\subsection{Remarks}
\label{s7.2}

In this subsection we collect miscellaneous remarks on the interplay between equilibrium and non-equilibrium states of BHs, the relation of our results to those in the literature, open issues and directions for future work.\vskip0.1cm

1. As we just discussed, we used a key property of BHs in general relativity (with vanishing cosmological constant): There is a unique 2-parameter family of stationary states --the Kerr solutions--  and, furthermore, the pair $(\RIH[S],\, \JIH[S])$, defined intrinsically on any IHS $\IH$, completely fixes the \emph{micro-state} of the Kerr-horizon. That is, it determine \emph{all} the multipole moments that provide a complete, invariant characterization of its geometry, encoded in the pair $(\qo_{ab}, \Do)$. This is in striking contrast to ordinary thermodynamical systems where there are \emph{infinitely many equilibrium micro-states} compatible with pre-specified values of extensive quantities (such as $(S,\, V$) that feature in the first law.

In spite of this key difference, there is a complete parallel between our \emph{final picture} and the standard thermodynamic one. In usual thermodynamics, the total energy $E$ of a system is assumed to be a function of the extensive parameters such as $(S,\, V)$, and the intensive parameters $(T,\, p)$ can be obtained by taking partial derivatives of $E(S,\,V)$ with respect to its arguments. In the final picture, the situation is the same for QLHs. Thus, $\MIH$ is a function only of the extensive variables $(A_{\IH},\, \JIH)$ and its partial derivatives with respect to $A_{\IH},\, \JIH$ yield the intensive parameters $\kappao$ and $\Omegao$. Similarly, $\MDH [S]$ is a function only of the extensive variables $(A_{DH}[S],\, \JIH[S])$, and its partial derivatives with respect to $A_{\DH}[S],\, \JIH [S]$ yield the intensive parameters $\kappao [S]$ and $\Omegao [S]$.

Finally we understand from experts that the BH uniqueness theorems of general relativity extend to a large number of diffeomorphism covariant, metric-based  theories of gravity, and also to supersymmetric generalizations of these theories, in which the issue has been analyzed in detail. More generally, when uniqueness fails, typically the new stationary black hole solutions are unstable. (Discussions at the Simons Workshop on ``50 years of the information loss issue" at Stony Brook in 2025, and at the ``Symphony of Space-time" conference at the Raman Research Institute in 2026.) However, these discussions often involve additional (often implicit) assumptions. If uniqueness does fail in physically interesting theories, one would have to revisit the discussion of section \ref{s5}, probably on a case by case basis.
\vskip0.1cm

2. There is a sizable literature on the fist law of mechanics of quasi-local horizons (see, e.g., \cite{Ashtekar:2000sz,Ashtekar:2000hw,Ashtekar:2003hk,Ashtekar:2004cn,Ashtekar:2001is}). Those discussions differ from ours in a number of respects that can be summarized as follows, using the terminology of this paper. 

In the case of IHSs, the universal structure and the symmetry group $\G$ were not known at the time when earlier papers were written. The starting point was finite dimensional symmetry groups that \emph{individual} IHS geometries can support --akin to focusing on groups of isometries \emph{individual} space-time metrics can admit, rather than on the BMS group $\B$ of asymptotic symmetries shared by all asymptotically flat space-times. Therefore, one did not have the charge $Q^{{}_{(\ko)}}_\IH$ associated with the dilation symmetry in $G$ that served as the starting point for our discussion of the first law in section \ref{s3.3}. Also, in the definition  $\to^a = \ko^a +\Omegao \varphio$, the surface gravity $\kappao$ of $\ko^a$ and the angular velocity $\Omegao$ were left to be general functionals of space-time metrics under consideration, and not determined by the projection map $\Pio$ introduced in the present paper. The first law arose as a necessary and sufficient condition for  $\to^a$ to admit a space-time extension, such that the diffeomorphisms it generates preserves the symplectic structure on the space of solutions admitting $\IH$ as an inner boundary. 

In the present work, all considerations are local to the IHS under consideration. The focus is on (i) the (infinite dimensional) symmetry group $\G$ preserving the universal structure of all $\NEH$ (or, equivalently, of all $\IH$) and charges associated with these symmetries; and, (ii) the use of BH uniqueness theorems to endow general IHSs with specific intensive parameters $(\kappao,\, \Omegao)$. One does not need to extend $\to^a$ or require that it induce canonical transformations. Instead, one simply observes that the charges satisfy the generalized Smarr relations (\ref{M-IH1}) as well as the identity (\ref{identity1}) on \emph{all} IHSs. The first law is then an immediate consequence. 

In the case of DHSs, there were several balance laws in the literature that can be regarded as generalizations of the BCH first law (\ref{BCH}) in that they provide explicit expressions relating changes in the Hawking mass, area of MTSs, and angular momentum on them, to fluxes across portions $\Delta\H$ of $\DH$. However, a key element was missing:  the analog of the $\DH$-mass $\MDH$ that is guaranteed to agree with the $\IH$-mass $\MIH$ when the BH reaches equilibrium. Among the candidate expressions for the first law, there was only one in which each term has the physical dimensions of energy as in (\ref{BCH}). But in that law the left side referred to the Hawking mass $M_{\rm Haw}$ --rather than $\MDH$ as in (\ref{dh1law1})-- which does not in general agree with the $\MIH$ when the BH reaches equilibrium. Thus, while our present analysis made crucial use of the infrastructure that was provided by previous work, especially \cite{Ashtekar:2003hk}, it provides a more satisfactory generalization of (\ref{BCH}) to fully dynamical situations.\vskip0.1cm

3. DHSs are routinely used to investigate BH dynamics in a wide variety of situations, including binary black hole evolution and gravitational collapse (for a recent review, see \cite{Ashtekar:2025LivRev}), as well as 
BH evaporation (see,e.g.\cite{Ashtekar:2005cj,Sawayama:2005mw,Ashtekar_2011a,Ashtekar_2011b,ashtekar2023regularblackholesloop,Agullo_2024,Varadarajan:2024clw}). While this domain of application is vastly larger than that encompassed by perturbation theory around a given stationary BHs, it can probably be extended even further. For, while DHs are well-suited to model both the progenitors and the remnant during most of the binary BH evolution, as we mentioned in section \ref{s3}, the QLH fails to be a DHS for a short period very near the merger. In this short phase QLH-dynamics is very interesting but also very complicated (for a summary, see section 4.3 of \cite{Ashtekar:2005ez}). A detailed analytical understanding of this phase seems difficult but would bring rich rewards, as it will provide thermodynamical description of the \emph{entire} evolution, including the merger process itself. Work is already in progress on this issue \cite{metidieri2026}.

 \vskip0.1cm

4. As we discussed in section \ref{s1.1}, because of their teleological nature, EHs cannot be used to represent dynamical BHs in classical GR, especially in thermodynamical considerations. In the quantum evaporation process, it is unclear whether the full space-time even admits an EH! For example, in his talk at the GR-17 conference in Dublin, Hawking emphasized that ``a true event horizon never forms". As we noted above, results of this paper lend strong support to the numerous investigations that now use DHSs $\DH$ instead, both in classical GR and to study the quantum evaporation of BHs.

For spherically symmetric BHs --used nearly universally in the investigations of BH evaporation so far-- the space-time admits a unique spherically symmetric DHS, which we can use for thermodynamical considerations. However, in generic dynamical situations, we do not yet have a fully satisfactory answer to the basic questions: What is a BH? and, what characterizes its boundary? (For a summary of the present status see, e.g., see section 7 of \cite{Ashtekar:2025LivRev}.) In particular, as noted at the end of section \ref{s4}, in each of its dynamical phases, such a BH generically admits many `interweaving' DHSs $\DH$ \cite{Ashtekar:2005ez}, and so far there is no natural procedure to single out a unique one.%
\footnote{For a possible direction to signal out a preferred DHS, see part IV of \cite{Gourgoulhon_2006}.}
However, a key point is that \emph{each} $\DH$ provides us with a detailed, self-consistent  description of the evolution of the BH, and all our thermodynamical considerations also hold for each $\DH$. In particular, the generalized first and second laws of section \ref{s6.1} and \ref{s6.2} hold on all of them. When the BH reaches equilibrium, represented by an IHS, all these DHSs asymptote to that IHS and their thermodynamic parameters agree with those of the IHS \cite{Ashtekar:2013qta}. Thus, there is consistency between the dynamical phase represented by any of the DHSs, and the final equilibrium phase. 

As suggested in \cite{Ashtekar:2025LivRev}, there may be a small but \emph{inherent} ambiguity in the notion of the boundary of a dynamical BH, and therefore also in the notion of its entropy --similar to the inherent ambiguity in the notion of angular momentum at $\scrip$ in dynamical situations. For angular momentum, one just has to accept that at $\scrip$ the familiar special relativistic notion of angular momentum has to be generalized because, in the presence of gravitational waves, the BMS group does not admit a canonical Poincar\'e subgroup. Similarly, in fully dynamical contexts, the notion of BH entropy may undergo a generalization. Now, at $\scrip$, thanks to the BMS group, one can systematize the ambiguity in the notion of angular momentum. Is there perhaps an interesting group that relates the various DHSs associated with a given dynamical phase of a BH? If so, perhaps the entropy has a well-defined transformation property as one passes from one DHS to another, and the presence of the group, again, systematizes the ambiguity.

\section*{Acknowledgments}

AA acknowledges stimulating discussions on quasi-local horizons with %Ivan Booth, Miguel Campiglia, Greg Galloway, Sean Hayward, Jose Luis Jaramillo, Jerzy Lewandowski, Tomasz Pawlowski, Jose Senovilla, Simone Speziale, and especially
a large number of colleagues over the years, especially Badri Krishnan. Presentation of the material benefited from discussions at the Simons workshop \emph{50 years of black hole information paradox} at Stony Brook, the conference \emph{Cosmological Olentzero} at Bilbao and an International Loop Quantum Gravity Seminar, particularly the subsequent correspondence with Jose Luis Jaramillo and Simone Speziale. This work was supported in part by the Atherton and Eberly funds of Penn State.  D.E.P. acknowledges support via Penn State's Bunton-Waller Award and the Walker Fellowship from its Applied Research Laboratory.

\begin{appendix}

\section{Angular momentum of QLHs in absence of a rotational Killing field}
\label{a1}

In the main body of this paper, we restricted ourselves to axisymmetric isolated horizons in order to define their angular momentum. However, this restriction is not necessary: there is a well-defined notion of angular momentum on all NEHSs $\NEH$ and therefore, in particular, on IHSs. We did not use this notion in section \ref{s3} because the construction involves a substantial detour which would have been a distraction from our focus on thermodynamical considerations.  In this Appendix we  summarize the construction. The discussion is divided into two parts. In the first we recall from \cite{akkl1} the universal structure and the symmetry group $\G$ shared by all NEHSs. Associated with each symmetry, there is an $\NEH$-charge \cite{Ashtekar:2021kqj}. In the second part we use the Lorentz symmetries in $\G$ to define angular momentum $\JNEH$ using a strategy introduced by Korzynski \cite{Korzynski:2007hu}. The strategy does not need the full $\NEH$-structure; \emph{it also provides an angular momentum on each MTS of a DHS $\DH$}. It is referred to as $\JDH [S]$ in the main text and used in sections \ref{s5.1} and \ref{s6.1}.

\subsection{Universal structure and symmetries of $\NEH$} 
\label{a1.1}

There are close parallels between null infinity, $\scrip$, and NEHSs, $\NEH$ \cite{Ashtekar2_2024}. Just as $\scrip$ serves as the outer (future) boundary of asymptotically flat space-times,  $\NEH$ serves as the inner boundary of a large class of space-times. Both are null 3-surfaces with topology $S^2\times R$ and ruled by null geodesics. Recall that the BMS group $\B$ arises as the group of symmetries that preserves the universal structure at $\scrip$ that is shared by \emph{all} asymptotically flat space-times. Although most of these space-times do not carry any isometries, they all share asymptotic symmetries. Similarly, one can isolate the universal structure on NEHs shared by \emph{all} space-times that admit a $\NEH$ as an inner boundary, and identify the symmetry group $\G$ that preserves this universal structure. In fact the structure of $\G$ is very similar to that of the BMS group at $\scrip$. We briefly review this construction here (for details, see \cite{akkl1}).

To isolate the universe structure, let us first note that, without loss of generality, one can restrict oneself to null normals $\ko^a$ that are affinely parametrized geodesic null vector fields on any given $\NEH$. This class is natural because, since the acceleration $\kappao$ vanishes, the restriction does not single out any scale. It can be chosen on \emph{any} NEH, \emph{including the non-extremal KHs, such as the $\KH$ of a Schwarzschild BH.} Eq. (\ref{0law}) implies that these normals also satisfy $\Lie_{\ko} \omegao_a\, \=\,0$. There is a large rescaling freedom among them,\, $\ko^a \rightarrow e^{f}\,\ko^a$ where $f$ is any function satisfying $\Lie_{\ko} f \= 0$. But this freedom can be essentially exhausted by requiring that $\omegao_a$ be divergence-free $\qo^{ab} \Do_a \omegao_b \=0$. (This condition is well-defined because, since $\omegao_a \ko^a\, \=0\,$ and $\Lie_{\ko} \omega_a\, \=\, 0$, it is insensitive to the freedom of adding a term of the type $V^{(a} \ko^{b)}$ to $\qo^{ab}$.) Then the permissible rescaling freedom is reduced to $\ko^a \rightarrow e^c \ko^a$ where $c$ is a constant \cite{Ashtekar:2001jb}. Thus, every NEHS is endowed with a preferred equivalence class $[\ko^a]$ of geodesic null normals where one is related to another by a rescaling by a positive constant.  

Next, let us consider the $\NEH$-geometry. The (degenerate) metric $\qo_{ab}$ and the derivative operator $\Do$ vary from one $\NEH$ to another and therefore are not part of the universal structure. They encode information about the specific $\NEH$ under consideration, such as its shape and spin structure (and are thus analogous to fields like $\Psi_4^\circ$ and Bondi news $N_{ab}$ at $\scrip$ that vary from one asymptotically flat space-time to another.) However, the  metric $\qo_{ab}$ on any given $\NEH$ is conformally related to  \emph{unit, round, 2-sphere metrics} $\ro_{ab}$ via  $\ro_{ab} \,\=\, \psio^2\, \qo_{ab}$.% 
\footnote{$\qo_{ab}$ and $\ro_{ab}$ are pull--backs to $\NEH$ of metrics defined on the 2-sphere $S_\NEH$ of generators of $\NEH$. As is common, for simplicity, we will not distinguish between covariant fields on $S_\NEH$ and their lifts to $\NEH$.}
The conformal factor $\psio$ is determined by the scalar curvature of $\qo_{ab}$ and, for any given $\qo_{ab}$, there is precisely a 3-parameter family of $\ro_{ab}$. Among themselves, these round metrics are related by  $\ro_{ab}^\prime \,\=\,  \alpha^2\,\ro_{ab}$, where the conformal factor $\alpha$ is given by
\be \label{rel1} \alpha^{-1} = \alpha_0 + \sum_{i=1}^{3}\alpha_i \hat{r}^i \quad {\rm with}\quad \hat{r}^i\, \equiv \,(\sin\theta\cos\varphi,\,\sin\theta\sin\varphi,\, \cos\theta)\,. \ee 
Here $(\theta,\,\varphi)$ are the spherical coordinates of $\ro_{ab}$, and $\alpha_0, \, \alpha_i$ are real constants subject to $-\alpha_0 +\sum_{i=1}^3 (\alpha_i)^2 =1$. Note that while the conformal factor $\psio$ relating  $\qo_{ab}$ and $\ro_{ab}$ varies from one $\qo_{ab}$ to another, the relation \emph{between the round metrics themselves} is universal. For reasons motivated by considerations of multipoles, on any given $\NEH$ equipped with $(\qo_{ab},\, [\ko^a])$, it is natural to associate with each round metric $\ro_{ab}$ an equivalence class of null normal $[\ello^a] = \psio^{-1} [\ko^a]$ where, as before, two are equivalent if they are related via a rescaling by a positive constant.  Again, while the relation between the null normals $[\ko^a]$ and $[\ello^a]$ varies from one $\NEH$ to another, that 
between any two $[\ello^a]$ is universal: $[\ello^{\prime\,a}] = \alpha^{-1} [\ello^a]$. Thus, the universal structure on NEHSs consists of pairs $[\ro_{ab},\,\ello^a]$ on $\NEH$ satisfying $\ro_{ab}\ello^a \=0$ and $\Lie_{\ello} \ro_{ab} \=0$,  where any 2 are related by $[\ro_{ab}^\prime,\,\ello^{\prime\,a}] = [\alpha^2\, \ro_{ab},\, \alpha^{-1} \ello^a]$, where $\alpha$ is given by (\ref{rel1}). \vskip0.1cm

This is very similar to the universal structure at $\scrip$ which can be taken to be pairs $[\ro_{ab},\, \no^a]$ defining Bondi conformal frames, where any two are again related by $[\ro_{ab}^\prime,\,\no^{\prime\,a}] = [\alpha^2\, \ro_{ab},\, \alpha^{-1} \no^a]$. As at $\scrip$, the $\NEH$-universal structure is specified on abstractly defined manifold $\NEH^\star$, topologically $S^2\times R$. In particular, since there is no intrinsic derivative operator $\Do$ on $\NEH^\star$ induced by any 4-metric, the vector fields $\ello^a$ that rule $\NEH^\star$ do not have a `geodesic' connotation, nor do they define any 1-form $\omegao_a$. The subgroup of diffeomorphisms on $\NEH^\star$ that preserves this structure is the NEH-symmetry group $\G$. Given a concrete $\NEH$, there is a diffeomorphism from $\NEH$ to $\NEH^\star$ that sends the pairs $(\ro_{ab},\,[\ello^a])$ on $\NEH$ to those fixed on $\NEH^\star$. This diffeomorphism is unique up to an element of $\G$. \vskip0.1cm

Thanks to the similarity with $\scrip$, the structure of $\G$ is the same as that of the BMS group $\B$ at $\scrip$. In particular, $\G$ is a semi-direct product of the Lorentz group $\mathcal{L}$ with an infinite dimensional group $\mathcal{V}$ generated by `vertical' vector fields on ${\NEH}^\star$, proportional to $[\ello^a]$. But because we have \emph{equivalence classes} $[\ello^a]$ in the universal structure of $\NEH^\star$ --in place of vector fields $\no^a$ on $\scrip$-- $\mathcal{V}$  is a 1-dimensional extension of the group $\mathcal{S}$ of supertranslations. The extra generator is a dilation, $\dilaton^{a} = k\, \vo\, \ello^a$, where $\vo$ is an affine parameter of $\ello^a$, and $k$ a constant. It generates diffeomorphisms on $\NEH^\star$ that rescale each $\ello^a$ by a constant $k$, preserving only the equivalence class $[\ello^a]$, and leaving each metric $\ro_{ab}$ untouched. If a given $\NEH$ happens to be a Killing horizon with surface gravity $\kappao >0$, then the  restriction of the Killing vector field to $\KH$ is the dilation $\dilaton^a = \kappao\, \vo\, \ello^a$. Thus, the 1-dimensional enlargement of the BMS group $\B$ on $\NEH$ is inevitable: while the restriction of a stationary Killing field to $\scrip$ is a BMS translation, its restriction to $\NEH$ is the dilation. $\G$ also has several other interesting features that can be found in \cite{akkl1}. \vskip0.1cm

We will conclude this discussion of the $\NEH$-universal structure and the symmetry group with some conceptual clarifications. Since any given $\NEH$ is a sub-manifold of space-time, it directly inherits from $g_{ab}$ and its $\nabla$, a (degenerate) metric $\qo_{ab}$ and a derivative operator $\Do$. To begin with, there is a large freedom in the choice null normals $\ko^a$; both $\ko^a$ and $f \ko^a$ are null normals for any positive but otherwise arbitrary function $f$ on $\NEH$. Each of these null normals is a geodesic vector field. To arrive at the universal structure shared by all NEHSs, one restricts oneself to null normals $\ko^a$ that are affinity parametrized geodesic vector fields (w.r.t. $\Do$). This is analogous to choosing divergence-free conformal frames at $\scrip$ which also make the null normal $\no^a$ to $\scrip$ an affinely parametrized geodesic vector field. Furthermore, by imposing the additional condition $\qo^{ab} \Do_a \omegao_b\,\=\,0$, one narrows the choice of null normals to a small equivalence class $[\ko^a]$ where any two are related via a rescaling by a positive \emph{constant}. This is possible because one begins with null normals $\ko^a$ with vanishing surface gravity $\kappao$, i.e., with extremal $\NEH$. On non-extremal NEHSs, there is no natural procedure to select the small canonical equivalence class $[\ko]$ where any two are equivalent if and only if one is a rescaling of the other by a positive constant.
 
Nonetheless, the restriction to the null normals that endow $(\NEH,\, \qo_{ab},\, \Do)$ with the structure of an extremal horizon structure may seem confusing at first because of one's familiarity with Killing horizons $\KH$. For example, one thinks of the Schwarzschild KH $\KH$ as being necessarily non-extremal, with $\kappao = 1/4M$. But whether a given $(\NEH,\, \qo_{ab},\, \Do)$ is extremal or not depends on which null normals $\ko^a$ one chooses to emphasize. In the Schwarzschild case, one routinely uses the null normal on $\KH$ that coincides there with the static Killing field $\to^a$ that is unit at infinity. But, as explained already in section \ref{s1}, we need to focus on structures available on the horizon itself, without reference to the exterior region. From this intrinsic perspective, geodesic null normals are more natural on any $\NEH$.%  
\footnote{In particular, for test quantum fields on the Schwarzschild space-time,  the affine parameter $\vo$ of the extremal null normals is used to define the Unruh vacuum that is best suited to mimic collapsing situations.}
In this setup, since the standard Killing field $\to^a$ of the Schwarzschild metric is a space-time \emph{symmetry}, in particular it preserves the $\NEH$-universal structure. Therefore it does play an important role, but as the generator of a distinguished $\NEH$-symmetry in $\G$, just as it is the generator of a distinguished BMS symmetry at $\scrip$. (The only difference is that while it belongs to the supertranslation subgroup of $\B$ at  $\scrip,$ it is a dilation, rather than a supertranslation in $\G$.)\goodbreak

\subsection{Angular Momentum}
\label{a1.2} 

On an axisymmetric $\NEH$ with $\varphio^a$ as the rotational Killing field of $\qo_{ab}$, there is a natural notion of angular momentum: It can be can be expressed using the 1-form $\omegao_a$ or $\Im [\Psi_2]$ as:
\be \label{J-NEH1} J^{(\varphio)}  \=\, -\f{1}{8\pi G} {\oint_S} \omegao_a \,\varphio^a \, \rmd^2 V\, \=\, -\f{1}{4\pi G} \oint_S \mathring{f}\, \Im [\Psi_2]\, \rmd^2V,\ee
where, in the second step, we have used the fact that, any divergence-free vector field $\varphio^a$ on $S$ has the form $\varphio^a \= \epsilono^{ab}\, D_b \mathring{f}$ \cite{Ashtekar:2000sz}. Since $2\,\Im\, \Psi_2\, = \epsilono^{ab}\,\Do_{[a} \omegao_{b]}$, the second expression in terms of $\Im [\Psi_2]$ makes it manifest that $J^{(\varphio)}$ depends only on the exterior derivative of the pull-back of $\omegao_a$ to the cross-section $S$, and is thus unaffected by rescalings of $\ko^a$. Finally, the value of $J^{(\varphio)}$ is independent of the choice of $S$. On the Kerr horizon one obtains $J^{(\varphio)} = Ma$, as expected. 

However, recall that one can define an angular momentum charge at $\scrip$ using Lorentz subgroups of the BMS group $\B$ even in the absence of a rotational Killing field. Similarly, one can define an angular momentum charge on $\NEH$ even when the physical metric $\qo_{ab}$ is \emph{not} axisymmetric by using generators of Lorentz subgroups of $\G$. In fact the procedure is simpler because there is no gravitational radiation on $\NEH$. Korzynski has spelled it out in \cite{Korzynski:2007hu}. It can be summarized as follows. 

Lorentz subgroups of $\G$ are intimately intertwined with groups of  conformal isometries of the round metrics $\ro_{ab}$ introduced in section \ref{a1.1}. Fix any cross-section $S$ of $\NEH$ and consider the pull-backs of unit round metrics $\ro_{ab}$ to $S$.  Now, there is a unique Lorentz subgroup $\L$ of $\G$ that leaves $S$ invariant. Its generators are just the six conformal Killing fields of any of the pulled-back $\ro_{ab}$ on $S$. Fix a $\ro_{ab}$ and denote its three Killing fields $\phio^a_{(i)}$ ($i=1,2,3$). They provide the three components $L_{(i)}$ of angular momentum on $S$:
\be  \label{L-NEH} L_{(i)}  \=\, -\f{1}{8\pi G} {\oint_S} \omegao_a \,\phio^a_{(i)} \, \rmd^2 V\,.\ee
Similarly the three proper conformal Killing fields $\etao^a_{(i)}$ (of the fixed $\ro_{ab}$) provide the three components $K_{(i)}$ of the boost angular momentum: 
\be \label{K-NEH} K_{(i)}  \=\, -\f{1}{8\pi G} {\oint_S} \omegao_a \,\etao^a_{(i)} \, \rmd^2 V\,.\ee
Let us first ensure that $L_{(i)}$ and $K_{(i)}$ are well-defined. The vector fields $\phio^a_{(i)}$ and $\etao^a_{(i)}$ are 
divergence-free with respect to the $\ro_{ab}$ we fixed, but not with respect to the \emph{physical} metric on $S$ (which determines the area element $\rmd^2 V$). Therefore, the values of the integrals would change if we were to replace $\omegao_a$ by $\omegao_a + D_a f$, i.e. if we replace $\ko^a$ by $e^f\,\ko^a$. However, thanks to our condition $\qo^{ab} \Do_a \omegao_b \,\=\, 0$,  the rescaling freedom is only by a constant, under which $\omegao_a$ remains unchanged. Thus $L_{(i)}$ and $K_{(i)}$ are indeed well-defined.

Our task is to extract from $L_{(i)}$ and $K_{(i)}$ a single number $J$,\, to be interpreted as the angular momentum of $\NEH$.  Now, the decomposition of the Lorentz angular momentum into $L_{(i)}$ and $K_{(i)}$ depends on the choice of a specific round metric $\ro_{ab}$ in the 3-parameter family. However, Korzynski \cite{Korzynski:2007hu} has shown that it is possible to extract, in an invariant manner, a preferred \emph{rotational vector field} $\varphio^a$ among the six Lorentz generators using just the Lorentz Casimirs \,$A := \vec{L}^2 - \vec{K}^2$ and $B := \vec{K}\cdot\vec{L}$.% 
\footnote{The construction assumes a genericity condition that both Casimirs do not vanish simultaneously. Therefore it does not provide a preferred rotation in the case when $\qo_{ab}$ is spherical symmetric. But in this case $J$ vanishes identically.}
It is a rotation in the sense that: (i) it has closed orbits, with exactly two zeros; and, (ii) the range of its affine parameter is $[0, \, 2\pi)$. The $\varphio^a$ picked out by this procedure depends on $\qo_{ab}$ and $\omegao_a$ --i.e. on the specifics of the geometry of the $\NEH$ under consideration--  just as one would physically expect. The component of $\vec{L}$ it defines is the angular momentum of $\NEH$ 
\be \label{J-NEH3} \JNEH  \=\, -\f{1}{8\pi G} {\oint_S} \omegao_a \,\varphio^a \, \rmd^2 V\, ,\ee
where we have dropped the suffix $(\varphio^a)$ on $J$ to emphasize the fact that $\varphio^a$ is not a pre-specified Killing field, as in (\ref{J-NEH1}), but constructed from the geometry of the given $\NEH$ itself. Thus, $\JNEH$ is the angular momentum of a generic $\NEH$ which need not be axisymmetric. In the special case that the physical metric $\t{q}_{ab}$ on $S$ happens to be axisymmetric, the $\varphio^a$ selected by this procedure coincides with the axial Killing field, and $\JNEH$ equals $J^{(\varphio)}$ of (\ref{J-NEH2}). In particular, if this $\NEH$ were an axisymmetric $\IH$, it would yield the expression (\ref{J-IH1}) of section \ref{s3.2}. %Finally, this procedure can be motivated by the steps one follows to define angular momentum of a system of $N$ non-interacting particles in Minkowski space \cite{Korzynski:2007hu}.
\vskip0.05cm 

However, there is a potential pitfall: we fixed a cross-section $S$ of $\NEH$ at the very beginning of the construction. A priori $\JNEH$ could depend on the choice of the cross-section $S$, in which case it would not represent an observable associated with the full $\NEH$ but only with $S$. Now, any two cross-sections $S$ and $S^\prime$ are related by a supertranslation and all supermomentum charges \emph{vanish identically} on an NEH \cite{Ashtekar:2021kqj}. Consequently, the value of $J$ does not depend on the initial choice of $S$. This feature can be made manifest by working entirely on the 2-sphere $\h{S}$ of orbits of the vector fields $[\ko^a]$. Because $\omegao_a \ko^a\, \=\, 0$ and $\Lie_{\ello} \omegao \,\=\,0$, $\omegao_a$ on $\NEH$ is the lift of  a 1-form $\h\omega_a$ on $\h{S}$ and one can carry out the entire discussion using $\hat{S}$ in place of $S$, and express $\JNEH$ as an integral on $\h{S}$. However, while this construction would then single out a rotational vector field $\h\varphi^a$ on $\h{S}$, its lift to $\NEH$ would not be unique; there is a freedom to add a supertranslation to $\varphio^a$.  Thus, on a given $\NEH$ the rotational symmetry vector field $\varphio^a$ picked by the Korzynski construction is tied to the choice of $S$. However,   different choices of cross-sections yield the same $J$ because, as mentioned above, all supertranslation charges vanish. To summarize, this construction assigns any given $\NEH$ is assigned a unique $\JNEH$. We could have used it in section \ref{s3.2} in the discussion of the first law to obtain a first law on all IHSs including those that are not axisymmetric. The relation of this $\JIH$ to other definitions in the literature is the same (indeed, it was obtained using Hamiltonian considerations in \cite{Ashtekar:2021kqj}). \vskip0.1cm

\emph{Remark:} An alternate definition of quasi-local angular momentum associated with any 2-sphere $S$ was recently proposed by Racz \cite{Racz2025}. The key step again involves singling out a preferred rotational vector field $\phi^a$ on $S$, using the intrinsic metric $\t{q}_{ab}$ thereon. However, the procedure differs from that of Korzynski in that $\phi^a$ is now required to be divergence-free with respect to $\t{q}_{ab}$. This strategy has the advantage that it is no longer necessary to fix the rescaling freedom in $(k^a,\, \uk^a)$ by imposing the condition $\qo^{ab} \Do_a \omegao_b \,\=\, 0$. On the other hand, the resulting angular momentum $J_{S}^{(\phi)}$ is no longer linear in the vector field $\phi^a$ and in fact depends on it non-locally. Since in Hamiltonian treatments, observables such as the angular momentum $J_S^{(\phi)}$ arise as `momentum maps' which are linear in the corresponding vector fields $\phi^a$, this alternate definition of angular momentum is not suitable for our purposes. 

\section{An alternative approach based on field equations.}
\label{a2}

In the main body of this paper we used a phase space approach to define various charges. Their properties, together with the projection map $\Pi$ of section \ref{s5.1} then led us to the generalizations of the first and second laws. Recall, however, that the original BCH derivation used only Einstein's equations --in fact just the constraints since their analysis assumed stationarity-- and made no reference to phase spaces and symplectic structure. Similarly at $\scrip$, Bondi, Newman, Penrose and others \cite{BMS,Newman:1961qr,Penrose:1965am} used only the field equations to arrive at charges at null infinity; symplectic structure considerations came much later \cite{Ashtekar:1981bq,AshtekarReula,Wald:1999wa}. We used the Hamiltonian methods in the main body to make closer contact with the recent literature on black hole thermodynamics, and also because one can regard the Hamiltonian framework as the imprint left by the quantum theory on classical physics. However, while this framework does provide charges, in the presence of boundaries they are often \emph{not} the generators of canonical transformations 
\footnote{This point is often glossed over. It becomes especially relevant when the space-time extensions of boundary symmetries are field dependent \cite{Barnich:2001jy,Barnich:2011mi}. As a result, the Noether charge $\kappao A/8\pi G$ in \cite{PhysRevD.108.044069,hollands2024entropydynamicalblackholes,visser2025dynamicalentropychargedblack} is not a generator of a canonical transformation. Interestingly, in our framework, the horizon charge $\kappao A/8\pi G$ associated with the $\IH$-symmetry $\ko^a$ is, and so is $\JIH$. But $\Omegao \JIH$ associated with $\Omegao\varphio$ is not (because $\Omegao$ is field-dependent), and therefore the charge $\MIH$ associated with $\to^a$ is also not a generator of a canonical transformation.}
and therefore not as useful in the passage to the quantum theory as one might think a priori. 
In this Appendix we will provide an alternative derivation of the first and second laws of black hole thermodynamics using only the balance laws that emerge directly from Einstein's equations. Hamiltonian methods make a crucial use of field equations in any case. Therefore, one can regard this alternative as a more natural generalization of the BCH analysis that did not have the `extra baggage' of symplectic structures and especially their potentials. 

Let us begin with the extension of the BCH analysis to IHSs discussed in section \ref{s3.3}. As in the \cite{Bardeen:1973gs,LesHouches:1973} one can use Komar integrals, but now associated with symmetries of the IHS $\IH$ itself, rather than space-time Killing fields. Let us fix the overall constant in front of the Komar integral will be adapted to yield the correct ADM energy. Thus, for a symmetry vector field $\xi^a$, we will set 
\be \label{komar} Q_{{}_{\rm K}}^{{}^{(\xi)}}\,[S] = - \f{1}{8\pi G}\, {\oint_S}  \epsilon_{ab}{}^{cd}\,\, \nabla_c \xi_d \,\,\rmd S^{ab}\, . \ee
To evaluate the right side, one requires the knowledge of $\nabla_a \xi_b$ on $\IH$. However, for any generator $\xi^a$ of the symmetry group $\G$, the derivative is completely determined by the universal structure and the restriction of the vector field $\xi^a$ to $\IH$. Using the normalization given in Eq.(\ref{komar}) and the canonical extension, one can evaluate the Komar integrals associated with $\ko^a$ and $\varphio^a$. They are given by:
\be Q_{\rm K}^{(\ko)}\, \=\, \f{1}{4\pi G}\, \kappao\, A, \qquad{\rm and} \qquad Q_{\rm K}^{(\varphio)} \,=\, -2 \JIH\,. \ee
Since the Komar integral is linear in the vector field, and since $\Omegao$ is (field dependent but) constant on any given $\IH$, we have $Q_{\rm K}^{(\Omegao\varphio)} \,=\, -2\Omegao \JIH$, and hence the Komar charge associated with the $\IH$-symmetry $\to^a = \ko^a - \Omegao \varphio^a$ is
\be \label{IH-smarr} Q_{\rm K}^{(\to)}\, \=\, \f{1}{4\pi G}\, \kappao\, A + 2 \Omegao \JIH \ee
Thus, the Smarr relation holds on a general $\IH$, even when there is matter and gravitational radiation outside $\IH$, so long as all quantities are defined \emph{at the IHS} without reference to infinity. Finally, because of our choices, $\kappao = \kappao_{\rm kerr} (\RIH,\,\JIH)$ and $\Omegao = \Omegao_{\rm kerr} (\RIH, \JIH)$ of Eqs (\ref{kerr-kappa}) and (\ref{kerr-Omega}), we have the identity (\ref{identity1}). This identity and (\ref{IH-smarr}) imply that the first law generalizes to all IHSs:
\be \label{ih1law}\delta\MIH \,\=\, \f{\kappao}{8\pi G}\, \delta A\,+\, \Omegao\, \delta\JIH \, .\ee
As noted in section \ref{s3.3}, axisymmetry is not essential to define $\JIH$; on a generic $\IH$ one can use the construction summarized in Appendix \ref{a1.2}. Then $\phio^a$ is only a symmetry generator of the $\NEH$-symmetry group $\G$, rather than a symmetry of the specific $\IH$-geometry under consideration. Nonetheless, since the first-order extension of $\varphio^a$ away from $\IH$ we used to define the Komar integral refers to all symmetry generators of $\G$,\, $\JIH$ is well defined and the argument sketched above goes through unaltered. Thus, the first law (\ref{ih1law}) holds on generic IHSs.\vskip0.1cm

Let us now turn to the dynamical situations discussed in section \ref{s6.1}. In this case, following \cite{Ashtekar:2003hk} one can use the constraint equations to define balance laws on $\DH$ (just as Bondi et al used Einstein's equations at $\scrip$ to arrive at the balance laws for the Bondi 4-momentum). On a generic $\DH$, Korzynski's construction \cite{Korzynski:2007hu} provides us with a rotational vector field $\varphio^a$ on any given MTS $S$, and the diffeomorphism constraint associates to it an angular momentum charge
\be \JDH [S] \, =\, -\f{1}{8\pi G}\, \oint_S   \t\omega_a \varphio^a\, \rmd^2V, \qquad{\rm with} \qquad \t\omega_a = \t{q}_{a}{}^b\, k_{bc}\, \h{r}^c \, ,\ee
where $\t{q}_{ab}$ the intrinsic metric of $S$; $k_{ab}$, the extrinsic curvature of $\DH$; and $\h{r}^a$, the unit normal to $S$ within $\DH$.  If $\DH$ were to join on to an $\IH$ at a cross-section $S$ because the infalling fluxes vanishes at $S$ (and to its immediate future or past), this charge would agree with the $\IH$-angular momentum discussed above. The constraint equation also provides a balance law (\ref{JDH2}) between any two MTSs of the DHS:
\be \JDH^{{}^{(\varphi)}} [S_2]\, -\, \JDH^{{}^{(\varphi)}}[S_1] \, \,\=\, -\int_{\Delta\H} \, \Big[T_{ab}\varphi^a \hat{\tau}^b\,+\, \f{1}{16\pi G} \big(k^{ab} - kq^{ab}\big)\, \Lie_{\varphi} q_{ab} \Big]\, \rmd^3 V\,\, \equiv\,\, \mathcal{F}^{{}^{(\varphi)}} [\Delta\H]\, .  
\ee

Next, Ref. \cite{Ashtekar:2003hk} also provided equations governing the changes in the Hawking mass and the area of MTSs due to infalling fluxes.  They arose as special cases of the balance law provided by the linear combination $(G_{ab} - 8\pi G T_{ab})\,\xi^{\prime\, a} \h\tau^b =0$ of constraints, with $\xi^{\prime\,a} = \sqrt{2}\, |DR|\, f(R) k^a$, for suitable choices of $f(R)$. Although they have the flavor of the first law, they are not direct generalizations of the BCH first law (or of the first law for IHSs). What was missing was the appropriate choice of $\xi^{\prime\,a}$ (or $f(R)$) with the property that the associated charge $Q_{\rm \DH}^{(\xi^\prime)} [S]$ agrees with $Q_{\rm K}^{(\ko)}\, = \,
\f{\kappao A}{4\pi G}$ on $\IH$,\, on any $S$ at which the $\DH$ joins on to $\IH$ at $S$. In section \ref{s5.2} we showed that the desired $\xi^{\prime\,a}$ can be found using the projection map $\Pi$ of section \ref{s5.1}, and is in fact unique. The full power of the projection map was not appreciated in \cite{Ashtekar:2003hk} nor in the subsequent literature on QLHs. Discussion \ref{s5.2} implies that the desired $\xi^{\prime\,a}$ corresponds to
\be f(R) = \f{1}{2\pi} \f{\rmd (\kappao A)}{\rmd R} \qquad {\hbox{\rm for which}}\qquad Q_{\DH}^{(\xi^\prime)}[S] = \f{(\kappao\, A)}{4\pi G}\,[S]\, .\ee
Thanks to the projection map $\Pi$, the right side is the value of $Q^{(\ko)}_{\rm K}$ at the equilibrium state $\e$ corresponding to $S$.%
\footnote{The factor of $2$ difference between $\xi^{\prime\,a}$ and $\xi^a$ of Eq.(\ref{f(R)}) lies in the fact that we are now asking for equality of the $\DH$-charge with the Komar integral $Q_{\rm K}^{(\ko)}$ on $\IH$, rather than with the $\ko$-phase space charge on IHSs which is half the Komar integral.} 
Following the procedure used for IHSs and using the projection map $\Pi$, it is natural to define a vector field $t^{\prime\,a} = \xi^{\prime\,a} - \Omegao \varphi^a$ on any given $\DH$. Then, since the charges (and fluxes) obtained from constraint equations are linear in the associated vector fields on $\DH$, the charge associated with $t^{\prime\,a}$ on any MTS $S$ is given by
\be Q_{\DH}^{(t^\prime)}[S] \, \=\, \f{1}{4\pi G}\, (\kappao\, A)\,[S] + 2 (\Omegao \JDH)\, [S] \, =: \MDH\ee 
where the notation $\MDH$ mimics the $\MIH$ used on $\IH$. Therefore, denoting the difference {\smash{$Q[S_2] - Q[S_1]$}} between values of charges $Q$ evaluated on $S_2$ and $S_1$ by $\Delta [Q]$, we have the first law on DHSs:
\be \label{dh1law6} \Delta\, \big[M_{{}_{\DH}}\big]\, \,\=\,\Delta\, \big[\f{\kappao A}{4\pi G}\big]\, +\, 2\,\Delta  \big[\Omegao \JDH \big]\,. \ee
One can bring out the physical content of this equation by using the fact that the constraint equation {\smash{$(G_{ab} - 8\pi G\, T_{ab})\,t^{\prime\,a}\, \h\tau^b\, \=\,0$}} implies that  given any two MTSs $S_1$ and $S_2$ on $\DH$, there is  a balance law relating $\Delta\,[{\MDH}]$ to fluxes across the portion of  $\DH$ bounded by the two MTSs:
\ba \label{balance4} \Delta\,[{\MDH}] &\,\=\,&  \,\int_{\Delta\H}\f{1}{8\pi G}\,\big[|DR|\, f(R)\,\,(|\t\sigma|^2+2|\t\zeta|^2)\,+\, (k^{ab}- k q^{ab})\, \Lie_{\Omegao\varphi}\,q_{ab}\,\,\big] +\,  T_{ab}\, t^a\h\tau^b \rmd^3V \nonumber\\
&\=&\,\,\, \Ftprimegws [\Delta\H]\, +\, \Ftprimematt [\Delta\H] \,\,\,\hat\equiv\,\,\, \mathfrak{F}^{{}^{(t^\prime)}} [\Delta \H].
\ea
Therefore, the first law can be rewritten as:
\be \, \mathfrak{F}^{{}^{(t^\prime)}} [\Delta \H]\,\=\, \Delta\,\big[\f{\kappao A}{4\pi G}\big]\, +\, 2 \Delta\, \big[\Omegao \JDH \big]\,.\ee
This is an \emph{active} version of the first law that informs us how $\kappao A$ and $\Omegao \JDH$ change due to  
physical fluxes falling across \emph{finite} portions  $\Delta \H$ of $\DH$ due to \emph{physical} processes. By contrast,  the BCH form of the first law --as well as its generalization to IHSs discussed above-- refers to infinitesimal changes in the observables associated with KHs/IHSs under passive motions in the space of equilibrium states. A key difference between the two is that while the intensive variables $\kappao$ and $\Omegao$ are outside the infinitesimal variations $\delta$ in (\ref{ih1law}) on IHSs,  on the DHSs, they are inside $\Delta$ and cannot be taken out because they have a complicated `time dependence' on $\DH$. However, as  discussed in section \ref{s6.1}, if one restricts oneself to infinitesimally separated MTSs on $\DH$,  one obtains:
\be \delta M_{\DH} \,=\, \f{\kappao}{8\pi\,G}\, \delta A\, +\, \Omegao\, \delta J_{\DH}\,, \ee
thanks to the properties of the projection map $\Pi$ from the space of non-equilibrium states $\N$ to the space of equilibrium states $\E$. Thus the infinitesimal form of the first law on DHSs has the same form as it has on IHSs, but the infinitesimal changes are now caused by physical processes in the immediate vicinity of the MTS at which the infinitesimal variation is taken.

To summarize, the BCH first law was established using Killing horizons. Its  extension to IHSs as well as DHSs can be carried out without any reference to the Hamiltonian framework, the associated symplectic structure or canonical transformations. One can just use Einstein's field equations. This discussion will be helpful to experts in geometric analysis and mathematical relativists with expertise on horizons, who are not as familiar with the intricacies associated with Hamiltonian formulations of GR.

\end{appendix}

\bigskip\bigskip

\bibliography{apspaper}{}

@article{Senovilla_2011,
   title={Trapped surfaces},
   volume={20},
   ISSN={1793-6594},
   url={http://dx.doi.org/10.1142/S0218271811020354},
   DOI={10.1142/s0218271811020354},
   number={11},
   journal={International Journal of Modern Physics D},
   publisher={World Scientific Pub Co Pte Lt},
   author={Senovilla, J.M. M.},
   year={2011},
   month=Oct, 
   pages={2139?2168}, 
}

@article{Clement_2013,
doi = {10.1088/0264-9381/30/6/065017},
url = {https://doi.org/10.1088/0264-9381/30/6/065017},
year = {2013},
month = {mar},
publisher = {IOP Publishing},
volume = {30},
number = {6},
pages = {065017},
author = {Clement, Maria E Gabach and Jaramillo, Jose Luis and Reiris, Martin},
title = {Proof of the area-angular momentum-charge inequality for axisymmetric black holes},
journal = {Classical and Quantum Gravity},
}

@article{Clement_2012,
   title={Black hole area angular momentum charge inequality in dynamical nonvcuum space-times},
   volume={86},
   ISSN={1550-2368},
   url={http://dx.doi.org/10.1103/PhysRevD.86.064021},
   DOI={10.1103/physrevd.86.064021},
   number={6},
   journal={Physical Review D},
   publisher={American Physical Society (APS)},
   author={Gabach Clement, Maria E. and Jaramillo, Jose Luis},
   year={2012},
}

@article{Booth_2007,
   title={Isolated, slowly evolving, and dynamical trapping horizons: Geometry and mechanics from surface deformations},
   volume={75},
   ISSN={1550-2368},
   url={http://dx.doi.org/10.1103/PhysRevD.75.084019},
   DOI={10.1103/physrevd.75.084019},
   number={8},
   journal={Physical Review D},
   publisher={American Physical Society (APS)},
   author={Booth, Ivan and Fairhurst, Stephen},
   year={2007},
   month=Apr }

@article{Gourgoulhon_2006,
   title={Area evolution, bulk viscosity, and entropy principles for dynamical horizons},
   volume={74},
   ISSN={1550-2368},
   url={http://dx.doi.org/10.1103/PhysRevD.74.087502},
   DOI={10.1103/physrevd.74.087502},
   number={8},
   journal={Physical Review D},
   publisher={American Physical Society (APS)},
   author={Gourgoulhon, Eric and Jaramillo, José Luis},
   year={2006},
}

@article{Racz2025,
   title={Quasilocal spin-angular momentum and the construction of axial vector fields},
   volume={112},
   ISSN={2470-0029},
   url={http://dx.doi.org/10.1103/ys28-6k15},
   DOI={10.1103/ys28-6k15},
   number={6},
   journal={Physical Review D},
   publisher={American Physical Society (APS)},
   author={Rácz, István},
   year={2025},
 }

@article{Gourgoulhon_2005,
   title={Generalized Damour-Navier-Stokes equation applied to trapping horizons},
   volume={72},
   ISSN={1550-2368},
   url={http://dx.doi.org/10.1103/PhysRevD.72.104007},
   DOI={10.1103/physrevd.72.104007},
   number={10},
   journal={Physical Review D},
   publisher={American Physical Society (APS)},
   author={Gourgoulhon, Eric},
   year={2005},
}

@misc{metidieri2026,
      title={Universality in the transition from inspiral to plunge: High-accuracy analytic solutions and Catastrophe theory}, 
      author={Ariadna Ribes Metidieri and Béatrice Bonga and Badri Krishnan and José Luis Jaramillo},
      year={2026},
      eprint={2606.13786},
      archivePrefix={arXiv},
      primaryClass={gr-qc},
      url={https://arxiv.org/abs/2606.13786}, 
}

@misc{Rojas_2025,
title={Singularity theory and strong gravity: Caustics and wavefronts as probes into black holes}, 
author={Rojas,O.M.},
Year={2025},
doi={10.70675/99066addz7219z43fbz81fez85eec7ff32cb},
}

@article{Varadarajan:2024clw,
    author = "Varadarajan, Madhavan",
    title = "{Spherical collapse and black hole evaporation}",
    eprint = "2406.09176",
    archivePrefix = "arXiv",
    primaryClass = "gr-qc",
    doi = "10.1103/PhysRevD.111.026005",
    journal = "Phys. Rev. D",
    volume = "111",
    number = "2",
    pages = "026005",
    year = "2025"
}

@article{Ashtekar:1981bq,
	Author = {Ashtekar, A. and Streubel, M.},
	Doi = {10.1098/rspa.1981.0109},
	Journal = {Proc. Roy. Soc. Lond. A},
	Pages = {585--607},
	Title = {{Symplectic Geometry of Radiative Modes and Conserved Quantities at Null Infinity}},
	Volume = {376},
	Year = {1981}
}

@article{Barnich:2011mi,
	Archiveprefix = {arXiv},
	Author = {Barnich, Glenn and Troessaert, Cedric},
	Doi = {10.1007/JHEP12(2011)105},
	Eprint = {hepth/1106.0213},
	Journal = {JHEP},
	Pages = {105},
	Primaryclass = {hep-th},
	Reportnumber = {ULB-TH-11-10},
	Title = {{BMS charge algebra}},
	Volume = {12},
	Year = {2011},
}

@article{Barnich:2001jy,
	Archiveprefix = {arXiv},
	Author = {Barnich, Glenn and Brandt, Friedemann},
	Doi = {10.1016/S0550-3213(02)00251-1},
	Eprint = {hep-th/0111246},
	Journal = {Nucl. Phys.},
	Pages = {3-82},
	Primaryclass = {hep-th},
	Reportnumber = {ULB-TH-01-19, MPI-MIS-94-2001},
	Slaccitation = {%%CITATION = HEP-TH/0111246;%%},
	Title = {{Covariant theory of asymptotic symmetries, conservation laws and central charges}},
	Volume = {B633},
	Year = {2002}
}

@inproceedings{AshtekarReula,
	Author = {Ashtekar, A and Bombelli, L and Reula, O},
	Booktitle = {Analysis, geometry and mechanics: 200 years after Lagrange},
	Editor = {Francaviglia, M and Holm, D},
	Publisher = {North-Holland},
	Title = {The covariant phase space of asymptotically flat gravitational fields},
	Year = {1991}}

@article{BMS,
	Author = {Bondi, H. and van der Burg, M. G. J. and Metzner, A. W. K.},
	Doi = {10.1098/rspa.1962.0161},
	Journal = {Proc. Roy. Soc. Lond.},
	Pages = {21-52},
	Slaccitation = {%%CITATION = PRSLA,A269,21;%%},
	Title = {{Gravitational waves in general relativity. 7. Waves from axisymmetric isolated systems}},
	Volume = {A269},
	Year = {1962}
}

@article{Chakraborty_2015,
   title={Thermodynamical interpretation of the geometrical variables associated with null surfaces},
   volume={92},
   ISSN={1550-2368},
   url={http://dx.doi.org/10.1103/PhysRevD.92.104011},
   DOI={10.1103/physrevd.92.104011},
   number={10},
   journal={Physical Review D},
   publisher={American Physical Society (APS)},
   author={Chakraborty, Sumanta and Padmanabhan, T.},
   year={2015},
   month=nov }

@article{Jacobson_1995,
   title={Thermodynamics of Spacetime: The Einstein Equation of State},
   volume={75},
   ISSN={1079-7114},
   url={http://dx.doi.org/10.1103/PhysRevLett.75.1260},
   DOI={10.1103/physrevlett.75.1260},
   number={7},
   journal={Physical Review Letters},
   publisher={American Physical Society (APS)},
   author={Jacobson, Ted},
   year={1995},
   month=aug, pages={1260-1263} }

@article{Hawking_1996,
   title={The gravitational Hamiltonian, action, entropy and surface terms},
   volume={13},
   ISSN={1361-6382},
   url={http://dx.doi.org/10.1088/0264-9381/13/6/017},
   DOI={10.1088/0264-9381/13/6/017},
   number={6},
   journal={Classical and Quantum Gravity},
   publisher={IOP Publishing},
   author={Hawking, S W and Horowitz, Gary T},
   year={1996},
   month=jun, pages={1487-1498} }

@article{PhysRevD.108.044069,
  title = {Second law from the Noether current on null hypersurfaces},
  author = {Rignon-Bret, Antoine},
  eprint={2303.07262},
      archivePrefix={arXiv},
      primaryClass={gr-qc},
      url={https://arxiv.org/abs/2303.07262}, 
      journal = {Phys. Rev. D},
  volume = {108},
  issue = {4},
  pages = {044069},
  numpages = {28},
  year = {2023},
  month = {Aug},
  publisher = {American Physical Society},
  doi = {10.1103/PhysRevD.108.044069},
}

@misc{rinconramirez2026maximumentropyconjectureblack,
      title={A Maximum Entropy Conjecture for Black Hole Mergers}, 
      author={Monica Rincon-Ramirez and Nathan K. Johnson-McDaniel and Eugenio Bianchi and Ish Gupta and Vaishak Prasad and B. S. Sathyaprakash},
      year={2026},
      eprint={2601.22388},
      archivePrefix={arXiv},
      primaryClass={gr-qc},
      url={https://arxiv.org/abs/2601.22388}, 
}

@article{Emperan-SGP,
	author = {Emperan,Roberto},
	journal = {50 Years the black hole information paradox, Simons Center for Geometry and Physics, 2025  Stonybrook workshop},
	title = {Is the black hole still there when we look?},
	year = {https://scgp.stonybrook.edu/video\_portal/video.php?id=7351}}

@misc{pookkolb2020horizonsbinaryblackhole,
      title={Horizons in a binary black hole merger II: Fluxes, multipole moments and stability}, 
      author={Daniel Pook-Kolb and Ofek Birnholtz and Jose Luis Jaramillo and Badri Krishnan and Erik Schnetter},
      year={2020},
      eprint={2006.03940},
      archivePrefix={arXiv},
      primaryClass={gr-qc},
      url={https://arxiv.org/abs/2006.03940}, 
}

@article{Agullo_2024,
   title={Entangled pairs in evaporating black holes without event horizons},
   volume={110},
   ISSN={2470-0029},
   url={http://dx.doi.org/10.1103/PhysRevD.110.085002},
   DOI={10.1103/physrevd.110.085002},
   number={8},
   journal={Physical Review D},
   publisher={American Physical Society (APS)},
   author={Agullo, Ivan and Calizaya Cabrera, Paula and Elizaga Navascu\'es, Beatriz},
   year={2024},
   month=oct }

@article{Ashtekar_2011a,
   title={Surprises in the Evaporation of 2D Black Holes},
   volume={106},
   ISSN={1079-7114},
   url={http://dx.doi.org/10.1103/PhysRevLett.106.161303},
   DOI={10.1103/physrevlett.106.161303},
   number={16},
   journal={Physical Review Letters},
   publisher={American Physical Society (APS)},
   author={Ashtekar, Abhay and Pretorius, Frans and Ramazanoglu, Fethi M.},
   year={2011},
   month=apr }

@article{Ashtekar_2011b,
   title={Evaporation of two-dimensional black holes},
   volume={83},
   ISSN={1550-2368},
   url={http://dx.doi.org/10.1103/PhysRevD.83.044040},
   DOI={10.1103/physrevd.83.044040},
   number={4},
   journal={Physical Review D},
   publisher={American Physical Society (APS)},
   author={Ashtekar, Abhay and Pretorius, Frans and Ramazanoglu, Fethi M.},
   year={2011},
   month=feb }

@misc{hollands2024entropydynamicalblackholes,
      title={The Entropy of Dynamical Black Holes}, 
      author={Stefan Hollands and Robert M. Wald and Victor G. Zhang},
      year={2024},
      eprint={2402.00818},
      archivePrefix={arXiv},
      primaryClass={hep-th},
      url={https://arxiv.org/abs/2402.00818}, }

@misc{visser2025dynamicalentropychargedblack,
      title={Dynamical entropy of charged black objects}, 
      author={Manus R. Visser and Zihan Yan},
      year={2025},
      eprint={2510.20747},
      archivePrefix={arXiv},
      primaryClass={hep-th},
      url={https://arxiv.org/abs/2510.20747}, 
}

@article{aps-letter,
  title = {Thermodynamics of Black Holes, Far from Equilibrium},
  author = {Ashtekar, Abhay and Paraizo, Daniel E. and Shu, Jonathan},
  journal = {Phys. Rev. Lett.},
  volume = {136},
  issue = {25},
  pages = {251405 (Editors Suggestion)},
  numpages = {6},
  year = {2026},
  month = {Jun},
  publisher = {American Physical Society},
  doi = {10.1103/3c1r-v8f1},
  url = {https://link.aps.org/doi/10.1103/3c1r-v8f1}
}

@Article{Ashtekar:2025LivRev,
     author    = "Ashtekar, Abhay and Krishnan, Badri",
     title     = "{Quasi-Local Horizons: Recent Developments}",
     journal   = "Living Rev. Rel.",
     volume    = "28",
     year      = "2025",
     pages     = "8",
     DOI       ={10.1007/s41114-025-00061-4},
     eprint    = "2502.11825",
     archivePrefix = "arXiv",
     SLACcitation  = "%%CITATION = GR-QC/0407042;%%"
}

@article{Pook-Kolb:2018igu,
title = {Existence and stability of marginally trapped surfaces in black-hole spacetimes},
  author = {Pook-Kolb, Daniel and Birnholtz, Ofek and Krishnan, Badri and Schnetter, Erik},
  journal = {Phys. Rev. D},
  volume = {99},
  issue = {6},
  pages = {064005},
  numpages = {21},
  year = {2019},
  month = {Mar},
  publisher = {American Physical Society},
  doi = {10.1103/PhysRevD.99.064005},
  url = {https://link.aps.org/doi/10.1103/PhysRevD.99.064005}
}

@article{PhysRevLett.123.171102,
  title = {Interior of a Binary Black Hole Merger},
  author = {Pook-Kolb, Daniel and Birnholtz, Ofek and Krishnan, Badri and Schnetter, Erik},
  journal = {Phys. Rev. Lett.},
  volume = {123},
  issue = {17},
  pages = {171102},
  numpages = {6},
  year = {2019},
  month = {Oct},
  publisher = {American Physical Society},
  doi = {10.1103/PhysRevLett.123.171102},
  url = {https://link.aps.org/doi/10.1103/PhysRevLett.123.171102}
}

@article{PhysRevD.100.084044,
  title = {Self-intersecting marginally outer trapped surfaces},
  author = {Pook-Kolb, Daniel and Birnholtz, Ofek and Krishnan, Badri and Schnetter, Erik},
  journal = {Phys. Rev. D},
  volume = {100},
  issue = {8},
  pages = {084044},
  numpages = {14},
  year = {2019},
  month = {Oct},
  publisher = {American Physical Society},
  doi = {10.1103/PhysRevD.100.084044},
  url = {https://link.aps.org/doi/10.1103/PhysRevD.100.084044}
}

@Book{LesHouches:1973,
      author         = "DeWitt, Cecile and DeWitt, Bryce S.", 
      title          = "Black Holes: 23rd session of summer school of Les Houches", 
      year           = 1973,
      publisher      = "Gordon Breach (New York)"
}

@inproceedings{bc,
author             = {Carter, Brandon},
editor             = {DeWitt, Cecile and DeWitt, Bryce S.},
title              = {Black hole equilibrium states},
booktitle          = {Black holes},
series             = {Les Houches},
year               = {1972},
pages              = {57--214}
}

@article{Andersson:2007fh,
      author         = "Andersson, Lars and Mars, Marc and Simon, Walter",
      title          = "{Stability of marginally outer trapped surfaces and
                        existence of marginally outer trapped tubes}",
      journal        = "Adv.Theor.Math.Phys.",
      volume         = "12",
      year           = "2008",
      eprint         = "0704.2889",
      archivePrefix  = "arXiv",
      primaryClass   = "gr-qc",
      SLACcitation   = "%%CITATION = ARXIV:0704.2889;%%",
}

@Article{Ashtekar:2000sz,
     author    = "Ashtekar, Abhay and others",
          title     = "{Isolated horizons and their applications}",
     journal   = "Phys. Rev. Lett.",
     volume    = "85",
     year      = "2000",
     pages     = "3564-3567",
     eprint    = "gr-qc/0006006",
     archivePrefix = "arXiv",
     doi       = "10.1103/PhysRevLett.85.3564",
     SLACcitation  = "%%CITATION = GR-QC/0006006;%%"
}

@Article{Ashtekar:2001is,
     author    = "Ashtekar, Abhay and Beetle, Christopher and Lewandowski,
                  Jerzy",
     title     = "{Mechanics of Rotating Isolated Horizons}",
     journal   = "Phys. Rev.",
     volume    = "D64",
     year      = "2001",
     pages     = "044016",
     eprint    = "gr-qc/0103026",
     archivePrefix = "arXiv",
     doi       = "10.1103/PhysRevD.64.044016",
     SLACcitation  = "%%CITATION = GR-QC/0103026;%%"
}

@Article{Ashtekar:2001jb,
     author    = "Ashtekar, Abhay and Beetle, Christopher and Lewandowski,
                  Jerzy",
     title     = "{Geometry of Generic Isolated Horizons}",
     journal   = "Class. Quant. Grav.",
     volume    = 19,
     year      = 2002,
     pages     = "1195-1225",
     eprint    = "gr-qc/0111067",
     archivePrefix = "arXiv",
     doi       = "10.1088/0264-9381/19/6/311",
     SLACcitation  = "%%CITATION = GR-QC/0111067;%%"
}

@Article{Ashtekar:2004gp,
     author    = "Ashtekar, Abhay and Engle, Jonathan and Pawlowski, Tomasz
                  and Van Den Broeck, Chris",
     title     = "{Multipole moments of isolated horizons}",
     journal   = "Class. Quant. Grav.",
     volume    = "21",
     year      = "2004",
     pages     = "2549-2570",
     eprint    = "gr-qc/0401114",
     archivePrefix = "arXiv",
     doi       = "10.1088/0264-9381/21/11/003",
     SLACcitation  = "%%CITATION = GR-QC/0401114;%%"
}

@Article{Ashtekar:2000hw,
     author    = "Ashtekar, Abhay and Fairhurst, Stephen and Krishnan, Badri
                  ",
     title     = "{Isolated horizons: Hamiltonian evolution and the first
                  law}",
     journal   = "Phys. Rev.",
     volume    = "D62",
     year      = "2000",
     pages     = "104025",
     eprint    = "gr-qc/0005083",
     archivePrefix = "arXiv",
     doi       = "10.1103/PhysRevD.62.104025",
     SLACcitation  = "%%CITATION = GR-QC/0005083;%%"
}

@Article{Ashtekar:2004cn,
     author    = "Ashtekar, Abhay and Krishnan, Badri",
     title     = "{Isolated and dynamical horizons and their applications}",
     journal   = "Living Rev. Rel.",
     volume    = "7",
     year      = "2004",
     pages     = "10",
     eprint    = "gr-qc/0407042",
     archivePrefix = "arXiv",
     SLACcitation  = "%%CITATION = GR-QC/0407042;%%"
}

@Article{Ashtekar:2002ag,
     author    = "Ashtekar, Abhay and Krishnan, Badri",
     title     = "{Dynamical horizons: Energy, angular momentum, fluxes and
                  balance laws}",
     journal   = "Phys. Rev. Lett.",
     volume    = "89",
     year      = "2002",
     pages     = "261101",
     eprint    = "gr-qc/0207080",
     archivePrefix = "arXiv",
     doi       = "10.1103/PhysRevLett.89.261101",
     SLACcitation  = "%%CITATION = GR-QC/0207080;%%"
}

@Article{Ashtekar:2003hk,
     author    = "Ashtekar, Abhay and Krishnan, Badri",
     title     = "{Dynamical horizons and their properties}",
     journal   = "Phys. Rev.",
     volume    = "D68",
     year      = "2003",
     pages     = "104030",
     eprint    = "gr-qc/0308033",
     archivePrefix = "arXiv",
     doi       = "10.1103/PhysRevD.68.104030",
     SLACcitation  = "%%CITATION = GR-QC/0308033;%%"
}

@article{PhysRevD.36.1065,
  title = {Origin of {H}awking radiation},
  author = {Haji\v{c}ek, Petr},
  journal = {Phys. Rev. D},
  volume = {36},
  issue = {4},
  pages = {1065--1079},
  numpages = {0},
  year = {1987},
  month = {Aug},
  publisher = {American Physical Society},
  doi = {10.1103/PhysRevD.36.1065},
  url = {https://link.aps.org/doi/10.1103/PhysRevD.36.1065}
}

@Article{Bardeen:1973gs,
     author    = "Bardeen, James M. and Carter, Brandon and Hawking, Stephen W.",
     title     = "{The Four laws of black hole mechanics}",
     journal   = "Commun. Math. Phys.",
     volume    = "31",
     year      = "1973",
     pages     = "161-170",
     doi       = "10.1007/BF01645742",
     SLACcitation  = "%%CITATION = CMPHA,31,161;%%"
}

@Article{Bekenstein:1973ur,
     author    = "Bekenstein, Jacob D.",
     title     = "{Black holes and entropy}",
     journal   = "Phys. Rev.",
     volume    = "D7",
     year      = "1973",
     pages     = "2333-2346",
     doi       = "10.1103/PhysRevD.7.2333",
     SLACcitation  = "%%CITATION = PHRVA,D7,2333;%%"
}

@Article{Bekenstein:1974ax,
     author    = "Bekenstein, Jacob D.",
     title     = "{Generalized second law of thermodynamics in black hole
                  physics}",
     journal   = "Phys. Rev.",
     volume    = "D9",
     year      = "1974",
     pages     = "3292-3300",
     doi       = "10.1103/PhysRevD.9.3292",
     SLACcitation  = "%%CITATION = PHRVA,D9,3292;%%"
}

@Article{Booth:2003ji,
     author    = "Booth, Ivan and Fairhurst, Stephen",
     title     = "{The first law for slowly evolving horizons}",
     journal   = "Phys. Rev. Lett.",
     volume    = "92",
     year      = "2004",
     pages     = "011102",
     eprint    = "gr-qc/0307087",
     archivePrefix = "arXiv",
     doi       = "10.1103/PhysRevLett.92.011102",
     SLACcitation  = "%%CITATION = GR-QC/0307087;%%"
}

@article{Geroch:1978ub,
      author         = "Geroch, Robert P. and Horowitz, G.T.",
      title          = "{Asymptotically simple does not imply asymptotically
                        Minkowskian}",
      journal        = "Phys.Rev.Lett.",
      volume         = "40",
      pages          = "203-206",
      doi            = "10.1103/PhysRevLett.40.203",
      year           = "1978",
      SLACcitation   = "%%CITATION = PRLTA,40,203;%%",
}

@Article{Hawking:1974sw,
     author    = "Hawking, Stephen W.",
     title     = "{Particle Creation by Black Holes}",
     journal   = "Commun. Math. Phys.",
     volume    = "43",
     year      = "1975",
     pages     = "199-220",
     doi       = "10.1007/BF02345020",
     SLACcitation  = "%%CITATION = CMPHA,43,199;%%"
}

@article{Hayward:1993wb,
      author         = "Hayward, Sean A.",
      eprint          = "gr-qc/9303006", 
      archivePrefix =  "arXiv",
      title          = "{General laws of black hole dynamics}",
      journal        = "Phys. Rev.",
      volume         = "D49",
      pages          = "6467-6474",
      doi            = "10.1103/PhysRevD.49.6467",
      year           = "1994",
      SLACcitation   = "%%CITATION = PHRVA,D49,6467;%%",
}

@Article{Lewandowski:1999zs,
     author    = "Lewandowski, Jerzy",
     title     = "{Spacetimes Admitting Isolated Horizons}",
     journal   = "Class. Quant. Grav.",
     volume    = "17",
     year      = "2000",
     pages     = "L53-L59",
     eprint    = "gr-qc/9907058",
     archivePrefix = "arXiv",
     doi       = "10.1088/0264-9381/17/4/101",
     SLACcitation  = "%%CITATION = GR-QC/9907058;%%"
}

@article{Senovilla_2003,
   title={On the existence of horizons in spacetimes with vanishing curvature invariants},
   volume={2003},
   ISSN={1029-8479},
   url={http://dx.doi.org/10.1088/1126-6708/2003/11/046},
   DOI={10.1088/1126-6708/2003/11/046},
   number={11},
   journal={JHEP},
   publisher={Springer Science and Business Media LLC},
   author={Senovilla, Jos\'e M. M},
   year={2003},
   month=nov, pages={046} 
}

@Article{Wald:1993nt,
     author    = "Wald, Robert M.",
     title     = "{Black hole entropy is the Noether charge}",
     journal   = "Phys. Rev.",
     volume    = "D48",
     year      = "1993",
     pages     = "3427-3431",
     eprint    = "gr-qc/9307038",
     archivePrefix = "arXiv",
     doi       = "10.1103/PhysRevD.48.R3427",
}

@Article{Wald:1999wa,
     author    = "Wald, Robert M. and Zoupas, Andreas",
     title     = "{A General Definition of "Conserved Quantities" in General
                  Relativity and Other Theories of Gravity}",
     journal   = "Phys. Rev.",
     volume    = "D61",
     year      = "2000",
     pages     = "084027",
     eprint    = "gr-qc/9911095",
     archivePrefix = "arXiv",
     doi       = "10.1103/PhysRevD.61.084027",
     SLACcitation  = "%%CITATION = GR-QC/9911095;%%"
}

@Article{Newman:1961qr,
     author    = "Newman, Ezra and Penrose, Roger",
     title     = "{An Approach to gravitational radiation by a method of spin
                  coefficients}",
     journal   = "J. Math. Phys.",
     volume    = "3",
     year      = "1962",
     pages     = "566-578",
     SLACcitation  = "%%CITATION = JMAPA,3,566;%%"
}

@article{Galloway:1999ny,
    author = "Galloway, Gregory J.",
    title = "{Maximum principles for null hypersurfaces and null splitting theorems}",
    eprint = "math/9909158",
    archivePrefix = "arXiv",
    doi = "10.1007/s000230050006",
    journal = "Annales Henri Poincare",
    volume = "1",
    pages = "543--567",
    year = "2000"
}

@Article{Ashtekar:2005ez,
     author    = "Ashtekar, Abhay and Galloway, Gregory J.",
     title     = "{Some uniqueness results for dynamical horizons}",
     journal   = "Adv. Theor. Math. Phys.",
     volume    = "9",
     year      = "2005",
     pages     = "1-30",
     eprint    = "gr-qc/0503109",
     archivePrefix = "arXiv",
     SLACcitation  = "%%CITATION = GR-QC/0503109;%%"
}

@Article{Ashtekar:2005cj,
     author    = "Ashtekar, Abhay and Bojowald, Martin",
     title     = "{Black hole evaporation: A paradigm}",
     journal   = "Class. Quant. Grav.",
     volume    = "22",
     year      = "2005",
     pages     = "3349-3362",
     eprint    = "gr-qc/0504029",
     archivePrefix = "arXiv",
     doi       = "10.1088/0264-9381/22/16/014",
     SLACcitation  = "%%CITATION = GR-QC/0504029;%%"
}

@Article{Bartnik:2005qj,
     author    = "Bartnik, Robert and Isenberg, James",
     title     = "{Spherically symmetric dynamical horizons}",
     journal   = "Class. Quant. Grav.",
     volume    = "23",
     year      = "2006",
     pages     = "2559-2570",
     eprint    = "gr-qc/0512091",
     archivePrefix = "arXiv",
     doi       = "10.1088/0264-9381/23/7/020",
     SLACcitation  = "%%CITATION = GR-QC/0512091;%%"
}

@Article{Booth:2005qc,
     author    = "Booth, Ivan",
     title     = "{Black hole boundaries}",
     journal   = "Can. J. Phys.",
     volume    = "83",
     year      = "2005",
     pages     = "1073-1099",
     eprint    = "gr-qc/0508107",
     archivePrefix = "arXiv",
     doi       = "10.1139/p05-063",
     SLACcitation  = "%%CITATION = GR-QC/0508107;%%"
}

@article{Dain:2011pi,
    author = "Dain, Sergio and Reiris, Martin",
    title = "{Area - Angular momentum inequality for axisymmetric black holes}",
    eprint = "1102.5215",
    archivePrefix = "arXiv",
    primaryClass = "gr-qc",
    doi = "10.1103/PhysRevLett.107.051101",
    journal = "Phys. Rev. Lett.",
    volume = "107",
    pages = "051101",
    year = "2011"
}

@Article{Dain:2010qr,
     author    = "Dain, Sergio",
     title     = "{Extreme throat initial data set and horizon area-angular
                  momentum inequality for axisymmetric black holes}",
     journal   = "Phys. Rev.",
     volume    = "D82",
     year      = "2010",
     pages     = "104010",
     eprint    = "1008.0019",
     archivePrefix = "arXiv",
     primaryClass  =  "gr-qc",
     doi       = "10.1103/PhysRevD.82.104010",
     SLACcitation  = "%%CITATION = 1008.0019;%%"
}

@Article{Gourgoulhon:2005ng,
     author    = "Gourgoulhon, Eric and Jaramillo, Jose Luis",
     title     = "{A 3+1 perspective on null hypersurfaces and isolated
                  horizons}",
     journal   = "Phys. Rept.",
     volume    = "423",
     year      = "2006",
     pages     = "159-294",
     eprint    = "gr-qc/0503113",
     archivePrefix = "arXiv",
     doi       = "10.1016/j.physrep.2005.10.005",
     SLACcitation  = "%%CITATION = GR-QC/0503113;%%"
}

@Article{Hayward:2009ji,
     author    = "Hayward, Sean A.",
     title     = "{Involute, minimal, outer and increasingly trapped
                  spheres}",
     journal   = "Phys. Rev.",
     volume    = "D81",
     year      = "2010",
     pages     = "024037",
     eprint    = "0905.3950",
     archivePrefix = "arXiv",
     primaryClass  =  "gr-qc",
     doi       = "10.1103/PhysRevD.81.024037",
     SLACcitation  = "%%CITATION = 0905.3950;%%"
}

@article{Jaramillo:2011pg,
      author         = "Jaramillo, Jose Luis and Reiris, Martin and Dain, Sergio",
      title          = "{Black hole Area-Angular momentum inequality in
                        non-vacuum spacetimes}",
      journal        = "Phys.Rev.",
      volume         = "D84",
      pages          = "121503",
      doi            = "10.1103/PhysRevD.84.121503",
      year           = "2011",
      eprint         = "1106.3743",
      archivePrefix  = "arXiv",
      primaryClass   = "gr-qc",
      SLACcitation   = "%%CITATION = ARXIV:1106.3743;%%",
}

@Article{Korzynski:2007hu,
     author    = "Korzynski, Mikolaj",
     title     = "{Quasi--local angular momentum of non--symmetric isolated
                  and dynamical horizons from the conformal decomposition of
                  the metric}",
     journal   = "Class. Quant. Grav.",
     volume    = "24",
     year      = "2007",
     pages     = "5935-5944",
     eprint    = "0707.2824",
     archivePrefix = "arXiv",
     primaryClass  =  "gr-qc",
     doi       = "10.1088/0264-9381/24/23/015",
     SLACcitation  = "%%CITATION = 0707.2824;%%"
}

@misc{aa-nk,
      title={Gravitational Wave Tomography}, 
      author={Ashtekar, Abhay and Khera, Neev},
      year={2025},
      eprint={in preparation},
      archivePrefix={arXiv},
      primaryClass={gr-qc},
      url={}, 
}

@Article{Sawayama:2005mw,
     author    = "Sawayama, Shintaro",
     title     = "{Dynamical horizon of evaporating black hole in Vaidya
                  spacetime. or:  Evaporating dynamical horizon with Hawking
                  effect in Vaidya spacetime}",
     journal   = "Phys. Rev.",
     volume    = "D73",
     year      = "2006",
     pages     = "064024",
     eprint    = "gr-qc/0509048",
     archivePrefix = "arXiv",
     doi       = "10.1103/PhysRevD.73.064024",
     SLACcitation  = "%%CITATION = GR-QC/0509048;%%"
}

@Article{Schnetter:2005ea,
     author    = "Schnetter, Erik and Krishnan, Badri",
     title     = "{Non-symmetric trapped surfaces in the Schwarzschild and
                  Vaidya spacetimes}",
     journal   = "Phys. Rev.",
     volume    = "D73",
     year      = "2006",
     pages     = "021502",
     eprint    = "gr-qc/0511017",
     archivePrefix = "arXiv",
     doi       = "10.1103/PhysRevD.73.021502",
     SLACcitation  = "%%CITATION = GR-QC/0511017;%%"
}

@article{Ashtekar:2013qta,
      author         = "Ashtekar, Abhay and Campiglia, Miguel and Shah, Samir",
      title          = "{Dynamical Black Holes: Approach to the Final State}",
      journal        = "Phys. Rev.",
      volume         = "D88",
      year           = "2013",
      number         = "6",
      pages          = "064045",
      doi            = "10.1103/PhysRevD.88.064045",
      eprint         = "1306.5697",
      archivePrefix  = "arXiv",
      primaryClass   = "gr-qc",
      reportNumber   = "IGC-13-06-2",
      SLACcitation   = "%%CITATION = ARXIV:1306.5697;%%"
}

@article{Jaramillo:2011zw,
      author         = "Jaramillo, Jose Luis",
      title          = "{An introduction to local Black Hole horizons in the 3+1
                        approach to General Relativity}",
      journal        = "Int. J. Mod. Phys.",
      volume         = "D20",
      year           = "2011",
      pages          = "2169",
      doi            = "10.1142/S0218271811020366",
      eprint         = "1108.2408",
      archivePrefix  = "arXiv",
      primaryClass   = "gr-qc",
      SLACcitation   = "%%CITATION = ARXIV:1108.2408;%%"
}

@article{Chrusciel:1996tw,
      author         = "Chrusciel, Piotr T. and Galloway, Gregory J.",
      title          = "{`Nowhere' differentiable horizons}",
      journal        = "Commun. Math. Phys.",
      volume         = "193",
      year           = "1998",
      pages          = "449-470",
      doi            = "10.1007/s002200050336",
      eprint         = "gr-qc/9611032",
      archivePrefix  = "arXiv",
      primaryClass   = "gr-qc",
      SLACcitation   = "%%CITATION = GR-QC/9611032;%%"
}

@article{Prasad:2020xgr,
    author = "Prasad, Vaishak and Gupta, Anshu and Bose, Sukanta and Krishnan, Badri and Schnetter, Erik",
    title = "{News from horizons in binary black hole mergers}",
    eprint = "2003.06215",
    archivePrefix = "arXiv",
    primaryClass = "gr-qc",
    reportNumber = "LIGO Preprint number LIGO-P2000098",
    doi = "10.1103/PhysRevLett.125.121101",
    journal = "Phys. Rev. Lett.",
    volume = "125",
    number = "12",
    pages = "121101",
    year = "2020"
}

@article{Prasad:2021dfr,
    author = "Prasad, Vaishak and Gupta, Anshu and Bose, Sukanta and Krishnan, Badri",
    title = "{Tidal deformation of dynamical horizons in binary black hole mergers}",
    eprint = "2106.02595",
    archivePrefix = "arXiv",
    primaryClass = "gr-qc",
    reportNumber = "This manuscript has been assigned the LIGO Preprint number
  LIGO-P2100109",
    doi = "10.1103/PhysRevD.105.044019",
    journal = "Phys. Rev. D",
    volume = "105",
    number = "4",
    pages = "044019",
    year = "2022"
}

@article{akkl1,
	archiveprefix = {arXiv},
	author = {Abhay Ashtekar and Neev Khera and Maciej Kolanowski and Jerzy Lewandowski},
	doi = {10.1007/JHEP01(2022)028},
	eprint = {2111.07873},
	journal = {JHEP},
	number = {1},
	primaryclass = {gr-qc},
	title = {Non-Expanding horizons: Multipoles and the Symmetry Group},
	volume = {2022},
	year = {2022}
	}

@article{Ashtekar:2021kqj,
   title={Charges and fluxes on (perturbed) non-expanding horizons},
   volume={2022},
   ISSN={1029-8479},
   url={http://dx.doi.org/10.1007/JHEP02(2022)066},
   DOI={10.1007/jhep02(2022)066},
   number={2},
   journal={Journal of High Energy Physics},
   publisher={Springer Science and Business Media LLC},
   author={Ashtekar, Abhay and Khera, Neev and Kolanowski, Maciej and Lewandowski, Jerzy},
   year={2022},
   month=feb }

@article{Ashtekar_2020,
   title={Black Hole Evaporation: A Perspective from Loop Quantum Gravity},
   eprint = "2001.08833",
    archivePrefix = "arXiv",
    primaryClass = "gr-qc",
    volume={6},
   ISSN={2218-1997},
   url={http://dx.doi.org/10.3390/universe6020021},
   DOI={10.3390/universe6020021},
   number={2},
   journal={Universe},
   publisher={MDPI AG},
   author={Ashtekar, Abhay},
   year={2020},
   month=jan, pages={21} 
   }

@article{senovilla2012stabilityoperatormotscore,
    author = "Senovilla, Jos\'e M. M.",
    editor = "Bi\v{c}\'ak, Ji\v{r}\'\i{} and Ledvinka, Tom\'a\v{s}",
    title = "{On the Stability Operator for {MOTS} and the 'Core' of Black Holes}",
    eprint = "1210.3731",
    archivePrefix = "arXiv",
    primaryClass = "gr-qc",
    doi = "10.1007/978-3-319-06761-2_27",
    journal = "Springer Proc. Phys.",
    volume = "157",
    pages = "215--222",
    year = "2014"
}

@article{Hawking:1971vc,
    author = "Hawking, Stephen W.",
    title = "{Black holes in general relativity}",
    doi = "10.1007/BF01877517",
    journal = "Commun. Math. Phys.",
    volume = "25",
    pages = "152--166",
    year = "1972"
}

@article{Ashtekar2_2024,
   title={Null infinity as a weakly isolated horizon},
   volume={110},
   ISSN={2470-0029},
   url={http://dx.doi.org/10.1103/PhysRevD.110.044048},
   DOI={10.1103/physrevd.110.044048},
   number={4},
   journal={Physical Review D},
   publisher={American Physical Society (APS)},
   author={Ashtekar, Abhay and Speziale, Simone},
   year={2024},
   month=aug }

@article{Ashtekar3_2024,
   title={Null infinity and horizons: A new approach to fluxes and charges},
   volume={110},
   ISSN={2470-0029},
   url={http://dx.doi.org/10.1103/PhysRevD.110.044049},
   DOI={10.1103/physrevd.110.044049},
   number={4},
   journal={Physical Review D},
   publisher={American Physical Society (APS)},
   author={Ashtekar, Abhay and Speziale, Simone},
   year={2024},
   month=aug }

@misc{kehle2024extremalblackholeformation,
      title={Extremal black hole formation as a critical phenomenon}, 
      author={Christoph Kehle and Ryan Unger},
      year={2024},
      eprint={2402.10190},
      archivePrefix={arXiv},
      primaryClass={gr-qc},
      url={https://arxiv.org/abs/2402.10190}, 
}

@misc{gadioux2023creasescornerscausticsproperties,
      title={Creases, corners and caustics: properties of non-smooth structures on black hole horizons}, 
      author={Maxime Gadioux and Harvey S. Reall},
      year={2023},
      eprint={2303.15512},
      archivePrefix={arXiv},
      primaryClass={gr-qc},
      url={https://arxiv.org/abs/2303.15512}, 
}

@article{Booth:2021sow,
    author = "Booth, Ivan and Hennigar, Robie A. and Pook-Kolb, Daniel",
    title = "{Ultimate fate of apparent horizons during a binary black hole merger. I. Locating and understanding axisymmetric marginally outer trapped surfaces}",
    eprint = "2104.11343",
    archivePrefix = "arXiv",
    primaryClass = "gr-qc",
    doi = "10.1103/PhysRevD.104.084083",
    journal = "Phys. Rev. D",
    volume = "104",
    number = "8",
    pages = "084083",
    year = "2021"
}

@article{RibesMetidieri:2024tpk,
    author = "Ribes Metidieri, Ariadna and Bonga, B\'eatrice and Krishnan, Badri",
    title = "{Tidal deformations of slowly spinning isolated horizons}",
    eprint = "2403.17114",
    archivePrefix = "arXiv",
    primaryClass = "gr-qc",
    doi = "10.1103/PhysRevD.110.024069",
    journal = "Phys. Rev. D",
    volume = "110",
    number = "2",
    pages = "024069",
    year = "2024"
}

@article{Andersson:2005gq,
      author         = "Andersson, Lars and Mars, Marc and Simon, Walter",
      title          = "{Local existence of dynamical and trapping horizons}",
      journal        = "Phys.Rev.Lett.",
      volume         = "95",
      pages          = "111102",
      doi            = "10.1103/PhysRevLett.95.111102",
      year           = "2005",
      eprint         = "gr-qc/0506013",
      archivePrefix  = "arXiv",
      primaryClass   = "gr-qc",
      SLACcitation   = "%%CITATION = GR-QC/0506013;%%",
}

@article{Pook-Kolb:2021gsh,
    author = "Pook-Kolb, Daniel and Hennigar, Robie A. and Booth, Ivan",
    title = "{What Happens to Apparent Horizons in a Binary Black Hole Merger?}",
    eprint = "2104.10265",
    archivePrefix = "arXiv",
    primaryClass = "gr-qc",
    doi = "10.1103/PhysRevLett.127.181101",
    journal = "Phys. Rev. Lett.",
    volume = "127",
    number = "18",
    pages = "181101",
    year = "2021"
}

@article{Pook-Kolb:2021jpd,
    author = "Pook-Kolb, Daniel and Booth, Ivan and Hennigar, Robie A.",
    title = "{Ultimate fate of apparent horizons during a binary black hole merger. II. The vanishing of apparent horizons}",
    eprint = "2104.11344",
    archivePrefix = "arXiv",
    primaryClass = "gr-qc",
    doi = "10.1103/PhysRevD.104.084084",
    journal = "Phys. Rev. D",
    volume = "104",
    number = "8",
    pages = "084084",
    year = "2021"
}

@article{pc,
    author = "Chru\'sciel, P. T.",
    title = "{On the global structure of Robinson-Trautman space-times}",
    doi = "10.1098/rspa.1992.0019",
    journal = "Proc. Roy. Soc. Lond. A",
    volume = "436",
    pages = "299--316",
    year = "1992"
}

@article{Podolsky:2009an,
    author = "Podolsky, Jiri and Svitek, Otakar",
    title = "{Past horizons in Robinson-Trautman spacetimes with a cosmological constant}",
    eprint = "0911.5317",
    archivePrefix = "arXiv",
    primaryClass = "gr-qc",
    doi = "10.1103/PhysRevD.80.124042",
    journal = "Phys. Rev. D",
    volume = "80",
    pages = "124042",
    year = "2009"
}

@article{andersson1997,
    author = "Andersson, Lars and Galloway, Gregory J. and Howard, Ralph",
    title = "{A Strong Maximum Principle for Weak Solutions of Quasi-Linear Elliptic Equations with Applications to Lorentzian and Riemannian Geometry}",
    eprint = "dg-ga/9707015",
    archivePrefix = "arXiv",
    doi = "{}",
    journal = "Commun. Pure App. Math.",
    volume = "L1",
    pages = "581-624",
    year = "1998"
}

@book{afshordi2024blackholesinside2024,
      title={Black Holes Inside and Out 2024: visions for the future of black hole physics}, 
      eprint={2410.14414},
      pages= {173-197},
      archivePrefix={arXiv},
      primaryClass={gr-qc},
      editor={Afshordi, Niayesh and others},
      Publisher = {arXiv},
      year={2024},
      url={https://arxiv.org/abs/2410.14414} 
}

@incollection{ashtekar2023regularblackholesloop,
      author={Abhay Ashtekar and Javier Olmedo and Parampreet Singh},
      eprint={2301.01309},
      archivePrefix  = {arXiv},
      PrimaryClass = {gr-qc},
      title={Regular black holes from Loop Quantum Gravity}, 
      editor = {Bambi, C.},
      publisher = {Springer},
      booktitle = {Regular Black Holes: Towards a New Paradigm of Gravitational Collapse},
      url={https://arxiv.org/abs/2301.01309}, 
      year={2023}
}

@article{Mourier:2020mwa,
    author = "Mourier, Pierre and Jim\'enez Forteza, Xisco and Pook-Kolb, Daniel and Krishnan, Badri and Schnetter, Erik",
    title = "{Quasinormal modes and their overtones at the common horizon in a binary black hole merger}",
    eprint = "2010.15186",
    archivePrefix = "arXiv",
    primaryClass = "gr-qc",
    doi = "10.1103/PhysRevD.103.044054",
    journal = "Phys. Rev. D",
    volume = "103",
    number = "4",
    pages = "044054",
    year = "2021"
}

@article{Khera:2023oyf,
    author = "Khera, Neev and Ribes Metidieri, Ariadna and Bonga, B\'eatrice and Jim\'enez Forteza, Xisco and Krishnan, Badri and Poisson, Eric and Pook-Kolb, Daniel and Schnetter, Erik and Yang, Huan",
    title = "{Nonlinear Ringdown at the Black Hole Horizon}",
    eprint = "2306.11142",
    archivePrefix = "arXiv",
    primaryClass = "gr-qc",
    doi = "10.1103/PhysRevLett.131.231401",
    journal = "Phys. Rev. Lett.",
    volume = "131",
    number = "23",
    pages = "231401",
    year = "2023"
}

@article{Chen_2022,
   title={Multipole moments on the common horizon in a binary-black-hole simulation},
   volume={106},
   ISSN={2470-0029},
   url={http://dx.doi.org/10.1103/PhysRevD.106.124045},
   DOI={10.1103/physrevd.106.124045},
   number={12},
   journal={Physical Review D},
   publisher={American Physical Society (APS)},
   author={Chen, Yitian and Kumar, Prayush and Khera, Neev and Deppe, Nils and Dhani, Arnab and Boyle, Michael and Giesler, Matthew and Kidder, Lawrence E. and Pfeiffer, Harald P. and Scheel, Mark A. and Teukolsky, Saul A.},
   year={2022},
   month=dec }

@article{Penrose:1965am,
	author = {Penrose, Roger},
	doi = {10.1098/rspa.1965.0058},
	journal = {Proc. Roy. Soc. Lond.},
	pages = {159},
	slaccitation = {%%CITATION = PRSLA,A284,159;%%},
	title = {{Zero rest mass fields including gravitation: Asymptotic behavior}},
	volume = {A284},
	year = {1965}
}

@article{ashtekar2025blackholeevaporationloop,
   title={Black Hole Evaporation in Loop Quantum Gravity},
    author={Abhay Ashtekar},
    volume={57},
   ISSN={},
   url={https://arxiv.org/abs/2502.04252},
   DOI={},
   number={},
   journal={Gen. Rel. Grav.},
   publisher={Springer},
   year={2025},
   month={}, pages={} 
}

@misc{Galloway:2025,
  author       = {Galloway, Gregory},
  title        = {Uniqueness of foliation by MTSs on time-like DHSs},
  month        = {February},
  year         = {2025},
  journal={},
  doi          = {},
  url          = {},
  note         = {Personal Communication to A. Ashtekar}
}

@article{Bengtsson_2011,
   title={Region with trapped surfaces in spherical symmetry, its core, and their boundaries},
   volume={83},
   ISSN={1550-2368},
   url={http://dx.doi.org/10.1103/PhysRevD.83.044012},
   DOI={10.1103/physrevd.83.044012},
   number={4},
   journal={Physical Review D},
   publisher={American Physical Society (APS)},
   author={Bengtsson, Ingemar and Senovilla, Jos\'e M. M.},
   year={2011},
   month=feb 
   }

@article{wald2001,
  author       = {Wald, Robert M.},
  title        = {The Thermodynamics of Black Holes},
  journal      = {Living Rev. Rel.},
  volume       = {4},
  number       = {6},
  pages        = {},
  year         = {2001},
  doi          = {10.12942/lrr-2001-6}
}

@article{Senovilla:2022bsn,
    author = "Senovilla, Jos{\'e} M. M.",
    title = "{Ultra-massive spacetimes}",
    eprint = "2209.14585",
    archivePrefix = "arXiv",
    primaryClass = "gr-qc",
    reportNumber = "YITP-22-104",
    doi = "10.4171/pm/2095",
    journal = "Portug. Math.",
    volume = "80",
    number = "1",
    pages = "133--155",
    year = "2023"
}

\end{document}